%% file: Salcedo.tex
\documentclass[12pt,fancy,BStimes,allowbreaks]{BSpaper}

\input{prelim.tex}

\begin{document}

	\BStitle

	%%%%%%%%%%%%%%%%%%%%%%%%%%%%%%%%%%%%%%%%%%%%%%%%%%%%%%%%%%%%%%%%%%%%%%%%%%%%%%%%%%%%%%%%%%%%%%%
	\input{intro.tex} %\label{sec:intro} who uses intro section anyways?
	
	%%%%%%%%%%%%%%%%%%%%%%%%%%%%%%%%%%%%%%%%%%%%%%%%%%%%%%%%%%%%%%%%%%%%%%%%%%%%%%%%%%%%%%%%%%%%%%%
	\section{Motivating example: a prisoner's dilemma}\label{sec:PD}
	\input{prisoners.tex}

	%%%%%%%%%%%%%%%%%%%%%%%%%%%%%%%%%%%%%%%%%%%%%%%%%%%%%%%%%%%%%%%%%%%%%%%%%%%%%%%%%%%%%%%%%%%%%%%
	\section{Interdependent-choice equilibrium and Nash implementation}\label{sec:model}
	\input{model.tex}

	%%%%%%%%%%%%%%%%%%%%%%%%%%%%%%%%%%%%%%%%%%%%%%%%%%%%%%%%%%%%%%%%%%%%%%%%%%%%%%%%%%%%%%%%%%%%%%%
	\section{Credible threats and the problem of perfection}\label{sec:ref}
	\input{sequential.tex}		

	%%%%%%%%%%%%%%%%%%%%%%%%%%%%%%%%%%%%%%%%%%%%%%%%%%%%%%%%%%%%%%%%%%%%%%%%%%%%%%%%%%%%%%%%%%%%%%%
	\section{Quasi-sequential implementation}\label{sec:PBI}
	\input{QSI.tex}

	%%%%%%%%%%%%%%%%%%%%%%%%%%%%%%%%%%%%%%%%%%%%%%%%%%%%%%%%%%%%%%%%%%%%%%%%%%%%%%%%%%%%%%%%%%%%%%%
	\section{Summary and discussion}\label{sec:dis}
	\input{discussion.tex}

	%%%%%%%%%%%%%%%%%%%%%%%%%%%%%%%%%%%%%%%%%%%%%%%%%%%%%%%%%%%%%%%%%%%%%%%%%%%%%%%%%%%%%%%%%%%%%%%
	\BSbiblio

	%%%%%%%%%%%%%%%%%%%%%%%%%%%%%%%%%%%%%%%%%%%%%%%%%%%%%%%%%%%%%%%%%%%%%%%%%%%%%%%%%%%%%%%%%%%%%%%
	\BSappendix
	\section{Extensive form mechanisms}\label{sec:EFM}
\input{EFM.tex}

	%%%%%%%%%%%%%%%%%%%%%%%%%%%%%%%%%%%%%%%%%%%%%%%%%%%%%%%%%%%%%%%%%%%%%%%%%%%%%%%%%%%%%%%%%%%%%%%
	\section{Proofs}\label{sec:proof}
\input{proofs.tex}

\end{document}

%% file: prelim.tex
\BSthanks{
	I wrote this paper under the invaluable guidance and supervision of Ed Green.
	I wish to gratefully acknowledge the comments and suggestions from
	Kalyan Chaterjee, Nail Kashaev, Vijay Krishna, Bulat Gafarov and Bruno Sultanum,
	as well as Wiroy Shin and the attendants of 
		the 1st Prospects in Economic Research Conference at Penn State, 
		the 2013 Midwest Economic Theory Meeting at Michigan State University,
		the 24th International Game Theory Conference at Stonny Brook,
		and the 2013 Asian Meeting of the Econometric Society at National University Singapore.
	All remaining errors are my own.}
\title{ Implementation without commitment in moral hazard environments%
		}
\date{November 2012\\[0.5ex] {Last revised: \today}}
\setKW{Interdependent choices \sep Sequential implementation \sep Mediation \sep Solution concepts \sep Robust predictions}
\setJEL{C72 \sep D86} % D74, D82} 
\setabstract{
	Interdependent-choice equilibrium is defined as an extension of correlated equilibrium
	in which the mediator is able to choose the timing of her signals,
	and observe the actions taken by the players. 
	The set of interdependent-choice  equilibria is a nonempty, closed and convex polytope.
	It characterizes all the outcomes that can be implemented in single shot interactions 
	without repetition, side payments, binding contracts or any other form of delegation.}
\usepackage{pst-node,pst-plot}
	
	\newcommand{\ua}{u}\newcommand{\uc}{U}\newcommand{\uz}{v}\newcommand{\us}{V}
	\newcommand{\uw}{\ushort{w}}
	\newcommand{\action}[1]{\BSmath{\mathrm{#1}}}
	\newcommand{\C}{\action{D}}\newcommand{\NC}{\action{C}}
	\newcommand{\shirk}{\action{S}}\newcommand{\work}{\action{W}}
	\newcommand{\aT}{\action{T}}\newcommand{\aB}{\action{B}}\newcommand{\aL}{\action{L}}\newcommand{\aR}{\action{R}}
	\newcommand{\G}{V}
	\newcommand{\A}{\mathcal{A}}
	\newcommand{\Nodes}{X}\newcommand{\node}{x}\newcommand{\Decision}{Y}\newcommand{\decision}{y}\newcommand{\Terminal}{Z}\newcommand{\terminal}{z}
	\newcommand{\Moves}{M}\newcommand{\move}{m}\newcommand{\mover}{\iota}\newcommand{\nat}{0}
	\renewcommand{\H}{\mathcal{H}}\newcommand{\h}{H}	% Information sets
	\newcommand{\disZ}{\zeta} \newcommand{\pre}{\leq}
	\newcommand{\tra}{\tau}\newcommand{\trainv}{\tra^{-1}}
	\newcommand{\rep}[1][\decision]{{\approx_{#1}}}
	\newcommand{\Pivotal}{\BSmath{D}}
	\newcommand{\ordn}{\BSmath{n}}\newcommand{\post}[1][i]{\BSmath{\ordn^+(#1)}}\newcommand{\prev}[1][i]{\BSmath{\ordn^-(#1)}}
	\newcommand{\ICE}{\mathrm{ICE}}\newcommand{\SE}{\mathrm{S}}
	\newcommand{\CEac}{A^{\infty}}\newcommand{\CEmac}{\alpha^{\ICE}}\newcommand{\B}{\mathrm{T}}
	\newcommand{\SEac}{{A^{\mathrm{S}}}}
	\newcommand{\CR}{\mathrm{CR}}\newcommand{\CRR}{A^{\CR}}\newcommand{\keep}{{\mathrm{K}}}
	\newcommand{\FR}{\mathrm{FCR}}\newcommand{\FLR}{A^\FR}
	\newcommand{\R}{A^\mathrm{R}}

\newcommand{\pl}{\BSmath{i}}\newcommand{\plo}{\BSmath{{-\pl}}}
\newcommand{\Ac}{\BSmath{A}}\newcommand{\Aci}{\BSmath{\Ac'}}\newcommand{\Acs}{\BSmath{\Ac^*}}
\newcommand{\ac}{\BSmath{a}}\newcommand{\aci}{\BSmath{\ac'}}\newcommand{\acs}{\BSmath{\ac^*}}
\newcommand{\mac}{\BSmath{\alpha}}
\newcommand{\Con}{{\Lambda}}\newcommand{\con}{{\lambda}}

\newcommand{\strat}{\BSmath{s}}
\newcommand{\Mixed}{\BSmath{\Sigma}}\newcommand{\mixed}{\BSmath{\sigma}}\newcommand{\mixeds}{\BSmath{\mixed^*}}
\newcommand{\Chance}{\BSmath{\Mixed_\nat}}\newcommand{\chance}{\BSmath{\mixed_\nat}}\newcommand{\chances}{\BSmath{\chance^*}}

%% file: intro.tex
%auto-ignore

%%%%%%%%%%%%%%%%%%%%%%%%%%%%%%%%%%%%%%%%%%%%%%%%%%%%%%%%%%%%%%%%%%%%%%%%%%%%%%%%%%%%%%%%%%%%%%%%%%%%%%%%%%%%%%%%%%%%%%%%%%%%%%%%%%%%%%%%%
%%%%%%%%%%%%%%%%%%%%%%%%%%%%%%%%%%%%%%%%%%%%%%%%%%%%%%%%%%%%%%%%%%%%%%%%%%%%%%%%%%%%%%%%%%%%%%%%%%%%%%%%%%%%%%%%%%%%%%%%%%%%%%%%%%%%%%%%%
% Section: Introduction
% Project: Coordination in Strategic Environments
%  Author: Bruno Salcedo
% Version: 7.0
%    Date: 09/06/13
%%%%%%%%%%%%%%%%%%%%%%%%%%%%%%%%%%%%%%%%%%%%%%%%%%%%%%%%%%%%%%%%%%%%%%%%%%%%%%%%%%%%%%%%%%%%%%%%%%%%%%%%%%%%%%%%%%%%%%%%%%%%%%%%%%%%%%%%%
%%%%%%%%%%%%%%%%%%%%%%%%%%%%%%%%%%%%%%%%%%%%%%%%%%%%%%%%%%%%%%%%%%%%%%%%%%%%%%%%%%%%%%%%%%%%%%%%%%%%%%%%%%%%%%%%%%%%%%%%%%%%%%%%%%%%%%%%%

%%%%%%%%%%%%%%%%%%%%%%%%%%%%%%%%%%%%%%%%%%%%%%%%%%%%%%%%%%%%%%%%%%%%%%%%%%%%%%%%%%%%%%%%%%%%%%%%%%%%%%%%%%%%%%%%%%%%%%%%%%%%%%%%%%%%%%%%%
%%%%%%%%%%%%%%%%%%%%%%%%%%%%%%%%%%%%%%%%%%%%%%%%%%%%%%%%%%%%%%%%%%%%%%%%%%%%%%%%%%%%%%%%%%%%%%%%%%%%%%%%%%%%%%%%%%%%%%%%%%%%%%%%%%%%%%%%%
When agents make choices independently, standard notions of rationality postulate that
each agent forms a belief about his opponent's behavior, and then chooses a \emph{myopic} best response.
%%%
The story is quite difference once the choices of some agents may depend on the choices of others. 
%%%
In such settings, it is not sufficient for agents to consider the material consequences of their choices
taking the behavior of their opponents as given. 
%%%
Each agent must also consider the way that his opponents will react depending on his own choice. 
%%%
A particularly important form of reaction or counterfactual reasoning is reciprocity:
an agent may believe that others will be god to him \emph{if and only if} he is good to them. 

%%%%%%%%%%%%%%%%%%%%%%%%%%%%%%%%%%%%%%%%%%%%%%%%%%%%%%%%%%%%%%%%%%%%%%%%%%%%%%%%%%%%%%%%%%%%%%%%%%%%%%%%%%%%%%%%%%%%%%%%%%%%%%%%%%%%%%%%%
%%%%%%%%%%%%%%%%%%%%%%%%%%%%%%%%%%%%%%%%%%%%%%%%%%%%%%%%%%%%%%%%%%%%%%%%%%%%%%%%%%%%%%%%%%%%%%%%%%%%%%%%%%%%%%%%%%%%%%%%%%%%%%%%%%%%%%%%%
Choice interdependence is relevant in moral hazard environments with no Pareto efficient Nash equilibria.
%%%
With complete information, moral hazard is completely eliminated when agents can enforce complete contracts (Coase theorem),
or when they interact repeatedly and are patient enough (folk theorems).
%%%
This is possible because written contracts or publicly observed histories serve as coordination devices allowing for interdependence.
%%%
The purpose of this paper is to abstract the notion of choice interdependence from such settings,
and investigate the extent to which its power remains in settings without: 
\emph{commitment}, \emph{repetition}, or \emph{side payments}.%
	\footnote{Different papers ask related questions under different assumptions,
		usually allowing for different forms of commitment, repetition or transfers,
		or departing from mainstream notions of rationality.
		Relevant works are discussed in section \ref{ssec:lit}.}
%		Relevant works include 
%		\cite{
%			RapAl,hofstadter96,halpern10,halpern12,
%			MouVia,forgo10,okada93,jackson05,
%			kalai81,bade09,renou09,kalai10,
%			tennenholtz04,MonTen09,
%			levine07,figuieres,eisert99}. %dufwenberg04
%			See \S\ref{sec:lit} for further discussion.} 
%%% 
In short, I find that a lot but not all of it remains.

%%%%%%%%%%%%%%%%%%%%%%%%%%%%%%%%%%%%%%%%%%%%%%%%%%%%%%%%%%%%%%%%%%%%%%%%%%%%%%%%%%%%%%%%%%%%%%%%%%%%%%%%%%%%%%%%%%%%%%%%%%%%%%%%%%%%%%%%%
%%%%%%%%%%%%%%%%%%%%%%%%%%%%%%%%%%%%%%%%%%%%%%%%%%%%%%%%%%%%%%%%%%%%%%%%%%%%%%%%%%%%%%%%%%%%%%%%%%%%%%%%%%%%%%%%%%%%%%%%%%%%%%%%%%%%%%%%%
In order to answer this question, I follow the methodology from \cite{forges86}.
%%%
First, I introduce a class of canonical mechanisms called \emph{mediated games},
in which a non-strategic mediator manages the play through private recommendations. 
%%%
\emph{Interdependent-choice equilibrium} (ICE) is defined as the set of distributions that can result from
Nash incentive compatible mediated games.
%%%
The main result is that the class of mediated games is complete,
in that every equilibrium outcome of any mechanism consistent with the assumptions
of no commitment, no repetition and no side payments, is an ICE. 

%%%%%%%%%%%%%%%%%%%%%%%%%%%%%%%%%%%%%%%%%%%%%%%%%%%%%%%%%%%%%%%%%%%%%%%%%%%%%%%%%%%%%%%%%%%%%%%%%%%%%%%%%%%%%%%%%%%%%%%%%%%%%%%%%%%%%%%%%
%%%%%%%%%%%%%%%%%%%%%%%%%%%%%%%%%%%%%%%%%%%%%%%%%%%%%%%%%%%%%%%%%%%%%%%%%%%%%%%%%%%%%%%%%%%%%%%%%%%%%%%%%%%%%%%%%%%%%%%%%%%%%%%%%%%%%%%%%
Related literatures take as given either the sequential structure of choices, the information structure, or both.%
	\footnote{
		For example, papers on communication \citep{aumann87,myerson86,forges86},
		espionage games \citep{solan04}, and endogenous leadership \citep{vandamme}.}
%%%
In contrast, I allow the mediator to determine the timing of signals and the order in which actions are taken,%
	\footnote{
		Allowing the mediator to control the timing of \emph{choices} might be a strong assumption.
		%%%
		This lax approach helps to capture full potential of choice interdependence. 
		%%%
		The model can be adapted to settings in which the mediator can only choose the timing of her \emph{signals},
		see \S\ref{disc:alternative}.}
to observe the actions taken by the players, and to choose signals conditional on her observation. 
%%%
These enables some reciprocal strategies.
%%%
If a player deviates from the intended plan, and he is not the last player to move,
the mediator can recommend the remaining players to \emph{punish} him. 
%%%
As a result the set of implementable outcomes can be considerably large.
%%%
However, no ``folk theorem'' is obtained because it is only possible 
to generate incentives for early movers according to the chosen order. 

%%%%%%%%%%%%%%%%%%%%%%%%%%%%%%%%%%%%%%%%%%%%%%%%%%%%%%%%%%%%%%%%%%%%%%%%%%%%%%%%%%%%%%%%%%%%%%%%%%%%%%%%%%%%%%%%%%%%%%%%%%%%%%%%%%%%%%%%%
%%%%%%%%%%%%%%%%%%%%%%%%%%%%%%%%%%%%%%%%%%%%%%%%%%%%%%%%%%%%%%%%%%%%%%%%%%%%%%%%%%%%%%%%%%%%%%%%%%%%%%%%%%%%%%%%%%%%%%%%%%%%%%%%%%%%%%%%%
ICE is defined in terms of Nash incentive constraints, 
\emph{as if} the players could commit to punish deviations off the equilibrium path. 
%%%
In order to comply with the no-commitment requirement, it is thus important to take into account the problem of perfection. 
%%%
Sections \ref{sec:ref} and \ref{sec:PBI} provide sufficient conditions for sequential implementation, 
as well as complete characterizations of sequential implementation in $2\times 2$ environments,
and a new notion of \emph{quasi-sequential} implementation in arbitrary environments.
%%%
The conditions are tractable and have a relatively small impact.
%%%
For example, if there are no strictly dominated actions, every ICE
can be implemented as a sequential equilibrium of a mediated game. 

%% file: prisoners.tex
%auto-ignore

%%%%%%%%%%%%%%%%%%%%%%%%%%%%%%%%%%%%%%%%%%%%%%%%%%%%%%%%%%%%%%%%%%%%%%%%%%%%%%%%%%%%%%%%%%%%%%%%%%%%%%%%%%%%%%%%%%%%%%%%%%%%%%%%%%%%%%%%%
%%%%%%%%%%%%%%%%%%%%%%%%%%%%%%%%%%%%%%%%%%%%%%%%%%%%%%%%%%%%%%%%%%%%%%%%%%%%%%%%%%%%%%%%%%%%%%%%%%%%%%%%%%%%%%%%%%%%%%%%%%%%%%%%%%%%%%%%%
% Section: Prisoners Dilemma
% Project: Coordination in Strategic Environments
%  Author: Bruno Salcedo
% Version: 7.0
%    Date: 09/06/13
%%%%%%%%%%%%%%%%%%%%%%%%%%%%%%%%%%%%%%%%%%%%%%%%%%%%%%%%%%%%%%%%%%%%%%%%%%%%%%%%%%%%%%%%%%%%%%%%%%%%%%%%%%%%%%%%%%%%%%%%%%%%%%%%%%%%%%%%%
%%%%%%%%%%%%%%%%%%%%%%%%%%%%%%%%%%%%%%%%%%%%%%%%%%%%%%%%%%%%%%%%%%%%%%%%%%%%%%%%%%%%%%%%%%%%%%%%%%%%%%%%%%%%%%%%%%%%%%%%%%%%%%%%%%%%%%%%%
\cite{nishihara97,nishihara99} provides an illustrative example showing that 
cooperation in a prisoner's dilemma can sometimes be implemented without contracts,side payments or repetition.
%%%
The salient features of Nishihara's model are that players are uncertain about the order of choices,
and the information structure allows players to recognize and react to past defections.
%%%
This section describes a form to generate Nishinara's information structure using a typical mediation mechanism. 

%%%%%%%%%%%%%%%%%%%%%%%%%%%%%%%%%%%%%%%%%%%%%%%%%%%%%%%%%%%%%%%%%%%%%%%%%%%%%%%%%%%%%%%%%%%%%%%%%%%%%%%%%%%%%%%%%%%%%%%%%%%%%%%%%%%%%%%%%
%%%%%%%%%%%%%%%%%%%%%%%%%%%%%%%%%%%%%%%%%%%%%%%%%%%%%%%%%%%%%%%%%%%%%%%%%%%%%%%%%%%%%%%%%%%%%%%%%%%%%%%%%%%%%%%%%%%%%%%%%%%%%%%%%%%%%%%%%
Suppose that two suspects of a crime are arrested.
%%%
The DA has enough evidence to convict them of a misdemeanor but requires a written confession to convict them for the alleged crime.
%%%
The DA then offers each prisoner a sentence reduction in exchange for a confession.
%%%
Each of the prisoners has to choose whether to behave cooperatively (\NC) by remaining silent or to defect (\C) by confessing.
%%%
Their preferences are summarized by the payoff matrix in Figure \ref{fig:PD}, where $B<b<g<G$.

%%%%%%%%%%%%%%%%%%%%%%%%%%%%%%%%%%%%%%%%%%%%%%%%%%%%%%%%%%%%%%%%%%%%%%%%%%%%%%%%%%%%%%%%%%%%%%%%%%%%%%%%%%%%%%%%%%%%%%%%%%%%%%%%%%%%%%%%%
%%%%%%%%%%%%%%%%%%%%%%%%%%%%%%%%%%%%%%%%%%%%%%%%%%%%%%%%%%%%%%%%%%%%%%%%%%%%%%%%%%%%%%%%%%%%%%%%%%%%%%%%%%%%%%%%%%%%%%%%%%%%%%%%%%%%%%%%%
\begin{figure}[htb]
\centering
\begin{tabular}{c|cc}
	      &   \NC     &    \C     \\
	\hline\\[-2ex]
	\NC & $g \:,\: g$ & $B \:,\: G$ \\[1ex]
	\C  & $G \:,\: B$ & $b \:,\: b$
\end{tabular}
\caption{Payoff matrix for the prisoner's dilemma}\label{fig:PD}
\end{figure}

%%%%%%%%%%%%%%%%%%%%%%%%%%%%%%%%%%%%%%%%%%%%%%%%%%%%%%%%%%%%%%%%%%%%%%%%%%%%%%%%%%%%%%%%%%%%%%%%%%%%%%%%%%%%%%%%%%%%%%%%%%%%%%%%%%%%%%%%%
%%%%%%%%%%%%%%%%%%%%%%%%%%%%%%%%%%%%%%%%%%%%%%%%%%%%%%%%%%%%%%%%%%%%%%%%%%%%%%%%%%%%%%%%%%%%%%%%%%%%%%%%%%%%%%%%%%%%%%%%%%%%%%%%%%%%%%%%%
In the story told, there is no reason to assume that players will have to make a decision at exactly the same time.
%%%
Also, even if the prisoners cannot directly communicate with each other,
it is by no means clear that their choices need to be independent.
%%%
Different forms of interdependence could either arise naturally or be artificially constructed.
%%%
Even so, implementing cooperation remains a non-trivial task because the decision to confess cannot be delegated,
and the legal obligation to confess prevents the enforcement of contracts that would bind them to remain silent.
%%%
However, they could hire a lawyer who would schedule  and be present in all the negotiations with the DA
and instruct him as follows:

%%%%%%%%%%%%%%%%%%%%%%%%%%%%%%%%%%%%%%%%%%%%%%%%%%%%%%%%%%%%%%%%%%%%%%%%%%%%%%%%%%%%%%%%%%%%%%%%%%%%%%%%%%%%%%%%%%%%%%%%%%%%%%%%%%%%%%%%%
%%%%%%%%%%%%%%%%%%%%%%%%%%%%%%%%%%%%%%%%%%%%%%%%%%%%%%%%%%%%%%%%%%%%%%%%%%%%%%%%%%%%%%%%%%%%%%%%%%%%%%%%%%%%%%%%%%%%%%%%%%%%%%%%%%%%%%%%%
\begin{quote}
	``You must randomize uniformly the order of our meetings. 
	%%%
	If the DA offers us a (prisoner's dilemma) deal you must always recommend that we do not confess,
	unless one of us has already confessed, in which case you must recommend that we do confess.
	%%%
	Other than those recommendations, you must not provide us with any additional information.''
\end{quote}

%%%%%%%%%%%%%%%%%%%%%%%%%%%%%%%%%%%%%%%%%%%%%%%%%%%%%%%%%%%%%%%%%%%%%%%%%%%%%%%%%%%%%%%%%%%%%%%%%%%%%%%%%%%%%%%%%%%%%%%%%%%%%%%%%%%%%%%%%
%%%%%%%%%%%%%%%%%%%%%%%%%%%%%%%%%%%%%%%%%%%%%%%%%%%%%%%%%%%%%%%%%%%%%%%%%%%%%%%%%%%%%%%%%%%%%%%%%%%%%%%%%%%%%%%%%%%%%%%%%%%%%%%%%%%%%%%%%
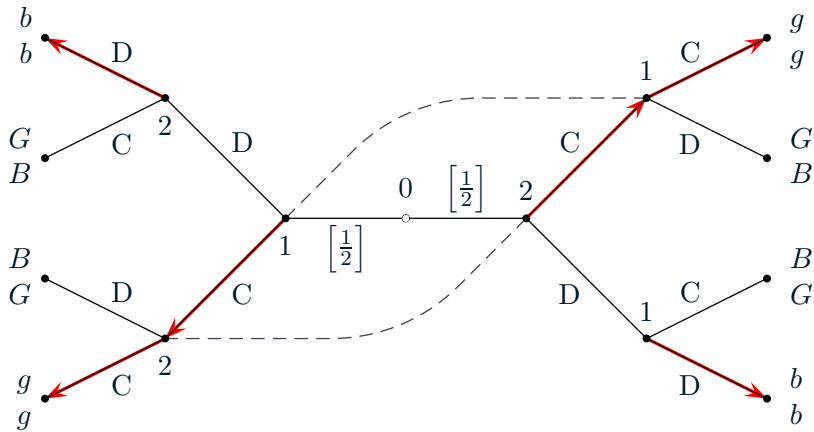
\begin{figure}[htb]
\centering
\psset{unit=8mm}
\begin{pspicture}(0,-1)(14,7)
	%\psgrid%
	% Equilibrium path 
	\psset{linewidth=1.2pt,linecolor=red,arrows=->}
		\psline(5,3)(3,1)
		\psline(3,1)(1,0)
		\psline(9,3)(11,5)
		\psline(11,5)(13,6)
		\psline(3,5)(1,6)
		\psline(11,1)(13,0)
	\psreset%
	% Branches
		\psline(1,6)(3,5)(1,4)
		\psline(1,2)(3,1)(1,0)
		\psline(13,6)(11,5)(13,4)
		\psline(13,2)(11,1)(13,0)
		\psline(11,5)(9,3)(11,1)
		\psline(3,5)(5,3)(3,1)
		\psline(5,3)(9,3)
	% Information sets
	\psset{linearc=3,linestyle=dashed,linecolor=BSCdark}
		\psline(5,3)(7,5)(11,5)
		\psline(9,3)(7,1)(3,1)
	\psreset%
	% Nodes
	\psset{fillcolor=white}
		\psdots[dotstyle=o](7,3)
		\psdots(1,6)(1,4)(1,2)(1,0)(3,5)(3,1)(5,3)
		\psdots(9,3)(11,5)(11,1)(13,6)(13,4)(13,2)(13,0)
	\psreset%
	% Labels
	\small
	\rput(7,3.5){{$0$}}
	\rput[t](6,2.9){{$\left[\frac{1}{2}\right]$}}
	\rput[b](8,3.1){{$\left[\frac{1}{2}\right]$}}
	\rput[t](5,2.7){{$1$}}
	\rput[t](3,4.7){{$2$}}
	\rput[t](3,0.7){{$2$}}
	\rput[b](9,3.3){{$2$}}
	\rput[b](11,1.3){{$1$}}
	\rput[b](11,5.3){{$1$}}
	\rput[lb](4.1,4.1){{\C}}
	\rput[lb](2.1,5.6){{\C}}
	\rput[lb](2.1,1.6){{\C}}
	\rput[tr](9.9,1.9){{\C}}
	\rput[tr](11.9,4.4){{\C}}
	\rput[tr](11.9,0.4){{\C}}
	\rput[br](9.9,4.1){{\NC}}
	\rput[br](11.9,1.6){{\NC}}
	\rput[br](11.9,5.6){{\NC}}
	\rput[tl](4.1,1.9){{\NC}}
	\rput[tl](2.1,4.4){{\NC}}
	\rput[tl](2.1,0.4){{\NC}}
	\rput[r](1,6){ $\begin{array}{c} b\\ b\end{array}$}
	\rput[r](1,4){ $\begin{array}{c} G\\ B\end{array}$}
	\rput[r](1,2){ $\begin{array}{c} B\\ G\end{array}$}
	\rput[r](1,0){ $\begin{array}{c} g\\ g\end{array}$}
	\rput[l](13,6){ $\begin{array}{c} g\\ g\end{array}$}
	\rput[l](13,4){ $\begin{array}{c} G\\ B\end{array}$}
	\rput[l](13,2){ $\begin{array}{c} B\\ G\end{array}$}
	\rput[l](13,0){ $\begin{array}{c} b\\ b\end{array}$}
\end{pspicture}
\caption{A sequential mechanism for the prisoner's dilemma}
\label{fig:PDmech}
\end{figure}

%%%%%%%%%%%%%%%%%%%%%%%%%%%%%%%%%%%%%%%%%%%%%%%%%%%%%%%%%%%%%%%%%%%%%%%%%%%%%%%%%%%%%%%%%%%%%%%%%%%%%%%%%%%%%%%%%%%%%%%%%%%%%%%%%%%%%%%%%
%%%%%%%%%%%%%%%%%%%%%%%%%%%%%%%%%%%%%%%%%%%%%%%%%%%%%%%%%%%%%%%%%%%%%%%%%%%%%%%%%%%%%%%%%%%%%%%%%%%%%%%%%%%%%%%%%%%%%%%%%%%%%%%%%%%%%%%%%
The resulting situation can be described by the extensive form game in Figure \ref{fig:PDmech}.
%%%
In the event that the first prisoner to move confesses, the second prisoner will be informed of this choice before making his own.
%%%
If the first player decides to cooperate, the second prisoner will remain uninformed about which of the two following events is true:
%%%
(\emph{i}) the event in which he is the first prisoner to receive the offer, 
%%%
and (\emph{ii}) the event in which he is the second one and his accomplice remained silent.

%%%%%%%%%%%%%%%%%%%%%%%%%%%%%%%%%%%%%%%%%%%%%%%%%%%%%%%%%%%%%%%%%%%%%%%%%%%%%%%%%%%%%%%%%%%%%%%%%%%%%%%%%%%%%%%%%%%%%%%%%%%%%%%%%%%%%%%%%
%%%%%%%%%%%%%%%%%%%%%%%%%%%%%%%%%%%%%%%%%%%%%%%%%%%%%%%%%%%%%%%%%%%%%%%%%%%%%%%%%%%%%%%%%%%%%%%%%%%%%%%%%%%%%%%%%%%%%%%%%%%%%%%%%%%%%%%%%
The strategies represented with arrows support full cooperation and constitute an equilibrium as long as $G-g \leq g-b$.
%%%
That is, as long as the benefit that a player can obtain from unilaterally deviating from (\NC,\NC)
is less or equal to the inefficiency of both players confessing.
%%%
This is possible because, along the equilibrium path, each prisoner assigns sufficient probability to the event in which:
%%%
(\emph{i}) if he cooperates,  then his accomplice will remain uninformed and will also cooperate;
%%%
and (\emph{ii}) if he confesses, then his accomplice will learn of his defection and will punish him by also confessing.

%% file: model.tex
%auto-ignore

%%%%%%%%%%%%%%%%%%%%%%%%%%%%%%%%%%%%%%%%%%%%%%%%%%%%%%%%%%%%%%%%%%%%%%%%%%%%%%%%%%%%%%%%%%%%%%%%%%%%%%%%%%%%%%%%%%%%%%%%%%%%%%%%%%%%%%%%%
%%%%%%%%%%%%%%%%%%%%%%%%%%%%%%%%%%%%%%%%%%%%%%%%%%%%%%%%%%%%%%%%%%%%%%%%%%%%%%%%%%%%%%%%%%%%%%%%%%%%%%%%%%%%%%%%%%%%%%%%%%%%%%%%%%%%%%%%%
% Section: Nash implementation
% Project: Coordination in Strategic Environments
%  Author: Bruno Salcedo
% Version: 7.1
%    Date: 09/07/13
%%%%%%%%%%%%%%%%%%%%%%%%%%%%%%%%%%%%%%%%%%%%%%%%%%%%%%%%%%%%%%%%%%%%%%%%%%%%%%%%%%%%%%%%%%%%%%%%%%%%%%%%%%%%%%%%%%%%%%%%%%%%%%%%%%%%%%%%%
%%%%%%%%%%%%%%%%%%%%%%%%%%%%%%%%%%%%%%%%%%%%%%%%%%%%%%%%%%%%%%%%%%%%%%%%%%%%%%%%%%%%%%%%%%%%%%%%%%%%%%%%%%%%%%%%%%%%%%%%%%%%%%%%%%%%%%%%%

%%%%%%%%%%%%%%%%%%%%%%%%%%%%%%%%%%%%%%%%%%%%%%%%%%%%%%%%%%%%%%%%%%%%%%%%%%%%%%%%%%%%%%%%%%%%%%%%%%%%%%%%%%%%%%%%%%%%%%%%%%%%%%%%%%%%%%%%%
%%%%%%%%%%%%%%%%%%%%%%%%%%%%%%%%%%%%%%%%%%%%%%%%%%%%%%%%%%%%%%%%%%%%%%%%%%%%%%%%%%%%%%%%%%%%%%%%%%%%%%%%%%%%%%%%%%%%%%%%%%%%%%%%%%%%%%%%%
The environment is described by a tuple  $E=(I,A,u)$.
%%%
It represents a situation in which players $i \in I = \{1,2,\ldots,n\}$ are to make decisions.
%%%
For exposition purposes I assume $n=2$, see \S\ref{disc:many} for hte general case. 
%%%
Each player $i$ is to choose and perform one and only one action from a finite set $A_i = \{a_i,a'_i,\ldots\}$.
%%%
$i$'s preferences over action profiles are represented by $u_i:A\rightarrow\Real$.%
	\footnote{I employ the notation $-i$ for $i$'s opponent, $a\in A =\cart_i A_i$ for action profiles,
				$\alpha\in\Delta(A)$ for joint distributions, $\alpha_i\in\Delta(A_i)$ for marginal distributions, 
				and $\alpha(\blank|a_i)\in\Delta(A_{-i})$ for conditional distributions.}

%%%%%%%%%%%%%%%%%%%%%%%%%%%%%%%%%%%%%%%%%%%%%%%%%%%%%%%%%%%%%%%%%%%%%%%%%%%%%%%%%%%%%%%%%%%%%%%%%%%%%%%%%%%%%%%%%%%%%%%%%%%%%%%%%%%%%%%%%
%%%%%%%%%%%%%%%%%%%%%%%%%%%%%%%%%%%%%%%%%%%%%%%%%%%%%%%%%%%%%%%%%%%%%%%%%%%%%%%%%%%%%%%%%%%%%%%%%%%%%%%%%%%%%%%%%%%%%%%%%%%%%%%%%%%%%%%%%
Such description is only a \emph{partial} characterization of the environment.
%%%
It says nothing about the order in which choices will be made,
nor about the information that each player will have at the moment of making his choice.
In particular, it is not assumed that choices are independent or simultaneous.

%%%%%%%%%%%%%%%%%%%%%%%%%%%%%%%%%%%%%%%%%%%%%%%%%%%%%%%%%%%%%%%%%%%%%%%%%%%%%%%%%%%%%%%%%%%%%%%%%%%%%%%%%%%%%%%%%%%%%%%%%%%%%%%%%%%%%%%%%
%%%%%%%%%%%%%%%%%%%%%%%%%%%%%%%%%%%%%%%%%%%%%%%%%%%%%%%%%%%%%%%%%%%%%%%%%%%%%%%%%%%%%%%%%%%%%%%%%%%%%%%%%%%%%%%%%%%%%%%%%%%%%%%%%%%%%%%%%
%%%%%%%%%%%%%%%%%%%%%%%%%%%%%%%%%%%%%%%%%%%%%%%%%%%%%%%%%%%%%%%%%%%%%%%%%%%%%%%%%%%%%%%%%%%%%%%%%%%%%%%%%%%%%%%%%%%%%%%%%%%%%%%%%%%%%%%%%
%%%%%%%%%%%%%%%%%%%%%%%%%%%%%%%%%%%%%%%%%%%%%%%%%%%%%%%%%%%%%%%%%%%%%%%%%%%%%%%%%%%%%%%%%%%%%%%%%%%%%%%%%%%%%%%%%%%%%%%%%%%%%%%%%%%%%%%%%
%%%%%%%%%%%%%%%%%%%%%%%%%%%%%%%%%%%%%%%%%%%%%%%%%%%%%%%%%%%%%%%%%%%%%%%%%%%%%%%%%%%%%%%%%%%%%%%%%%%%%%%%%%%%%%%%%%%%%%%%%%%%%%%%%%%%%%%%%
%%%%%%%%%%%%%%%%%%%%%%%%%%%%%%%%%%%%%%%%%%%%%%%%%%%%%%%%%%%%%%%%%%%%%%%%%%%%%%%%%%%%%%%%%%%%%%%%%%%%%%%%%%%%%%%%%%%%%%%%%%%%%%%%%%%%%%%%%
%%%%%%%%%%%%%%%%%%%%%%%%%%%%%%%%%%%%%%%%%%%%%%%%%%%%%%%%%%%%%%%%%%%%%%%%%%%%%%%%%%%%%%%%%%%%%%%%%%%%%%%%%%%%%%%%%%%%%%%%%%%%%%%%%%%%%%%%%
%%%%%%%%%%%%%%%%%%%%%%%%%%%%%%%%%%%%%%%%%%%%%%%%%%%%%%%%%%%%%%%%%%%%%%%%%%%%%%%%%%%%%%%%%%%%%%%%%%%%%%%%%%%%%%%%%%%%%%%%%%%%%%%%%%%%%%%%%
%%%%%%%%%%%%%%%%%%%%%%%%%%%%%%%%%%%%%%%%%%%%%%%%%%%%%%%%%%%%%%%%%%%%%%%%%%%%%%%%%%%%%%%%%%%%%%%%%%%%%%%%%%%%%%%%%%%%%%%%%%%%%%%%%%%%%%%%%
%%%%%%%%%%%%%%%%%%%%%%%%%%%%%%%%%%%%%%%%%%%%%%%%%%%%%%%%%%%%%%%%%%%%%%%%%%%%%%%%%%%%%%%%%%%%%%%%%%%%%%%%%%%%%%%%%%%%%%%%%%%%%%% ssec: ICE
\subsection{Interdependent-choice equilibrium}\label{ssec:ICE}

%%%%%%%%%%%%%%%%%%%%%%%%%%%%%%%%%%%%%%%%%%%%%%%%%%%%%%%%%%%%%%%%%%%%%%%%%%%%%%%%%%%%%%%%%%%%%%%%%%%%%%%%%%%%%%%%%%%%%%%%%%%%%%%%%%%%%%%%%
%%%%%%%%%%%%%%%%%%%%%%%%%%%%%%%%%%%%%%%%%%%%%%%%%%%%%%%%%%%%%%%%%%%%%%%%%%%%%%%%%%%%%%%%%%%%%%%%%%%%%%%%%%%%%%%%%%%%%%%%%%%%%%%%%%%%%%%%%
\emph{Interdependent-choice equilibrium} (ICE) is defined in terms of a simple class of extensive form games in which 
a non-strategic mediator manages the players through private recommendations.
%%%
A (sequentially) \emph{mediated mechanism} is a tuple $(\alpha,\theta,B)$.
%%%
$\alpha\in\Delta(A)$ is a distribution over action profiles to be implemented.
%%%
$\theta: A\rightarrow \Delta(I)$ specifies a distribution over the order in which players will move,
conditional on the action profile to be implemented.
%%%
$\theta(i|a)$ is the probability that player $i$ will be the first player to move, conditional on $a$ being chosen.
%%%
$B=\cart_i B_i$ specifies actions that can be recommended as \emph{additional} credible threats. 
%%%
The \emph{effective} set of credible threats $B_i^* = B_i \cup \supp \alpha_i$ also includes the actions played along the equilibrium path.

%%%%%%%%%%%%%%%%%%%%%%%%%%%%%%%%%%%%%%%%%%%%%%%%%%%%%%%%%%%%%%%%%%%%%%%%%%%%%%%%%%%%%%%%%%%%%%%%%%%%%%%%%%%%%%%%%%%%%%%%%%%%%%%%%%%%%%%%%
%%%%%%%%%%%%%%%%%%%%%%%%%%%%%%%%%%%%%%%%%%%%%%%%%%%%%%%%%%%%%%%%%%%%%%%%%%%%%%%%%%%%%%%%%%%%%%%%%%%%%%%%%%%%%%%%%%%%%%%%%%%%%%%%%%%%%%%%%
The tuple $(\alpha,\theta,B)$ characterizes the extensive form game described as follows.
%%%
The game begins with the mediator privately choosing the action profile $a^*$ that she wants to implement (according to $\alpha$),
and the player $i^*$ to move first (according to $\theta(\blank|a^*)$).
%%%
She then ``visits'' each of the players one by one, visiting $i^*$ first and $-i^*$ second.
%%%
When visiting each player $i$, the mediator recommends an action $a_i^r$, and observes the action actually taken $a_i^p$.
%%%
At the moment of making their choices, the players do not possess any information other than the recommendation they receive. 
%%%
The mediator always recommends the intended action to the first player, i.e.\ $a^r_{i^*} = a^*_{i^*}$. 
%%%
She recommends the intended action to the second player if the first player complied,
and one of the worst available punishments in $B^*_{-i^*}$ otherwise, i.e.:
%%%
\begin{align*}
	\begin{array}{ccc}
	a^r_{-i^*} = a^*_{-i^*} &\text{  if }& a^p_{i^*} = a^*_{i^*} \\[1ex]
	a^r_{-i^*}\in\displaystyle\argmin_{a_{-i^*}\in B^*_{-i^*}} u_{i^*} \big(a^p_{i^*},a_{-i^*}\big)
		&\text{ if }& a^p_{i^*} \neq a^*_{i^*}
	\end{array}
\end{align*}

%%%%%%%%%%%%%%%%%%%%%%%%%%%%%%%%%%%%%%%%%%%%%%%%%%%%%%%%%%%%%%%%%%%%%%%%%%%%%%%%%%%%%%%%%%%%%%%%%%%%%%%%%%%%%%%%%%%%%%%%%%%%%%%%%%%%%%%%%
%%%%%%%%%%%%%%%%%%%%%%%%%%%%%%%%%%%%%%%%%%%%%%%%%%%%%%%%%%%%%%%%%%%%%%%%%%%%%%%%%%%%%%%%%%%%%%%%%%%%%%%%%%%%%%%%%%%%%%%%%%%%%%%%%%%%%%%%%
A mediated mechanism is incentive compatible if and only if following the mediator's recommendations constitutes a Nash equilibrium.
%%%
Since only Nash incentive compatibility is required,
there are no incentive constraints for the punishments (which occur off the equilibrium path). 
%%%
Incentive compatibility is thus characterized by requiring that for every player $i\in I$ and every pair of actions $a_i,a'_i\in A_i$:
%%%
\begin{align} \label{eqn:ICE}
	\sum_{a_{-i}\in A_{-i}} \alpha(a_i,a_{-i})
		\Big(
			u_i(a_i,a_{-i}) 
			- \big(1-\theta(i|a)\big) u_i(a'_i,a_{-i}) 
			- \theta(i|a)\uw_i(a_i'|B_{-i}^*) 
		\Big)\geq 0
\end{align}
%%%
where $\ushort{w}_i(a_i'|B_{-i}^*) \equiv \displaystyle\min_{a_{-i}\in B^*_{-i}} u_{i} (a'_{i},a_{-i})$.

%%%%%%%%%%%%%%%%%%%%%%%%%%%%%%%%%%%%%%%%%%%%%%%%%%%%%%%%%%%%%%%%%%%%%%%%%%%%%%%%%%%%%%%%%%%%%%%%%%%%%%%%%%%%%%%%%%%%%%%%%%%%%%%%%%%%%%%%%
%%%%%%%%%%%%%%%%%%%%%%%%%%%%%%%%%%%%%%%%%%%%%%%%%%%%%%%%%%%%%%%%%%%%%%%%%%%%%%%%%%%%%%%%%%%%%%%%%%%%%%%%%%%%%%%%%%%%%%%%%%%%%%%%%%%%%%%%%
\begin{definition}\label{defn:ICE}
	A distribution over action profiles $\alpha\in\Delta(A)$
	is an \emph{interdependent-choice equilibrium} with respect to a set of credible threats $B$,
	if and only if there exists some conditional ordering distribution $\theta$
	such that $(\alpha,\theta,B)$ is incentive compatible.
\end{definition}

%%%%%%%%%%%%%%%%%%%%%%%%%%%%%%%%%%%%%%%%%%%%%%%%%%%%%%%%%%%%%%%%%%%%%%%%%%%%%%%%%%%%%%%%%%%%%%%%%%%%%%%%%%%%%%%%%%%%%%%%%%%%%%%%%%%%%%%%%
%%%%%%%%%%%%%%%%%%%%%%%%%%%%%%%%%%%%%%%%%%%%%%%%%%%%%%%%%%%%%%%%%%%%%%%%%%%%%%%%%%%%%%%%%%%%%%%%%%%%%%%%%%%%%%%%%%%%%%%%%%%%%%%%%%%%%%%%%
Let $\ICE(B)$ denote the set of ICE with respect to $B$.
%%%
When $B=A$, I omit the reference to the set of credible threats, and simply say that $\alpha\in \ICE$ is an ICE.

%%%%%%%%%%%%%%%%%%%%%%%%%%%%%%%%%%%%%%%%%%%%%%%%%%%%%%%%%%%%%%%%%%%%%%%%%%%%%%%%%%%%%%%%%%%%%%%%%%%%%%%%%%%%%%%%%%%%%%%%%%%%%%%%%%%%%%%%%
%%%%%%%%%%%%%%%%%%%%%%%%%%%%%%%%%%%%%%%%%%%%%%%%%%%%%%%%%%%%%%%%%%%%%%%%%%%%%%%%%%%%%%%%%%%%%%%%%%%%%%%%%%%%%%%%%%%%%%%%%%%%%%%%%%%%%%%%%
The inequalities defining ICE  resemble those which define other solution concepts involving choice interdependence.
%%%
Setting $\theta(i|a)= \delta\in(0,1)$ and imposing additional restrictions on $\ushort{w}$,
results in the recursive characterization of  SPNE of repeated games due to \cite{APS}. 
%%%
If $\theta(i|a)=0$, then players cannot punish deviations, and the definition reduces to correlated equilibria. 
%%%
In the opposite extreme, if $\theta(i|a)=1$, then players are \emph{always} able to punish deviations,
which results in interim individual rationality.
%%%
Of course, it cannot be the case that $\theta(i|a)=1$ for every $i$, because $\theta(\blank|a)$ is a probability measure. 
%%%
This is the reason why ICE does not result in a folk theorem: it is only possible to generate incentives for the first player to move, 
and it cannot be the case that both players move before their opponent.

%%%%%%%%%%%%%%%%%%%%%%%%%%%%%%%%%%%%%%%%%%%%%%%%%%%%%%%%%%%%%%%%%%%%%%%%%%%%%%%%%%%%%%%%%%%%%%%%%%%%%%%%%%%%%%%%%%%%%%%%%%%%%%%%%%%%%%%%%
%%%%%%%%%%%%%%%%%%%%%%%%%%%%%%%%%%%%%%%%%%%%%%%%%%%%%%%%%%%%%%%%%%%%%%%%%%%%%%%%%%%%%%%%%%%%%%%%%%%%%%%%%%%%%%%%%%%%%%%%%%%%%%%%%%%%%%%%%
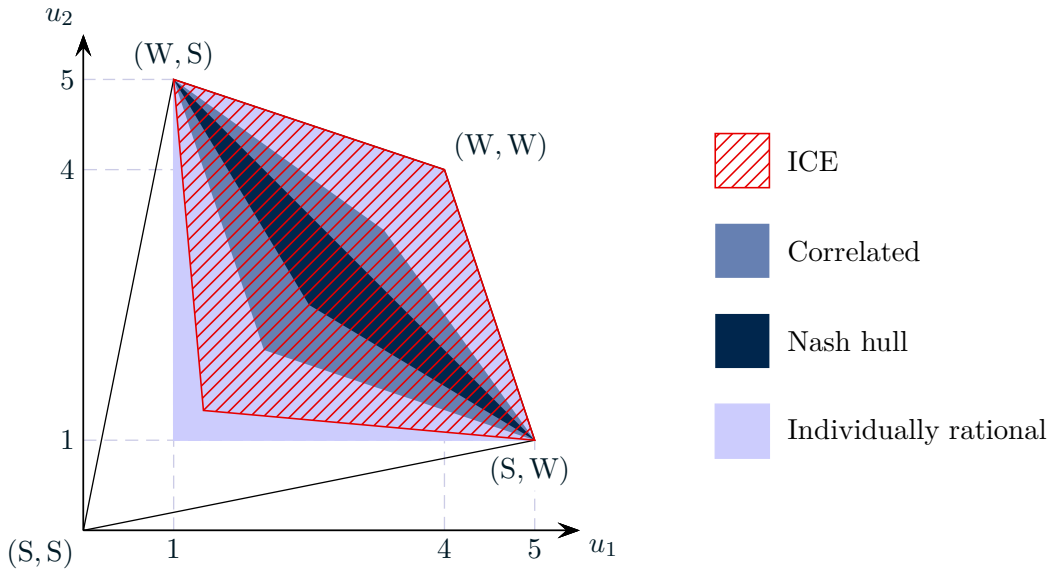
\begin{figure}[htb]
\centering
\psset{unit=12mm}
\begin{pspicture}(-0.5,-0.5)(10,6)
	\small%
	%\psgrid%
	% Payoff grid
	\psset{linecolor=BSClight,linestyle=dashed}
		\psline(0,1)(5,1)\psline(0,5)(1,5)\psline(0,4)(4,4)\psline(1,0)(1,5)\psline(4,0)(4,4)\psline(5,0)(5,1)
	\psreset%
	% Axes
	\psaxes[labels=none,ticks=none]{->}(0,0)(0,0)(5.5,5.5)
	% Payoff sets
		% Individually rational
		\newrgbcolor{indrat}{0.80 0.80 0.99}
		\psset{fillstyle=solid,fillcolor=indrat,linecolor=indrat}
			\pspolygon(1,1)(1,5)(4,4)(5,1)(1,1)
			\pspolygon(7.0,0.8)(7.0,1.4)(7.6,1.4)(7.6,0.8)(7.0,0.8)
			\rput[l](7.8,1.1){Individually rational}
		\psreset%
		% Correlated (private)
		\newrgbcolor{private}{0.40 0.50 0.70}
		\psset{fillstyle=solid,fillcolor=private,linestyle=none}
			\pspolygon(2,2)(1,5)(3.33,3.33)(5,1)(2,2)
			\pspolygon(7.0,2.8)(7.0,3.4)(7.6,3.4)(7.6,2.8)(7.0,2.8)
			\rput[l](7.8,3.1){Correlated}
		\psreset%
		% Public Nash
		%\newrgbcolor{public}{0.05 0.10 0.25}
		\newrgbcolor{public}{0.00 0.15 0.30}
		\psset{fillstyle=solid,fillcolor=public,linestyle=none}
			\pspolygon(2.5,2.5)(1,5)(5,1)(2.5,2.5)
			\pspolygon(7.0,1.8)(7.0,2.4)(7.6,2.4)(7.6,1.8)(7.0,1.8)
			\rput[l](7.8,2.1){Nash hull}
		\psreset%
		% Feasible
			\psline(0,0)(1,5)(4,4)(5,1)(0,0)
		% Coordinated
		\newrgbcolor{coord}{0.90 0.00 0.00}
		\psset{fillstyle=hlines,hatchcolor=coord,linecolor=coord,linewidth=0.6pt,hatchsep=3pt,hatchwidth=0.6pt}
			\pspolygon(1.33,1.33)(1,5)(4,4)(5,1)(1.33,1.33)
			\pspolygon(7.0,3.8)(7.0,4.4)(7.6,4.4)(7.6,3.8)(7.0,3.8)
			\rput[l](7.8,4.1){ICE}
		\psreset%
	% Labels
	\rput[tl](5.6,-0.1){$u_1$} \rput[br](-0.1,5.6){$u_2$}
	\rput[r](-0.1,1){$1$} \rput[r](-0.1,4){$4$} \rput[r](-0.1,5){$5$}
	\rput[t](1,-0.1){$1$} \rput[t](4,-0.1){$4$} \rput[t](5,-0.1){$5$}
	\rput[tr](-0.1,-0.1){$(\shirk,\shirk)$}
	\rput[t](5,0.9){\pslab{$(\shirk,\work)$}}
	\rput[b](1.0,5.1){$(\work,\shirk)$}
	\rput[bl](4.1,4.1){$(\work,\work)$}
\end{pspicture}
\caption{Equilibrium payoffs for example \ref{eg:chicken}}\label{fig:chicken}
\end{figure}

%%%%%%%%%%%%%%%%%%%%%%%%%%%%%%%%%%%%%%%%%%%%%%%%%%%%%%%%%%%%%%%%%%%%%%%%%%%%%%%%%%%%%%%%%%%%%%%%%%%%%%%%%%%%%%%%%%%%%%%%%%%%%%%%%%%%%%%%%
%%%%%%%%%%%%%%%%%%%%%%%%%%%%%%%%%%%%%%%%%%%%%%%%%%%%%%%%%%%%%%%%%%%%%%%%%%%%%%%%%%%%%%%%%%%%%%%%%%%%%%%%%%%%%%%%%%%%%%%%%%%%%%%%%%%%%%%%%
From the previous analysis, it follows that the inequalities defining ICE are tighter than those of individual rationality,
and weaker than those of correlated equilibrium. 
%%%
Hence the set of ICE is always contained in the set of individually rational outcomes,
and contains the set of correlated equilibria (and is thus nonempty).
%%%
The following example adapted from \cite{aumann87} shows that the containments can be strict.

%%%%%%%%%%%%%%%%%%%%%%%%%%%%%%%%%%%%%%%%%%%%%%%%%%%%%%%%%%%%%%%%%%%%%%%%%%%%%%%%%%%%%%%%%%%%%%%%%%%%%%%%%%%%%%%%%%%%%%%%%%%%%%%%%%%%%%%%%
%%%%%%%%%%%%%%%%%%%%%%%%%%%%%%%%%%%%%%%%%%%%%%%%%%%%%%%%%%%%%%%%%%%%%%%%%%%%%%%%%%%%%%%%%%%%%%%%%%%%%%%%%%%%%%%%%%%%%%%%%%%%%%%%%%%%%%%%%
\begin{example}\label{eg:chicken}
	Two partners decide whether to work ($\work$) or shirk ($\shirk$) in a joint-venture,
	their payoffs are depicted in Figure \ref{fig:chicken}.
	%%%
	The figure also shows the sets of payoffs corresponding to individual rationality,
	Nash equilibrium with public randomization, correlated equilibrium, and ICE. 
	%%%
	In this example, all the Pareto efficient outcomes correspond to ICE,
	and all but one ICE are sequential equilibria of the mediated game.

	%%%%%%%%%%%%%%%%%%%%%%%%%%%%%%%%%%%%%%%%%%%%%%%%%%%%%%%%%%%%%%%%%%%%%%%%%%%%%%%%%%%%%%%%%%%%%%%%%%%%%%%%%%%%%%%%%%%%%%%%%%%%%%%%%%%%
	$(\work,\work)$ is not a Nash equilibrium because, whenever an agent is working, his opponent prefers to shirk. 
	%%%
	It is an ICE because, a player who considers shirking knows that with some probability, 
	his opponent will learn of this defection and react by also shirking. 
	%%%
	The payoff vector $(1,1)$ cannot be attained as an ICE, because it requires players to shirk with high probability. 
	%%%
	Since each player always prefers that his opponent works, this leaves too little room to punish deviations. 
\end{example}

%%%%%%%%%%%%%%%%%%%%%%%%%%%%%%%%%%%%%%%%%%%%%%%%%%%%%%%%%%%%%%%%%%%%%%%%%%%%%%%%%%%%%%%%%%%%%%%%%%%%%%%%%%%%%%%%%%%%%%%%%%%%%%%%%%%%%%%%%
%%%%%%%%%%%%%%%%%%%%%%%%%%%%%%%%%%%%%%%%%%%%%%%%%%%%%%%%%%%%%%%%%%%%%%%%%%%%%%%%%%%%%%%%%%%%%%%%%%%%%%%%%%%%%%%%%%%%%%%%%%%%%%%%%%%%%%%%%
The existence statement makes definition \ref{defn:ICE} appear complicated.
%%%
Alternatively, let $\Gamma(B)$ be the set of joint distributions  $\gamma\in\Delta(A\times I)$ satisfying the inequalities:
%%%
\begin{align} \label{eqn:ICEalt}
	\sum_{a_{-i}\in A_{-i}}
		\Big[
			\gamma(a) u(a)
			- \gamma(a,i) \uw(a'_i,B_{-i}^*)
			- \gamma(a,-i) u(a'_i,a_{-i})
		\Big]\geq 0
\end{align}
%%%
for all $i\in I$ and $a_i,a_i'\in A_i$.
%%%
$\ICE(B)$ is the projection of $\Gamma(B)$ over $\Delta(A)$.
%%%
Hence, when the corresponding support of all equilibria is guaranteed to be a subset of $B$ (e.g.\ when $B=A$),
the set of ICE with respect to $B$ is a simple object characterized by a finite set of affine inequalities.
%%%
A difficulty may arise otherwise, because $\uw$ may depend on on $B^*$, which in turn depends on $\gamma$.
%%%
Hence the incentive constraints may no longer be affine or even continuous.

%%%%%%%%%%%%%%%%%%%%%%%%%%%%%%%%%%%%%%%%%%%%%%%%%%%%%%%%%%%%%%%%%%%%%%%%%%%%%%%%%%%%%%%%%%%%%%%%%%%%%%%%%%%%%%%%%%%%%%%%%%%%%%%%%%%%%%%%%
%%%%%%%%%%%%%%%%%%%%%%%%%%%%%%%%%%%%%%%%%%%%%%%%%%%%%%%%%%%%%%%%%%%%%%%%%%%%%%%%%%%%%%%%%%%%%%%%%%%%%%%%%%%%%%%%%%%%%%%%%%%%%%%%%%%%%%%%%
%%%%%%%%%%%%%%%%%%%%%%%%%%%%%%%%%%%%%%%%%%%%%%%%%%%%%%%%%%%%%%%%%%%%%%%%%%%%%%%%%%%%%%%%%%%%%%%%%%%%%%%%%%%%%%%%%%%%%%%%%%%%%%%%%%%%%%%%%
%%%%%%%%%%%%%%%%%%%%%%%%%%%%%%%%%%%%%%%%%%%%%%%%%%%%%%%%%%%%%%%%%%%%%%%%%%%%%%%%%%%%%%%%%%%%%%%%%%%%%%%%%%%%%%%%%%%%%%%%%%%%%%%%%%%%%%%%%
%%%%%%%%%%%%%%%%%%%%%%%%%%%%%%%%%%%%%%%%%%%%%%%%%%%%%%%%%%%%%%%%%%%%%%%%%%%%%%%%%%%%%%%%%%%%%%%%%%%%%%%%%%%%%%%%%%%%%%%%%%%%%%%%%%%%%%%%%
%%%%%%%%%%%%%%%%%%%%%%%%%%%%%%%%%%%%%%%%%%%%%%%%%%%%%%%%%%%%%%%%%%%%%%%%%%%%%%%%%%%%%%%%%%%%%%%%%%%%%%%%%%%%%%%%%%%%%%%%%%%%%%%%%%%%%%%%%
%%%%%%%%%%%%%%%%%%%%%%%%%%%%%%%%%%%%%%%%%%%%%%%%%%%%%%%%%%%%%%%%%%%%%%%%%%%%%%%%%%%%%%%%%%%%%%%%%%%%%%%%%%%%%%%%%%%%%%%%%%%%%%%%%%%%%%%%%
%%%%%%%%%%%%%%%%%%%%%%%%%%%%%%%%%%%%%%%%%%%%%%%%%%%%%%%%%%%%%%%%%%%%%%%%%%%%%%%%%%%%%%%%%%%%%%%%%%%%%%%%%%%%%%%%%%%%%%%%%%%%%%%%%%%%%%%%%
%%%%%%%%%%%%%%%%%%%%%%%%%%%%%%%%%%%%%%%%%%%%%%%%%%%%%%%%%%%%%%%%%%%%%%%%%%%%%%%%%%%%%%%%%%%%%%%%%%%%%%%%%%%%%%%%%%%%%%%%%%%%%%%%%%%%%%%%%
%%%%%%%%%%%%%%%%%%%%%%%%%%%%%%%%%%%%%%%%%%%%%%%%%%%%%%%%%%%%%%%%%%%%%%%%%%%%%%%%%%%%%%%%%%%%%%%%%%%%%%%%%%%%%%%%%%%%%%%%%%%%%% ssec: Nash
\subsection{Nash implementation}\label{ssec:nash}

%%%%%%%%%%%%%%%%%%%%%%%%%%%%%%%%%%%%%%%%%%%%%%%%%%%%%%%%%%%%%%%%%%%%%%%%%%%%%%%%%%%%%%%%%%%%%%%%%%%%%%%%%%%%%%%%%%%%%%%%%%%%%%%%%%%%%%%%%
%%%%%%%%%%%%%%%%%%%%%%%%%%%%%%%%%%%%%%%%%%%%%%%%%%%%%%%%%%%%%%%%%%%%%%%%%%%%%%%%%%%%%%%%%%%%%%%%%%%%%%%%%%%%%%%%%%%%%%%%%%%%%%%%%%%%%%%%%
The purpose of the current work is to characterize the outcomes that can be implemented as equilibria 
without any form of \emph{repetition}, \emph{monetary transfers} or \emph{binding agreements} from the part of the players.
%%%
In order to formalize the meaning of implementation, one needs to define the largest class of mechanisms 
which are consistent with the partial characterization of the environment, and with such restrictions. 
%%%
The technical definition is relegated to appendix \ref{sec:EFM}.
%%%
The essential conditions can be informally described as follows. 

%%%%%%%%%%%%%%%%%%%%%%%%%%%%%%%%%%%%%%%%%%%%%%%%%%%%%%%%%%%%%%%%%%%%%%%%%%%%%%%%%%%%%%%%%%%%%%%%%%%%%%%%%%%%%%%%%%%%%%%%%%%%%%%%%%%%%%%%%
%%%%%%%%%%%%%%%%%%%%%%%%%%%%%%%%%%%%%%%%%%%%%%%%%%%%%%%%%%%%%%%%%%%%%%%%%%%%%%%%%%%%%%%%%%%%%%%%%%%%%%%%%%%%%%%%%%%%%%%%%%%%%%%%%%%%%%%%%
\newcounter{EFM}
\setcounter{EFM}{\value{definition}}
\begin{definition} \label{defn:EFMinformal} 
	An \emph{extensive form mechanism} (EFM) is any extensive form such that:
	\begin{enumerate}
		\item\label{req:A} Terminal nodes can be identified with action profiles from the environment.
		\item\label{req:B} Along every possible path of play, each player makes some move 
			which can be interpreted as choosing one of his actions from the environment.
		\item\label{req:C} At the moment of choosing his action,
			each player could have chosen \emph{any} other action from his original action space.
	\end{enumerate}
\end{definition}

%%%%%%%%%%%%%%%%%%%%%%%%%%%%%%%%%%%%%%%%%%%%%%%%%%%%%%%%%%%%%%%%%%%%%%%%%%%%%%%%%%%%%%%%%%%%%%%%%%%%%%%%%%%%%%%%%%%%%%%%%%%%%%%%%%%%%%%%%
%%%%%%%%%%%%%%%%%%%%%%%%%%%%%%%%%%%%%%%%%%%%%%%%%%%%%%%%%%%%%%%%%%%%%%%%%%%%%%%%%%%%%%%%%%%%%%%%%%%%%%%%%%%%%%%%%%%%%%%%%%%%%%%%%%%%%%%%%
Condition \ref{req:B} rules out delegation and binding agreements.
%%%
The ruling out of delegation is straightforward: each player has to
\emph{freely} choose which action to perform at some point of the game. 
%%%
The ruling out of binding agreements is a little more subtle. 
%%%
To understand what the requirement encompasses it is useful to consider the literature
on preplay negotiations dating back to \cite{kalai81}.
%%%
Kalai allows players publicly announce their intentions, and these announcements \emph{may}
become binding depending on the announcements of others. 
%%%
Announcing an action is not exactly the same as performing it, because the action
is only executed \emph{if} the announcement becomes binding.
%%% 
Otherwise the choice can be reverted and the player may choose a different action. 
%%%
In contrast, I require that some moves of the game should corresponds to
the \emph{irreversible} act of performing an action from the environment.

%%%%%%%%%%%%%%%%%%%%%%%%%%%%%%%%%%%%%%%%%%%%%%%%%%%%%%%%%%%%%%%%%%%%%%%%%%%%%%%%%%%%%%%%%%%%%%%%%%%%%%%%%%%%%%%%%%%%%%%%%%%%%%%%%%%%%%%%%
%%%%%%%%%%%%%%%%%%%%%%%%%%%%%%%%%%%%%%%%%%%%%%%%%%%%%%%%%%%%%%%%%%%%%%%%%%%%%%%%%%%%%%%%%%%%%%%%%%%%%%%%%%%%%%%%%%%%%%%%%%%%%%%%%%%%%%%%%
Condition \ref{req:C} rules out partial-commitment.
%%%
For example, consider a variation of the prisoners dilemma in which each prisoner chooses 
between confessing to a big crime, confessing to a minor crime or not confessing at all. 
%%%
I rule out the possibility that at some point of the game a player makes a move that will force him to confess,
and later on decides which crime to confess. 
%%%
Every action from the environment has to be available at the moment of making a decisive move. 

%%%%%%%%%%%%%%%%%%%%%%%%%%%%%%%%%%%%%%%%%%%%%%%%%%%%%%%%%%%%%%%%%%%%%%%%%%%%%%%%%%%%%%%%%%%%%%%%%%%%%%%%%%%%%%%%%%%%%%%%%%%%%%%%%%%%%%%%%
%%%%%%%%%%%%%%%%%%%%%%%%%%%%%%%%%%%%%%%%%%%%%%%%%%%%%%%%%%%%%%%%%%%%%%%%%%%%%%%%%%%%%%%%%%%%%%%%%%%%%%%%%%%%%%%%%%%%%%%%%%%%%%%%%%%%%%%%%
Condition \ref{req:A} implies that 
players preferences must be determined by the chosen action profile, and cannot depend on side payments or future events. 
%%%
This rules out monetary transfers and repetition.
%%%
Furthermore, it enables to identify outcomes of the mechanism with outcomes of the environment.
%%%
Therefore, once can say that a strategy profile for an EFM induces a distribution over action profiles.
%%%
This results in the following natural definition of implementation.

%%%%%%%%%%%%%%%%%%%%%%%%%%%%%%%%%%%%%%%%%%%%%%%%%%%%%%%%%%%%%%%%%%%%%%%%%%%%%%%%%%%%%%%%%%%%%%%%%%%%%%%%%%%%%%%%%%%%%%%%%%%%%%%%%%%%%%%%%
%%%%%%%%%%%%%%%%%%%%%%%%%%%%%%%%%%%%%%%%%%%%%%%%%%%%%%%%%%%%%%%%%%%%%%%%%%%%%%%%%%%%%%%%%%%%%%%%%%%%%%%%%%%%%%%%%%%%%%%%%%%%%%%%%%%%%%%%%
\newcounter{IMP}
\setcounter{IMP}{\value{definition}}
\begin{definition} \label{defn:IMPinformal}
	A distribution over action profiles is (Nash, sequentially, \ldots) \emph{implementable}
	if and only if it can be induced by a (Nash, sequential, \ldots )
	equilibrium of an extensive form mechanism. 
\end{definition}

%%%%%%%%%%%%%%%%%%%%%%%%%%%%%%%%%%%%%%%%%%%%%%%%%%%%%%%%%%%%%%%%%%%%%%%%%%%%%%%%%%%%%%%%%%%%%%%%%%%%%%%%%%%%%%%%%%%%%%%%%%%%%%%%%%%%%%%%%
%%%%%%%%%%%%%%%%%%%%%%%%%%%%%%%%%%%%%%%%%%%%%%%%%%%%%%%%%%%%%%%%%%%%%%%%%%%%%%%%%%%%%%%%%%%%%%%%%%%%%%%%%%%%%%%%%%%%%%%%%%%%%%%%%%%%%%%%%
Mediated games do not involve any form of repetition, monetary transfers or binding agreements from the part of the players.
%%%
The role of the mediator is simply to generate an information structure that allows for responsive strategies. 
%%%
Hence, the set of ICE describes a set of outcomes that can be \emph{Nash implemented} under these restrictions.
%%%
A natural question is whether there are other implementable distributions, and the answer is no. 

%%%%%%%%%%%%%%%%%%%%%%%%%%%%%%%%%%%%%%%%%%%%%%%%%%%%%%%%%%%%%%%%%%%%%%%%%%%%%%%%%%%%%%%%%%%%%%%%%%%%%%%%%%%%%%%%%%%%%%%%%%%%%%%%%%%%%%%%%
%%%%%%%%%%%%%%%%%%%%%%%%%%%%%%%%%%%%%%%%%%%%%%%%%%%%%%%%%%%%%%%%%%%%%%%%%%%%%%%%%%%%%%%%%%%%%%%%%%%%%%%%%%%%%%%%%%%%%%%%%%%%%%%%%%%%%%%%%
\begin{theorem} \label{thm:Nash}
	A distribution over action profiles is Nash implementable if and only if it is an interdependent-choice equilibrium. 
\end{theorem}

%%%%%%%%%%%%%%%%%%%%%%%%%%%%%%%%%%%%%%%%%%%%%%%%%%%%%%%%%%%%%%%%%%%%%%%%%%%%%%%%%%%%%%%%%%%%%%%%%%%%%%%%%%%%%%%%%%%%%%%%%%%%%%%%%%%%%%%%%
%%%%%%%%%%%%%%%%%%%%%%%%%%%%%%%%%%%%%%%%%%%%%%%%%%%%%%%%%%%%%%%%%%%%%%%%%%%%%%%%%%%%%%%%%%%%%%%%%%%%%%%%%%%%%%%%%%%%%%%%%%%%%%%%%%%%%%%%%
The proof follows the general approach introduced by \cite{forges86}.
%%%
The sufficiency is standard because mediated games are EFMs.
%%%
For necessity, given an EFM and an equilibrium,
it is possible to construct a mediated game that replicates the strategic features,
but gives players the minimal amount of information needed to follow the plan of action. 
%%%
By reducing their information, players' ability to deviate profitably also decreases.

%%%%%%%%%%%%%%%%%%%%%%%%%%%%%%%%%%%%%%%%%%%%%%%%%%%%%%%%%%%%%%%%%%%%%%%%%%%%%%%%%%%%%%%%%%%%%%%%%%%%%%%%%%%%%%%%%%%%%%%%%%%%%%%%%%%%%%%%%
%%%%%%%%%%%%%%%%%%%%%%%%%%%%%%%%%%%%%%%%%%%%%%%%%%%%%%%%%%%%%%%%%%%%%%%%%%%%%%%%%%%%%%%%%%%%%%%%%%%%%%%%%%%%%%%%%%%%%%%%%%%%%%%%%%%%%%%%%
%%%%%%%%%%%%%%%%%%%%%%%%%%%%%%%%%%%%%%%%%%%%%%%%%%%%%%%%%%%%%%%%%%%%%%%%%%%%%%%%%%%%%%%%%%%%%%%%%%%%%%%%%%%%%%%%%%%%%%%%%%%%%%%%%%%%%%%%%
%%%%%%%%%%%%%%%%%%%%%%%%%%%%%%%%%%%%%%%%%%%%%%%%%%%%%%%%%%%%%%%%%%%%%%%%%%%%%%%%%%%%%%%%%%%%%%%%%%%%%%%%%%%%%%%%%%%%%%%%%%%%%%%%%%%%%%%%%
%%%%%%%%%%%%%%%%%%%%%%%%%%%%%%%%%%%%%%%%%%%%%%%%%%%%%%%%%%%%%%%%%%%%%%%%%%%%%%%%%%%%%%%%%%%%%%%%%%%%%%%%%%%%%%%%%%%%%%%%%%%%%%%%%%%%%%%%%
%%%%%%%%%%%%%%%%%%%%%%%%%%%%%%%%%%%%%%%%%%%%%%%%%%%%%%%%%%%%%%%%%%%%%%%%%%%%%%%%%%%%%%%%%%%%%%%%%%%%%%%%%%%%%%%%%%%%%%%%%%%%%%%%%%%%%%%%%
%%%%%%%%%%%%%%%%%%%%%%%%%%%%%%%%%%%%%%%%%%%%%%%%%%%%%%%%%%%%%%%%%%%%%%%%%%%%%%%%%%%%%%%%%%%%%%%%%%%%%%%%%%%%%%%%%%%%%%%%%%%%%%%%%%%%%%%%%
%%%%%%%%%%%%%%%%%%%%%%%%%%%%%%%%%%%%%%%%%%%%%%%%%%%%%%%%%%%%%%%%%%%%%%%%%%%%%%%%%%%%%%%%%%%%%%%%%%%%%%%%%%%%%%%%%%%%%%%%%%%%%%%%%%%%%%%%%
%%%%%%%%%%%%%%%%%%%%%%%%%%%%%%%%%%%%%%%%%%%%%%%%%%%%%%%%%%%%%%%%%%%%%%%%%%%%%%%%%%%%%%%%%%%%%%%%%%%%%%%%%%%%%%%%%%%%%%%%%%%%%%%%%%%%%%%%%
%%%%%%%%%%%%%%%%%%%%%%%%%%%%%%%%%%%%%%%%%%%%%%%%%%%%%%%%%%%%%%%%%%%%%%%%%%%%%%%%%%%%%%%%%%%%%%%%%%%%%%%%%%%%%%%%%%%%%%%% ssec: Literature
\subsection{Comparison with related literature}\label{ssec:lit}

%%%%%%%%%%%%%%%%%%%%%%%%%%%%%%%%%%%%%%%%%%%%%%%%%%%%%%%%%%%%%%%%%%%%%%%%%%%%%%%%%%%%%%%%%%%%%%%%%%%%%%%%%%%%%%%%%%%%%%%%%%%%%%%%%%%%%%%%%
%%%%%%%%%%%%%%%%%%%%%%%%%%%%%%%%%%%%%%%%%%%%%%%%%%%%%%%%%%%%%%%%%%%%%%%%%%%%%%%%%%%%%%%%%%%%%%%%%%%%%%%%%%%%%%%%%%%%%%%%%%%%%%%%%%%%%%%%%
Choice interdependence is a common theme across a variety of otherwise distant literatures. 
%%%
Besides the well established literatures on repeated games and games with contracts, 
different literatures allow for different forms of implicit repetition, commitment or transfers.
%%%
The literature on counterfactual variations can be thought of as a reduced representation
of repeated games \citep{kalSta}.
%%%
Commitment can be traced back to \cite{MouVia} and \cite{kalai81},
which allow players to delegate choices to a mediator, or make preplay \emph{binding} announcements. 
%%%
With unrestricted commitment, one obtains folk theorems \citep{kalai10}.
%%%
Recent relevant works in this area include \cite{bade09,renou09,forgo10}.
%%%
For monetary transfers see \cite{jackson05}.

%%%%%%%%%%%%%%%%%%%%%%%%%%%%%%%%%%%%%%%%%%%%%%%%%%%%%%%%%%%%%%%%%%%%%%%%%%%%%%%%%%%%%%%%%%%%%%%%%%%%%%%%%%%%%%%%%%%%%%%%%%%%%%%%%%%%%%%%%
%%%%%%%%%%%%%%%%%%%%%%%%%%%%%%%%%%%%%%%%%%%%%%%%%%%%%%%%%%%%%%%%%%%%%%%%%%%%%%%%%%%%%%%%%%%%%%%%%%%%%%%%%%%%%%%%%%%%%%%%%%%%%%%%%%%%%%%%%
Other literatures  across different fields
allow for counterfactual reasoning without being too explicit about the source of interdependence. 
%%%
Seminal examples include \cite{RapAl} and \cite{howard71}, see also \cite{hofstadter96} and \cite{brams93}.
%%%
More recently, \cite{halpern10} and \cite{halpern12}, analize equilibrium and rationaliability with counterfactual beliefs. 
%%%
Once again, folk theorems hold if no further restrictions are imposed. 
%%%
These literatures have faced some criticism as some of the beliefs may not be consistent with free will, \citep{GibHar,lewis79}.

%%%%%%%%%%%%%%%%%%%%%%%%%%%%%%%%%%%%%%%%%%%%%%%%%%%%%%%%%%%%%%%%%%%%%%%%%%%%%%%%%%%%%%%%%%%%%%%%%%%%%%%%%%%%%%%%%%%%%%%%%%%%%%%%%%%%%%%%%
%%%%%%%%%%%%%%%%%%%%%%%%%%%%%%%%%%%%%%%%%%%%%%%%%%%%%%%%%%%%%%%%%%%%%%%%%%%%%%%%%%%%%%%%%%%%%%%%%%%%%%%%%%%%%%%%%%%%%%%%%%%%%%%%%%%%%%%%%
The present work serves as a bridge between these literatures, 
asking which couterfactual beliefs can arise from different sequential and informational structures,
but  without repetition or commitment.  
%%%
These restrictions are sufficient to rule out a folk theorem. 
%%%
Other works along these lines typically depart from classical notions of choice and rationality. 
%%%
For instance, \cite{eisert99} shows that cooperation is possible in a prisoners dilemma,
when players can condition their choices on certain \emph{quantum} randomization devices with \emph{entangled} states. 

%%%%%%%%%%%%%%%%%%%%%%%%%%%%%%%%%%%%%%%%%%%%%%%%%%%%%%%%%%%%%%%%%%%%%%%%%%%%%%%%%%%%%%%%%%%%%%%%%%%%%%%%%%%%%%%%%%%%%%%%%%%%%%%%%%%%%%%%%
%%%%%%%%%%%%%%%%%%%%%%%%%%%%%%%%%%%%%%%%%%%%%%%%%%%%%%%%%%%%%%%%%%%%%%%%%%%%%%%%%%%%%%%%%%%%%%%%%%%%%%%%%%%%%%%%%%%%%%%%%%%%%%%%%%%%%%%%%
\cite{tennenholtz04}'s program equilibrium generates choice interedependence 
for games \emph{between computer programms} by allowing them to read each other's code before making a decision. 
%%%
Similarly, \cite{levine07}'s self-referential equilibrium allows player's to receive a signal about
their opponent's intentions before making their own choice. 
%%%
\cite{KamKan09,KamKan12} and \cite{calcagno12}'s revision games are based on a similar logic,
but are more explicit about the information structure. 
%%%
I rule out such signals this kind of signals as they represent a from of commitment:
from the moment of \emph{deciding} which action to play, to the moment of actually \emph{performing} it. 
%%%
ICE can arise in settings where choices are instantaneous and players can hide their intentions.

%%%%%%%%%%%%%%%%%%%%%%%%%%%%%%%%%%%%%%%%%%%%%%%%%%%%%%%%%%%%%%%%%%%%%%%%%%%%%%%%%%%%%%%%%%%%%%%%%%%%%%%%%%%%%%%%%%%%%%%%%%%%%%%%%%%%%%%%%
%%%%%%%%%%%%%%%%%%%%%%%%%%%%%%%%%%%%%%%%%%%%%%%%%%%%%%%%%%%%%%%%%%%%%%%%%%%%%%%%%%%%%%%%%%%%%%%%%%%%%%%%%%%%%%%%%%%%%%%%%%%%%%%%%%%%%%%%%
%%%%%%%%%%%%%%%%%%%%%%%%%%%%%%%%%%%%%%%%%%%%%%%%%%%%%%%%%%%%%%%%%%%%%%%%%%%%%%%%%%%%%%%%%%%%%%%%%%%%%%%%%%%%%%%%%%%%%%%%%%%%%%%%%%%%%%%%%
%%%%%%%%%%%%%%%%%%%%%%%%%%%%%%%%%%%%%%%%%%%%%%%%%%%%%%%%%%%%%%%%%%%%%%%%%%%%%%%%%%%%%%%%%%%%%%%%%%%%%%%%%%%%%%%%%%%%%%%%%%%%%%%%%%%%%%%%%
%%%%%%%%%%%%%%%%%%%%%%%%%%%%%%%%%%%%%%%%%%%%%%%%%%%%%%%%%%%%%%%%%%%%%%%%%%%%%%%%%%%%%%%%%%%%%%%%%%%%%%%%%%%%%%%%%%%%%%%%%%%%%%%%%%%%%%%%%
%%%%%%%%%%%%%%%%%%%%%%%%%%%%%%%%%%%%%%%%%%%%%%%%%%%%%%%%%%%%%%%%%%%%%%%%%%%%%%%%%%%%%%%%%%%%%%%%%%%%%%%%%%%%%%%%%%%%%%%%%%%%%%%%%%%%%%%%%
%%%%%%%%%%%%%%%%%%%%%%%%%%%%%%%%%%%%%%%%%%%%%%%%%%%%%%%%%%%%%%%%%%%%%%%%%%%%%%%%%%%%%%%%%%%%%%%%%%%%%%%%%%%%%%%%%%%%%%%%%%%%%%%%%%%%%%%%%
%%%%%%%%%%%%%%%%%%%%%%%%%%%%%%%%%%%%%%%%%%%%%%%%%%%%%%%%%%%%%%%%%%%%%%%%%%%%%%%%%%%%%%%%%%%%%%%%%%%%%%%%%%%%%%%%%%%%%%%%%%%%%%%%%%%%%%%%%
%%%%%%%%%%%%%%%%%%%%%%%%%%%%%%%%%%%%%%%%%%%%%%%%%%%%%%%%%%%%%%%%%%%%%%%%%%%%%%%%%%%%%%%%%%%%%%%%%%%%%%%%%%%%%%%%%%%%%%%%%%%%%%%%%%%%%%%%%
%%%%%%%%%%%%%%%%%%%%%%%%%%%%%%%%%%%%%%%%%%%%%%%%%%%%%%%%%%%%%%%%%%%%%%%%%%%%%%%%%%%%%%%%%%%%%%%%%%%%%%%%%%%%%%%%%%%%%%%%%% END OF SECTION

%% file: sequential.tex
%auto-ignore

%%%%%%%%%%%%%%%%%%%%%%%%%%%%%%%%%%%%%%%%%%%%%%%%%%%%%%%%%%%%%%%%%%%%%%%%%%%%%%%%%%%%%%%%%%%%%%%%%%%%%%%%%%%%%%%%%%%%%%%%%%%%%%%%%%%%%%%%%
%%%%%%%%%%%%%%%%%%%%%%%%%%%%%%%%%%%%%%%%%%%%%%%%%%%%%%%%%%%%%%%%%%%%%%%%%%%%%%%%%%%%%%%%%%%%%%%%%%%%%%%%%%%%%%%%%%%%%%%%%%%%%%%%%%%%%%%%%
% Section: Sequential implementation
% Project: Coordination in Strategic Environments
%  Author: Bruno Salcedo
% Version: 7.1
%    Date: 09/07/13
%%%%%%%%%%%%%%%%%%%%%%%%%%%%%%%%%%%%%%%%%%%%%%%%%%%%%%%%%%%%%%%%%%%%%%%%%%%%%%%%%%%%%%%%%%%%%%%%%%%%%%%%%%%%%%%%%%%%%%%%%%%%%%%%%%%%%%%%%
%%%%%%%%%%%%%%%%%%%%%%%%%%%%%%%%%%%%%%%%%%%%%%%%%%%%%%%%%%%%%%%%%%%%%%%%%%%%%%%%%%%%%%%%%%%%%%%%%%%%%%%%%%%%%%%%%%%%%%%%%%%%%%%%%%%%%%%%%

%%%%%%%%%%%%%%%%%%%%%%%%%%%%%%%%%%%%%%%%%%%%%%%%%%%%%%%%%%%%%%%%%%%%%%%%%%%%%%%%%%%%%%%%%%%%%%%%%%%%%%%%%%%%%%%%%%%%%%%%%%%%%%%%%%%%%%%%%
%%%%%%%%%%%%%%%%%%%%%%%%%%%%%%%%%%%%%%%%%%%%%%%%%%%%%%%%%%%%%%%%%%%%%%%%%%%%%%%%%%%%%%%%%%%%%%%%%%%%%%%%%%%%%%%%%%%%%%%%%%%%%%%%%%%%%%%%%
A problem remains because, while ICE does not involve any form of explicit commitment from part of the players,
Nash incentive compatibility does not impose any restrictions off the equilibrium path.
%%%
ICE is defined \emph{as if} players where committed to punish deviations off the equilibrium path, 
regardless of whether doing so would be optimal if the circumstance were to arise. 
%%%
This section addresses this issue by investigating which ICE can be implemented as \emph{sequential equilibria} of EFM.
%%%
Section \S\ref{ssec:suff} provides tractable sufficient conditions for arbitrary environments,
and section \S\ref{ssec:Crat} provides a generic characterization for $2\times2$ environments. 

%%%%%%%%%%%%%%%%%%%%%%%%%%%%%%%%%%%%%%%%%%%%%%%%%%%%%%%%%%%%%%%%%%%%%%%%%%%%%%%%%%%%%%%%%%%%%%%%%%%%%%%%%%%%%%%%%%%%%%%%%%%%%%%%%%%%%%%%%
%%%%%%%%%%%%%%%%%%%%%%%%%%%%%%%%%%%%%%%%%%%%%%%%%%%%%%%%%%%%%%%%%%%%%%%%%%%%%%%%%%%%%%%%%%%%%%%%%%%%%%%%%%%%%%%%%%%%%%%%%%%%%%%%%%%%%%%%%
%%%%%%%%%%%%%%%%%%%%%%%%%%%%%%%%%%%%%%%%%%%%%%%%%%%%%%%%%%%%%%%%%%%%%%%%%%%%%%%%%%%%%%%%%%%%%%%%%%%%%%%%%%%%%%%%%%%%%%%%%%%%%%%%%%%%%%%%%
%%%%%%%%%%%%%%%%%%%%%%%%%%%%%%%%%%%%%%%%%%%%%%%%%%%%%%%%%%%%%%%%%%%%%%%%%%%%%%%%%%%%%%%%%%%%%%%%%%%%%%%%%%%%%%%%%%%%%%%%%%%%%%%%%%%%%%%%%
%%%%%%%%%%%%%%%%%%%%%%%%%%%%%%%%%%%%%%%%%%%%%%%%%%%%%%%%%%%%%%%%%%%%%%%%%%%%%%%%%%%%%%%%%%%%%%%%%%%%%%%%%%%%%%%%%%%%%%%%%%%%%%%%%%%%%%%%%
%%%%%%%%%%%%%%%%%%%%%%%%%%%%%%%%%%%%%%%%%%%%%%%%%%%%%%%%%%%%%%%%%%%%%%%%%%%%%%%%%%%%%%%%%%%%%%%%%%%%%%%%%%%%%%%%%%%%%%%%%%%%%%%%%%%%%%%%%
%%%%%%%%%%%%%%%%%%%%%%%%%%%%%%%%%%%%%%%%%%%%%%%%%%%%%%%%%%%%%%%%%%%%%%%%%%%%%%%%%%%%%%%%%%%%%%%%%%%%%%%%%%%%%%%%%%%%%%%%%%%%%%%%%%%%%%%%%
%%%%%%%%%%%%%%%%%%%%%%%%%%%%%%%%%%%%%%%%%%%%%%%%%%%%%%%%%%%%%%%%%%%%%%%%%%%%%%%%%%%%%%%%%%%%%%%%%%%%%%%%%%%%%%%%%%%%%%%%%%%%%%%%%%%%%%%%%
%%%%%%%%%%%%%%%%%%%%%%%%%%%%%%%%%%%%%%%%%%%%%%%%%%%%%%%%%%%%%%%%%%%%%%%%%%%%%%%%%%%%%%%%%%%%%%%%%%%%%%%%%%%%%%%%%%%%%%%%%%%%%%%%%%%%%%%%%
%%%%%%%%%%%%%%%%%%%%%%%%%%%%%%%%%%%%%%%%%%%%%%%%%%%%%%%%%%%%%%%%%%%%%%%%%%%%%%%%%%%%%%%%%%%%%%%%%%%%%%%%%%%%%%%%%%%%%%%% ssec: sufficient
\subsection{A sufficient condition for sequential implementation}\label{ssec:suff}

%%%%%%%%%%%%%%%%%%%%%%%%%%%%%%%%%%%%%%%%%%%%%%%%%%%%%%%%%%%%%%%%%%%%%%%%%%%%%%%%%%%%%%%%%%%%%%%%%%%%%%%%%%%%%%%%%%%%%%%%%%%%%%%%%%%%%%%%%
%%%%%%%%%%%%%%%%%%%%%%%%%%%%%%%%%%%%%%%%%%%%%%%%%%%%%%%%%%%%%%%%%%%%%%%%%%%%%%%%%%%%%%%%%%%%%%%%%%%%%%%%%%%%%%%%%%%%%%%%%%%%%%%%%%%%%%%%%
The sufficient condition for sequential implementation provided here is based on three observations. 
%%%
First, suppose that $\alpha$ is a interdependent-choice equilibrium with respect to $B=\emptyset$.
%%%
In the corresponding mediated game,  all the actions which are recommended as punishments are also 
recommended along the equilibrium path. 
%%%
This implies that every information set is reached with positive probability,
and following the mediator recommendations thus constitutes a sequential equilibrium. 

%%%%%%%%%%%%%%%%%%%%%%%%%%%%%%%%%%%%%%%%%%%%%%%%%%%%%%%%%%%%%%%%%%%%%%%%%%%%%%%%%%%%%%%%%%%%%%%%%%%%%%%%%%%%%%%%%%%%%%%%%%%%%%%%%%%%%%%%%
%%%%%%%%%%%%%%%%%%%%%%%%%%%%%%%%%%%%%%%%%%%%%%%%%%%%%%%%%%%%%%%%%%%%%%%%%%%%%%%%%%%%%%%%%%%%%%%%%%%%%%%%%%%%%%%%%%%%%%%%%%%%%%%%%%%%%%%%%
Secondly, given two extensive form mechanisms, one can construct a new mechanism in which nature randomizes between them.  
%%%
If the outcome of the randomization public, then incentives remain unchanged.
%%%
Let $A^\SE=\cart_{i\in I} A_i^\SE$ denote the set of  profiles of actions which are played
with positive probability in \emph{some} sequentially implementable distribution. 
%%%
The second observation implies that there exists a sequentially implementable distribution $\alpha^\SE$
such that $\alpha_i^\SE(a_i)>0$ for every  $_ia\in A_i^\SE$.
%%%
Also, together with the first observation, it implies that every $\alpha\in\ICE(A^\SE)$
can be approximated by sequentially implementable distributions. 

%%%%%%%%%%%%%%%%%%%%%%%%%%%%%%%%%%%%%%%%%%%%%%%%%%%%%%%%%%%%%%%%%%%%%%%%%%%%%%%%%%%%%%%%%%%%%%%%%%%%%%%%%%%%%%%%%%%%%%%%%%%%%%%%%%%%%%%%%
%%%%%%%%%%%%%%%%%%%%%%%%%%%%%%%%%%%%%%%%%%%%%%%%%%%%%%%%%%%%%%%%%%%%%%%%%%%%%%%%%%%%%%%%%%%%%%%%%%%%%%%%%%%%%%%%%%%%%%%%%%%%%%%%%%%%%%%%%
Finally, let $\R$ be the set of first-order rationalizable action profiles.
%%%
Sequential equilibrium does not impose any restrictions on the relative likelihood of different deviations.
%%%
Hence, any action in $\R$ can be used as a credible threat.
%%%
For instance, if $a_1$ is a best response to $a_2$, then one can construct a mediated game in which,
whenever $1$ is asked to play $a_1$, he will believe that it is because $2$ deviated by choosing $a_2$.

%%%%%%%%%%%%%%%%%%%%%%%%%%%%%%%%%%%%%%%%%%%%%%%%%%%%%%%%%%%%%%%%%%%%%%%%%%%%%%%%%%%%%%%%%%%%%%%%%%%%%%%%%%%%%%%%%%%%%%%%%%%%%%%%%%%%%%%%%
%%%%%%%%%%%%%%%%%%%%%%%%%%%%%%%%%%%%%%%%%%%%%%%%%%%%%%%%%%%%%%%%%%%%%%%%%%%%%%%%%%%%%%%%%%%%%%%%%%%%%%%%%%%%%%%%%%%%%%%%%%%%%%%%%%%%%%%%%
The preceding discussion suggests that every $\alpha\in\ICE(\R\cup A^\SE)$ is sequentially implementable. 
%%%%
However, there is no obvious way of finding $A^\SE$ and, unless $A^\SE\subseteq\R$, 
computing $\ICE(\R)$ can be difficult since the incentive constraints could be neither linear nor continuous.
%%%
Instead, I propose a recursive procedure  much in the spirit of the APS algorithm \citep{APS}, 
to find a set $\CEac\subseteq\SEac$ with the property that  every ICE equilibrium with respect to $\CEac$ is sequentially implementable.
%%%
Let $\A = \left\{ \times_{i\in I} A'_i \:|\: A'_i\subseteq A_i \right\}$ and define $\B:\A\rightarrow\A$ by:
%%%
\begin{align*}
	\B(A') = \cart_{i\in I}\left\{
			a_i\in A_i \:\big|\quad
			\exists \alpha\in\ICE(A'\cup\R)
			\quad\text{\BSCblack such that}\quad \alpha_i(a_i) >0
		\right\}
\end{align*}
%%%
In words, $\B(A')$ includes only those actions played with positive probability in some ICE with respect to $\Aci\cup\R$.
%%%
Now define the sequence $\{\Ac^n\}_{n\in\Natural}$ by $\Ac^1 = \Ac$ and $\Ac^{n+1} = \B(\Ac^n)$.

%%%%%%%%%%%%%%%%%%%%%%%%%%%%%%%%%%%%%%%%%%%%%%%%%%%%%%%%%%%%%%%%%%%%%%%%%%%%%%%%%%%%%%%%%%%%%%%%%%%%%%%%%%%%%%%%%%%%%%%%%%%%%%%%%%%%%%%%%
%%%%%%%%%%%%%%%%%%%%%%%%%%%%%%%%%%%%%%%%%%%%%%%%%%%%%%%%%%%%%%%%%%%%%%%%%%%%%%%%%%%%%%%%%%%%%%%%%%%%%%%%%%%%%%%%%%%%%%%%%%%%%%%%%%%%%%%%%
\begin{proposition}\label{prop:APS}
	$A^n$ is a decreasing sequence converging in finite time to a nonempty limit $\CEac$,
	and this limit satisfies $\CEac = \B(\CEac)$ and $A'\subseteq \CEac$ whenever $A'\subseteq \B(A')$.
\end{proposition}

%%%%%%%%%%%%%%%%%%%%%%%%%%%%%%%%%%%%%%%%%%%%%%%%%%%%%%%%%%%%%%%%%%%%%%%%%%%%%%%%%%%%%%%%%%%%%%%%%%%%%%%%%%%%%%%%%%%%%%%%%%%%%%%%%%%%%%%%%
%%%%%%%%%%%%%%%%%%%%%%%%%%%%%%%%%%%%%%%%%%%%%%%%%%%%%%%%%%%%%%%%%%%%%%%%%%%%%%%%%%%%%%%%%%%%%%%%%%%%%%%%%%%%%%%%%%%%%%%%%%%%%%%%%%%%%%%%%
$\ICE(\CEac\cup\R)$ is convex and $\CEac = \B(\CEac)$.
%%%
Hence there exists an ICE $\CEmac\in\ICE(\CEac)$ with support $\CEmac$.
%%%
As before, since all the information sets are reached with positive probability
in corresponding mediated game, $\CEmac$ is sequentially implementable.
%%%
Therefore $\CEac\subseteq \SEac$ and the following theorem follows from the previous discussion.%
	\footnote{
		For the theorem to hold, there must be at least three players,
		or  it must be feasible for Nature to make null choices. 
		%%%
		Otherwise, instead $\alpha$ being sequentially implementable, one can only guarantee that for every $\epsilon>0$,
		there exists a sequentially implementable distribution $\alpha'$ such that $\|\alpha - \alpha'\| < \epsilon$.
		The reason for this is made apparent in the last step of the proof. \label{foot:suff}}

%%%%%%%%%%%%%%%%%%%%%%%%%%%%%%%%%%%%%%%%%%%%%%%%%%%%%%%%%%%%%%%%%%%%%%%%%%%%%%%%%%%%%%%%%%%%%%%%%%%%%%%%%%%%%%%%%%%%%%%%%%%%%%%%%%%%%%%%%
%%%%%%%%%%%%%%%%%%%%%%%%%%%%%%%%%%%%%%%%%%%%%%%%%%%%%%%%%%%%%%%%%%%%%%%%%%%%%%%%%%%%%%%%%%%%%%%%%%%%%%%%%%%%%%%%%%%%%%%%%%%%%%%%%%%%%%%%%
\begin{theorem}\label{thm:SE}
	If $\alpha$ is an interdependent-choice equilibrium with respect to $\CEac\cup\R$, then it is sequentially implementable.
\end{theorem}

%%%%%%%%%%%%%%%%%%%%%%%%%%%%%%%%%%%%%%%%%%%%%%%%%%%%%%%%%%%%%%%%%%%%%%%%%%%%%%%%%%%%%%%%%%%%%%%%%%%%%%%%%%%%%%%%%%%%%%%%%%%%%%%%%%%%%%%%%
%%%%%%%%%%%%%%%%%%%%%%%%%%%%%%%%%%%%%%%%%%%%%%%%%%%%%%%%%%%%%%%%%%%%%%%%%%%%%%%%%%%%%%%%%%%%%%%%%%%%%%%%%%%%%%%%%%%%%%%%%%%%%%%%%%%%%%%%%
The fact that the iterative procedure is monotone implies that the conditions  from theorem \ref{thm:SE} are computationally tractable. 
%%%
In each stage of the process, if $\alpha \in \ICE(\Ac^n\cup\R)$ then $\supp(\alpha)\subseteq \Ac^n$.
%%%
Hence, the set of effective threats coincides with the set of additional threats, 
and $\ICE(\Ac^n\cup\R)$ is characterized by a finite set of affine inequalities.
%%%
With this in mind, finding $\CEac$ is much computationally easier than the APS algorithm. 
%%%
First, only pure actions are eliminated at each iteration,
which means that the elements of the sequence can be described with finite information. 
%%%
Also, $\B(\Ac^n)$ is defined by a \emph{decreasing} number of affine inequalities.
%%%
Finally, convergence is achieved in \emph{finitely many} iterations and  --since ICE is a permissive solution concept-- 
the required number of iterations should be small (if positive at all). 

%%%%%%%%%%%%%%%%%%%%%%%%%%%%%%%%%%%%%%%%%%%%%%%%%%%%%%%%%%%%%%%%%%%%%%%%%%%%%%%%%%%%%%%%%%%%%%%%%%%%%%%%%%%%%%%%%%%%%%%%%%%%%%%%%%%%%%%%%
%%%%%%%%%%%%%%%%%%%%%%%%%%%%%%%%%%%%%%%%%%%%%%%%%%%%%%%%%%%%%%%%%%%%%%%%%%%%%%%%%%%%%%%%%%%%%%%%%%%%%%%%%%%%%%%%%%%%%%%%%%%%%%%%%%%%%%%%%
The conditions are also very permissive. 
%%%
Since both the sequential structure and the off-path beliefs are design variables,
restricting attention to sequential equilibria has a relatively small impact.
%%%
For example, in environments with no strict dominance, $\CEac\cup\R = \R = \Ac$,
and thus Nash and sequential implementability coincide.
%%%
Hence the following corollary.

%%%%%%%%%%%%%%%%%%%%%%%%%%%%%%%%%%%%%%%%%%%%%%%%%%%%%%%%%%%%%%%%%%%%%%%%%%%%%%%%%%%%%%%%%%%%%%%%%%%%%%%%%%%%%%%%%%%%%%%%%%%%%%%%%%%%%%%%%
%%%%%%%%%%%%%%%%%%%%%%%%%%%%%%%%%%%%%%%%%%%%%%%%%%%%%%%%%%%%%%%%%%%%%%%%%%%%%%%%%%%%%%%%%%%%%%%%%%%%%%%%%%%%%%%%%%%%%%%%%%%%%%%%%%%%%%%%%
\begin{corollary}
	When there are no strictly dominated actions, 
	a distribution is sequentially implementable if and only if it is an interdependent-choice equilibrium.
\end{corollary}

%%%%%%%%%%%%%%%%%%%%%%%%%%%%%%%%%%%%%%%%%%%%%%%%%%%%%%%%%%%%%%%%%%%%%%%%%%%%%%%%%%%%%%%%%%%%%%%%%%%%%%%%%%%%%%%%%%%%%%%%%%%%%%%%%%%%%%%%%
%%%%%%%%%%%%%%%%%%%%%%%%%%%%%%%%%%%%%%%%%%%%%%%%%%%%%%%%%%%%%%%%%%%%%%%%%%%%%%%%%%%%%%%%%%%%%%%%%%%%%%%%%%%%%%%%%%%%%%%%%%%%%%%%%%%%%%%%%
\begin{figure}[htb]
	\centering
	\begin{tabular}{c|cccc}
		      &   \aL     &   \NC & \C  & \aR     \\
		\hline\\[-2ex]
		\aT  & $ 3 \:,\:  0$ & $ 0 \:,\:  k$ & $0 \:,\: 0$ & $0 \:,\: 3$\\[1ex]
		\NC  & $ k \:,\:  0$ & $ 6 \:,\:  6$ & $2 \:,\: 9$ & $k \:,\: 0$\\[1ex]
		\C   & $ 0 \:,\:  0$ & $ 9 \:,\:  2$ & $5 \:,\: 5$ & $0 \:,\: 0$\\[1ex]
		\aB  & $ 0 \:,\:  3$ & $ 0 \:,\:  k$ & $0 \:,\: 0$ & $3 \:,\: 0$
	\end{tabular}
	\caption{Payoff matrix for examples \ref{eg:A} and \ref{eg:B}.} \label{fig:egA}
\end{figure}

%%%%%%%%%%%%%%%%%%%%%%%%%%%%%%%%%%%%%%%%%%%%%%%%%%%%%%%%%%%%%%%%%%%%%%%%%%%%%%%%%%%%%%%%%%%%%%%%%%%%%%%%%%%%%%%%%%%%%%%%%%%%%%%%%%%%%%%%%
%%%%%%%%%%%%%%%%%%%%%%%%%%%%%%%%%%%%%%%%%%%%%%%%%%%%%%%%%%%%%%%%%%%%%%%%%%%%%%%%%%%%%%%%%%%%%%%%%%%%%%%%%%%%%%%%%%%%%%%%%%%%%%%%%%%%%%%%%
\begin{example}\label{eg:A}
	Consider the environment described by the payoff matrix in Figure \ref{fig:egA} with $k=2$.
	%%%
	The central part of the matrix corresponds to a prisoner's dilemma in which cooperation is not an ICE. 
	%%%
	Cooperation can be implemented using $\aT$ and $\aR$ as punishments.
	%%%
	Furthermore, following recommendations is a sequential equilibrium
	as long as the players consider $\aB$ and $\aL$ to be the most likely deviations.
	%%%
	However, $\aT$ and $\aR$ cannot be played with positive probability in any ICE.
	%%%
	Hence $\CEac = \{(\C,\C)\} \neq \{\C,\NC\}\times\{\C,\NC\} = \SEac$,
	and $(\NC,\NC)$ is sequentially implementable despite the fact that it is not an ICE with respect to $\SEac$.
\end{example}

%%%%%%%%%%%%%%%%%%%%%%%%%%%%%%%%%%%%%%%%%%%%%%%%%%%%%%%%%%%%%%%%%%%%%%%%%%%%%%%%%%%%%%%%%%%%%%%%%%%%%%%%%%%%%%%%%%%%%%%%%%%%%%%%%%%%%%%%%
%%%%%%%%%%%%%%%%%%%%%%%%%%%%%%%%%%%%%%%%%%%%%%%%%%%%%%%%%%%%%%%%%%%%%%%%%%%%%%%%%%%%%%%%%%%%%%%%%%%%%%%%%%%%%%%%%%%%%%%%%%%%%%%%%%%%%%%%%
It is possible to construct more complicated examples with admit distributions that can only be sequentially implemented 
using punishments outside of $\SEac\cup\R$.
%%%
This implies that the conditions from Theorem \ref{thm:SE} are not sufficient. 
%%%
I have not being able to find a \emph{tractable} characterization of the set of
sequentially implementable outcomes in general environments. 
%%%
Instead, the remainder of this section focuses on $2\times2$ environments,
and the following section analizes a notion of quasi-sequential implementation. 

%%%%%%%%%%%%%%%%%%%%%%%%%%%%%%%%%%%%%%%%%%%%%%%%%%%%%%%%%%%%%%%%%%%%%%%%%%%%%%%%%%%%%%%%%%%%%%%%%%%%%%%%%%%%%%%%%%%%%%%%%%%%%%%%%%%%%%%%%
%%%%%%%%%%%%%%%%%%%%%%%%%%%%%%%%%%%%%%%%%%%%%%%%%%%%%%%%%%%%%%%%%%%%%%%%%%%%%%%%%%%%%%%%%%%%%%%%%%%%%%%%%%%%%%%%%%%%%%%%%%%%%%%%%%%%%%%%%
%%%%%%%%%%%%%%%%%%%%%%%%%%%%%%%%%%%%%%%%%%%%%%%%%%%%%%%%%%%%%%%%%%%%%%%%%%%%%%%%%%%%%%%%%%%%%%%%%%%%%%%%%%%%%%%%%%%%%%%%%%%%%%%%%%%%%%%%%
%%%%%%%%%%%%%%%%%%%%%%%%%%%%%%%%%%%%%%%%%%%%%%%%%%%%%%%%%%%%%%%%%%%%%%%%%%%%%%%%%%%%%%%%%%%%%%%%%%%%%%%%%%%%%%%%%%%%%%%%%%%%%%%%%%%%%%%%%
%%%%%%%%%%%%%%%%%%%%%%%%%%%%%%%%%%%%%%%%%%%%%%%%%%%%%%%%%%%%%%%%%%%%%%%%%%%%%%%%%%%%%%%%%%%%%%%%%%%%%%%%%%%%%%%%%%%%%%%%%%%%%%%%%%%%%%%%%
%%%%%%%%%%%%%%%%%%%%%%%%%%%%%%%%%%%%%%%%%%%%%%%%%%%%%%%%%%%%%%%%%%%%%%%%%%%%%%%%%%%%%%%%%%%%%%%%%%%%%%%%%%%%%%%%%%%%%%%%%%%%%%%%%%%%%%%%%
%%%%%%%%%%%%%%%%%%%%%%%%%%%%%%%%%%%%%%%%%%%%%%%%%%%%%%%%%%%%%%%%%%%%%%%%%%%%%%%%%%%%%%%%%%%%%%%%%%%%%%%%%%%%%%%%%%%%%%%%%%%%%%%%%%%%%%%%%
%%%%%%%%%%%%%%%%%%%%%%%%%%%%%%%%%%%%%%%%%%%%%%%%%%%%%%%%%%%%%%%%%%%%%%%%%%%%%%%%%%%%%%%%%%%%%%%%%%%%%%%%%%%%%%%%%%%%%%%%%%%%%%%%%%%%%%%%%
%%%%%%%%%%%%%%%%%%%%%%%%%%%%%%%%%%%%%%%%%%%%%%%%%%%%%%%%%%%%%%%%%%%%%%%%%%%%%%%%%%%%%%%%%%%%%%%%%%%%%%%%%%%%%%%%%%%%%%%%%%%%%%%%%%%%%%%%%
%%%%%%%%%%%%%%%%%%%%%%%%%%%%%%%%%%%%%%%%%%%%%%%%%%%%%%%%%%%%%%%%%%%%%%%%%%%%%%%%%%%%%%%%%%%%%%%%%%%%%%%%%%%%%%%%%%%%%%%%%%%%%%% ssec: 2x2
\subsection{C-rationalizability and 2 by 2 environments}\label{ssec:Crat}

%%%%%%%%%%%%%%%%%%%%%%%%%%%%%%%%%%%%%%%%%%%%%%%%%%%%%%%%%%%%%%%%%%%%%%%%%%%%%%%%%%%%%%%%%%%%%%%%%%%%%%%%%%%%%%%%%%%%%%%%%%%%%%%%%%%%%%%%%
%%%%%%%%%%%%%%%%%%%%%%%%%%%%%%%%%%%%%%%%%%%%%%%%%%%%%%%%%%%%%%%%%%%%%%%%%%%%%%%%%%%%%%%%%%%%%%%%%%%%%%%%%%%%%%%%%%%%%%%%%%%%%%%%%%%%%%%%%
A natural restriction on the set of credible threats is that every punishment  should be rationalizable,
i.e.\ it should be a best response to some rational belief of the player performing it. %\citep{bernheim,pearce}
%%% 
The relevant notion of rationalizability must take into account choice-interdependence.%
	\footnote{This notion was developed independently by \cite{halpern12}, who additionally show that,
		when counterfactual beliefs are allowed, it is equivalent to rationality and common certainty of rationality.}  

%%%%%%%%%%%%%%%%%%%%%%%%%%%%%%%%%%%%%%%%%%%%%%%%%%%%%%%%%%%%%%%%%%%%%%%%%%%%%%%%%%%%%%%%%%%%%%%%%%%%%%%%%%%%%%%%%%%%%%%%%%%%%%%%%%%%%%%%%
%%%%%%%%%%%%%%%%%%%%%%%%%%%%%%%%%%%%%%%%%%%%%%%%%%%%%%%%%%%%%%%%%%%%%%%%%%%%%%%%%%%%%%%%%%%%%%%%%%%%%%%%%%%%%%%%%%%%%%%%%%%%%%%%%%%%%%%%%
At the moment of choosing his action, each player's belief about his opponent behavior may depend on his own choice.
%%%
Players thus maximize expected utility with respect to \emph{counterfactual beliefs}
$\lambda_i:A_i\rightarrow\Delta(A_{-i})$, 
%%%
where $\lambda_i(a_{-i}|a_i) \equiv [\lambda_i(a_i)](a_{-i})$ represents $i$'s assessed likelihood 
that his opponent will choose to play $a_{-i}$ \emph{if he plays} $a_i$.
%%%
Expected utility is denote by  $\uc_i(a_i,\lambda_i) = \sum_{a_{-i}\in A_{-i}} \ua_i(a_i,a_{-i})\cdot \lambda_i(a_{-i}|a_i)$.
%%%
Given an action space $A'\in\A$, $\Lambda_i(A')$ denotes the set of counterfactual beliefs
such that $\lambda_i(A_{-i}|a_i)=1$ for every $a_i\in A_i$.

%%%%%%%%%%%%%%%%%%%%%%%%%%%%%%%%%%%%%%%%%%%%%%%%%%%%%%%%%%%%%%%%%%%%%%%%%%%%%%%%%%%%%%%%%%%%%%%%%%%%%%%%%%%%%%%%%%%%%%%%%%%%%%%%%%%%%%%%%
%%%%%%%%%%%%%%%%%%%%%%%%%%%%%%%%%%%%%%%%%%%%%%%%%%%%%%%%%%%%%%%%%%%%%%%%%%%%%%%%%%%%%%%%%%%%%%%%%%%%%%%%%%%%%%%%%%%%%%%%%%%%%%%%%%%%%%%%%
\begin{definition}[Counterfactual rationalizability] {}\
	\begin{itemize}
	\item $a^*_i\in A_i$ is \emph{C-rationalizable} with respect to $A'\in\A$, if and only if
		there exists some $\lambda_i\in\Lambda_i(A')$ such that $a^*_i \in \argmax_{a_i \in A_i} \uc_i(a_i,\lambda_i)$.
	\item $A'\in\A$ is \emph{self-C-rationalizable} if and only if every action profile in $\Aci$
		consists of actions that are C-rationalizable with respect to $\Aci$.
	\item The set of \emph{C-rationalizable actions} $\CRR$
		is the largest self-C-rationalizable set. 
	\end{itemize}
\end{definition}

%%%%%%%%%%%%%%%%%%%%%%%%%%%%%%%%%%%%%%%%%%%%%%%%%%%%%%%%%%%%%%%%%%%%%%%%%%%%%%%%%%%%%%%%%%%%%%%%%%%%%%%%%%%%%%%%%%%%%%%%%%%%%%%%%%%%%%%%%
%%%%%%%%%%%%%%%%%%%%%%%%%%%%%%%%%%%%%%%%%%%%%%%%%%%%%%%%%%%%%%%%%%%%%%%%%%%%%%%%%%%%%%%%%%%%%%%%%%%%%%%%%%%%%%%%%%%%%%%%%%%%%%%%%%%%%%%%%
Let $\CR_i(A')$ denote the set of $i$'s actions that are C-rationalzable with respect to $A'$.
%%%
$\CRR$ is guaranteed to exist because $\CR(\blank)$ is $\subseteq$-monotone.
%%%
Consequently, the union of all self C-rationalizable sets is also self C-rationalizable. 
%%%
It is nonempty because it always contains the set of rationalizable action profiles. 
%%%
$\CRR$ can be found in a tractable way using the notion of absolute dominance,
which is analogous to the standard notion of strict dominance.

%%%%%%%%%%%%%%%%%%%%%%%%%%%%%%%%%%%%%%%%%%%%%%%%%%%%%%%%%%%%%%%%%%%%%%%%%%%%%%%%%%%%%%%%%%%%%%%%%%%%%%%%%%%%%%%%%%%%%%%%%%%%%%%%%%%%%%%%%
%%%%%%%%%%%%%%%%%%%%%%%%%%%%%%%%%%%%%%%%%%%%%%%%%%%%%%%%%%%%%%%%%%%%%%%%%%%%%%%%%%%%%%%%%%%%%%%%%%%%%%%%%%%%%%%%%%%%%%%%%%%%%%%%%%%%%%%%%
\begin{definition}
	Given two actions $a_i,a_i' \in A_i$, 
	$a_i$ \emph{absolutely dominates} $a_i'$ with respect to $A'\in\A$ if and only if
	$\min_{a_{-i}\in A_{-i}'} \ua_i(a_i,a_{-i}) > \max_{a_{-i}\in A_{-i}'} \ua_i(a_i',a_{-i})$.
\end{definition}

%%%%%%%%%%%%%%%%%%%%%%%%%%%%%%%%%%%%%%%%%%%%%%%%%%%%%%%%%%%%%%%%%%%%%%%%%%%%%%%%%%%%%%%%%%%%%%%%%%%%%%%%%%%%%%%%%%%%%%%%%%%%%%%%%%%%%%%%%
%%%%%%%%%%%%%%%%%%%%%%%%%%%%%%%%%%%%%%%%%%%%%%%%%%%%%%%%%%%%%%%%%%%%%%%%%%%%%%%%%%%%%%%%%%%%%%%%%%%%%%%%%%%%%%%%%%%%%%%%%%%%%%%%%%%%%%%%%
In other words, $a_i$ absolutely dominates $a_i'$, if and only if
the best possible payoff from playing $a_i'$ is strictly worse than the worst possible payoff from playing $a_i$.
%%%
Absolute dominance is much simpler than strict dominance in computational terms because a player can
conjecture different reactions for each alternative action, and hence mixed actions need not be considered.
%%% 
The following proposition ensures that $\CR(A')$  can be obtained by eliminating absolutely dominated actions,
and $\CRR$ can be found by repeating this process iteratively.

%%%%%%%%%%%%%%%%%%%%%%%%%%%%%%%%%%%%%%%%%%%%%%%%%%%%%%%%%%%%%%%%%%%%%%%%%%%%%%%%%%%%%%%%%%%%%%%%%%%%%%%%%%%%%%%%%%%%%%%%%%%%%%%%%%%%%%%%%
%%%%%%%%%%%%%%%%%%%%%%%%%%%%%%%%%%%%%%%%%%%%%%%%%%%%%%%%%%%%%%%%%%%%%%%%%%%%%%%%%%%%%%%%%%%%%%%%%%%%%%%%%%%%%%%%%%%%%%%%%%%%%%%%%%%%%%%%%
\begin{proposition}\label{prop:Crat}
	An action is C-rationalizable with respect to $\Aci$  if and only if it is not absolutely dominated in $\Aci$,
	and the iterated removal of all absolutely dominated actions is order independent and converges in finite time to $\CRR$.
\end{proposition}

%%%%%%%%%%%%%%%%%%%%%%%%%%%%%%%%%%%%%%%%%%%%%%%%%%%%%%%%%%%%%%%%%%%%%%%%%%%%%%%%%%%%%%%%%%%%%%%%%%%%%%%%%%%%%%%%%%%%%%%%%%%%%%%%%%%%%%%%%
%%%%%%%%%%%%%%%%%%%%%%%%%%%%%%%%%%%%%%%%%%%%%%%%%%%%%%%%%%%%%%%%%%%%%%%%%%%%%%%%%%%%%%%%%%%%%%%%%%%%%%%%%%%%%%%%%%%%%%%%%%%%%%%%%%%%%%%%%
In $2\times 2$ environments without repeated payoffs there are two possibilities. 
%%%
If there are no absolutely dominated actions, then there is an ICE with full support and hence $\CEac = \Ac$.
%%%
Otherwise, there is a unique ICE with respect to $\CEac$;
namely, a player chooses his unique dominant action  and his opponent chooses the unique best response to it. 
%%%
In view of Theorem \ref{thm:SE}, this results in the following characterization of
sequential implementation for $2\times2$ environments without repeated payoffs.

%%%%%%%%%%%%%%%%%%%%%%%%%%%%%%%%%%%%%%%%%%%%%%%%%%%%%%%%%%%%%%%%%%%%%%%%%%%%%%%%%%%%%%%%%%%%%%%%%%%%%%%%%%%%%%%%%%%%%%%%%%%%%%%%%%%%%%%%%
%%%%%%%%%%%%%%%%%%%%%%%%%%%%%%%%%%%%%%%%%%%%%%%%%%%%%%%%%%%%%%%%%%%%%%%%%%%%%%%%%%%%%%%%%%%%%%%%%%%%%%%%%%%%%%%%%%%%%%%%%%%%%%%%%%%%%%%%%
\begin{proposition}\label{prop:2by2}
	In generic $2\times 2$ environments, a distribution is sequentially implementable
	if and only if it is a interdependent-choice equilibrium with respect to the set of self-C-rationalizable action profiles.
\end{proposition}

%%%%%%%%%%%%%%%%%%%%%%%%%%%%%%%%%%%%%%%%%%%%%%%%%%%%%%%%%%%%%%%%%%%%%%%%%%%%%%%%%%%%%%%%%%%%%%%%%%%%%%%%%%%%%%%%%%%%%%%%%%%%%%%%%%%%%%%%%
%%%%%%%%%%%%%%%%%%%%%%%%%%%%%%%%%%%%%%%%%%%%%%%%%%%%%%%%%%%%%%%%%%%%%%%%%%%%%%%%%%%%%%%%%%%%%%%%%%%%%%%%%%%%%%%%%%%%%%%%%%%%%%%%%%%%%%%%%
%%%%%%%%%%%%%%%%%%%%%%%%%%%%%%%%%%%%%%%%%%%%%%%%%%%%%%%%%%%%%%%%%%%%%%%%%%%%%%%%%%%%%%%%%%%%%%%%%%%%%%%%%%%%%%%%%%%%%%%%%%%%%%%%%%%%%%%%%
%%%%%%%%%%%%%%%%%%%%%%%%%%%%%%%%%%%%%%%%%%%%%%%%%%%%%%%%%%%%%%%%%%%%%%%%%%%%%%%%%%%%%%%%%%%%%%%%%%%%%%%%%%%%%%%%%%%%%%%%%%%%%%%%%%%%%%%%%
%%%%%%%%%%%%%%%%%%%%%%%%%%%%%%%%%%%%%%%%%%%%%%%%%%%%%%%%%%%%%%%%%%%%%%%%%%%%%%%%%%%%%%%%%%%%%%%%%%%%%%%%%%%%%%%%%%%%%%%%%%%%%%%%%%%%%%%%%
%%%%%%%%%%%%%%%%%%%%%%%%%%%%%%%%%%%%%%%%%%%%%%%%%%%%%%%%%%%%%%%%%%%%%%%%%%%%%%%%%%%%%%%%%%%%%%%%%%%%%%%%%%%%%%%%%%%%%%%%%%%%%%%%%%%%%%%%%
%%%%%%%%%%%%%%%%%%%%%%%%%%%%%%%%%%%%%%%%%%%%%%%%%%%%%%%%%%%%%%%%%%%%%%%%%%%%%%%%%%%%%%%%%%%%%%%%%%%%%%%%%%%%%%%%%%%%%%%%%%%%%%%%%%%%%%%%%
%%%%%%%%%%%%%%%%%%%%%%%%%%%%%%%%%%%%%%%%%%%%%%%%%%%%%%%%%%%%%%%%%%%%%%%%%%%%%%%%%%%%%%%%%%%%%%%%%%%%%%%%%%%%%%%%%%%%%%%%%%%%%%%%%%%%%%%%%
%%%%%%%%%%%%%%%%%%%%%%%%%%%%%%%%%%%%%%%%%%%%%%%%%%%%%%%%%%%%%%%%%%%%%%%%%%%%%%%%%%%%%%%%%%%%%%%%%%%%%%%%%%%%%%%%%%%%%%%%%%%%%%%%%%%%%%%%%
%%%%%%%%%%%%%%%%%%%%%%%%%%%%%%%%%%%%%%%%%%%%%%%%%%%%%%%%%%%%%%%%%%%%%%%%%%%%%%%%%%%%%%%%%%%%%%%%%%%%%%%%%%%%%%%%%%%%%%%%%% END OF SECTION

%% file: QSI.tex
%auto-ignore

%%%%%%%%%%%%%%%%%%%%%%%%%%%%%%%%%%%%%%%%%%%%%%%%%%%%%%%%%%%%%%%%%%%%%%%%%%%%%%%%%%%%%%%%%%%%%%%%%%%%%%%%%%%%%%%%%%%%%%%%%%%%%%%%%%%%%%%%%
%%%%%%%%%%%%%%%%%%%%%%%%%%%%%%%%%%%%%%%%%%%%%%%%%%%%%%%%%%%%%%%%%%%%%%%%%%%%%%%%%%%%%%%%%%%%%%%%%%%%%%%%%%%%%%%%%%%%%%%%%%%%%%%%%%%%%%%%%
% Section: Perfect Bayesian implementation
% Project: Coordination in Strategic Environments
%  Author: Bruno Salcedo
% Version: 7.1
%    Date: 09/08/13
%%%%%%%%%%%%%%%%%%%%%%%%%%%%%%%%%%%%%%%%%%%%%%%%%%%%%%%%%%%%%%%%%%%%%%%%%%%%%%%%%%%%%%%%%%%%%%%%%%%%%%%%%%%%%%%%%%%%%%%%%%%%%%%%%%%%%%%%%
%%%%%%%%%%%%%%%%%%%%%%%%%%%%%%%%%%%%%%%%%%%%%%%%%%%%%%%%%%%%%%%%%%%%%%%%%%%%%%%%%%%%%%%%%%%%%%%%%%%%%%%%%%%%%%%%%%%%%%%%%%%%%%%%%%%%%%%%%

%%%%%%%%%%%%%%%%%%%%%%%%%%%%%%%%%%%%%%%%%%%%%%%%%%%%%%%%%%%%%%%%%%%%%%%%%%%%%%%%%%%%%%%%%%%%%%%%%%%%%%%%%%%%%%%%%%%%%%%%%%%%%%%%%%%%%%%%%
%%%%%%%%%%%%%%%%%%%%%%%%%%%%%%%%%%%%%%%%%%%%%%%%%%%%%%%%%%%%%%%%%%%%%%%%%%%%%%%%%%%%%%%%%%%%%%%%%%%%%%%%%%%%%%%%%%%%%%%%%%%%%%%%%%%%%%%%%
This section defines a form of \emph{quasi-seqeuntial equilibrium} (QSE),
and provides sufficient and necessary conditions for quasi-sequential (QS) implementation.
%%%
Sequential equilibria are QSE, and thus these conditions are also necessary for sequential implementation. 
%%%
The focus on QSE is partially motivated by the fact that it is the finer refinement
for which I can provide a complete characterization.
%%%
However, QSE might be an interesting solution concept in its own right,
see section \ref{disc:QSE}.

%%%%%%%%%%%%%%%%%%%%%%%%%%%%%%%%%%%%%%%%%%%%%%%%%%%%%%%%%%%%%%%%%%%%%%%%%%%%%%%%%%%%%%%%%%%%%%%%%%%%%%%%%%%%%%%%%%%%%%%%%%%%%%%%%%%%%%%%%
%%%%%%%%%%%%%%%%%%%%%%%%%%%%%%%%%%%%%%%%%%%%%%%%%%%%%%%%%%%%%%%%%%%%%%%%%%%%%%%%%%%%%%%%%%%%%%%%%%%%%%%%%%%%%%%%%%%%%%%%%%%%%%%%%%%%%%%%%
%%%%%%%%%%%%%%%%%%%%%%%%%%%%%%%%%%%%%%%%%%%%%%%%%%%%%%%%%%%%%%%%%%%%%%%%%%%%%%%%%%%%%%%%%%%%%%%%%%%%%%%%%%%%%%%%%%%%%%%%%%%%%%%%%%%%%%%%%
%%%%%%%%%%%%%%%%%%%%%%%%%%%%%%%%%%%%%%%%%%%%%%%%%%%%%%%%%%%%%%%%%%%%%%%%%%%%%%%%%%%%%%%%%%%%%%%%%%%%%%%%%%%%%%%%%%%%%%%%%%%%%%%%%%%%%%%%%
%%%%%%%%%%%%%%%%%%%%%%%%%%%%%%%%%%%%%%%%%%%%%%%%%%%%%%%%%%%%%%%%%%%%%%%%%%%%%%%%%%%%%%%%%%%%%%%%%%%%%%%%%%%%%%%%%%%%%%%%%%%%%%%%%%%%%%%%%
%%%%%%%%%%%%%%%%%%%%%%%%%%%%%%%%%%%%%%%%%%%%%%%%%%%%%%%%%%%%%%%%%%%%%%%%%%%%%%%%%%%%%%%%%%%%%%%%%%%%%%%%%%%%%%%%%%%%%%%%%%%%%%%%%%%%%%%%%
%%%%%%%%%%%%%%%%%%%%%%%%%%%%%%%%%%%%%%%%%%%%%%%%%%%%%%%%%%%%%%%%%%%%%%%%%%%%%%%%%%%%%%%%%%%%%%%%%%%%%%%%%%%%%%%%%%%%%%%%%%%%%%%%%%%%%%%%%
%%%%%%%%%%%%%%%%%%%%%%%%%%%%%%%%%%%%%%%%%%%%%%%%%%%%%%%%%%%%%%%%%%%%%%%%%%%%%%%%%%%%%%%%%%%%%%%%%%%%%%%%%%%%%%%%%%%%%%%%%%%%%%%%%%%%%%%%%
%%%%%%%%%%%%%%%%%%%%%%%%%%%%%%%%%%%%%%%%%%%%%%%%%%%%%%%%%%%%%%%%%%%%%%%%%%%%%%%%%%%%%%%%%%%%%%%%%%%%%%%%%%%%%%%%%%%%%%%%%%%%%%%%%%%%%%%%%
%%%%%%%%%%%%%%%%%%%%%%%%%%%%%%%%%%%%%%%%%%%%%%%%%%%%%%%%%%%%%%%%%%%%%%%%%%%%%%%%%%%%%%%%%%%%%%%%%%%%%%%%%%%%%%%%%%%%%%%%%%%%%%% ssec: QSE
\subsection{Quasi-sequential equilibrium} \label{ssec:QSE}
Sequential equilibrium is defined in terms of \emph{sequential rationality} and \emph{belief consistency}. 
%%%
Sequential rationality requires choices to be optimal at the interim stage for every information set in the game.
%%%
Off the equilibrium path, belief consistency requires players to update their beliefs in accordance with
some prior assessment of the relative likelihoods of different trembles or mistakes.
%%%
Furthermore, it requires that these prior assessments should be common to all players.
%%%
QSE imposes sequential rationality and requires beliefs to be consistent with trembles,
but allows players to disagree about which deviations are more likely.

%%%%%%%%%%%%%%%%%%%%%%%%%%%%%%%%%%%%%%%%%%%%%%%%%%%%%%%%%%%%%%%%%%%%%%%%%%%%%%%%%%%%%%%%%%%%%%%%%%%%%%%%%%%%%%%%%%%%%%%%%%%%%%%%%%%%%%%%%
%%%%%%%%%%%%%%%%%%%%%%%%%%%%%%%%%%%%%%%%%%%%%%%%%%%%%%%%%%%%%%%%%%%%%%%%%%%%%%%%%%%%%%%%%%%%%%%%%%%%%%%%%%%%%%%%%%%%%%%%%%%%%%%%%%%%%%%%%
For two player environments, it is useful to allow Nature to assign zero probability to some of its available moves.
%%%
This is because, when faced with a null event, a player can believe that it was Nature who made a  mistake
instead of necessarily believing that an opponent deviated from the equilibrium.%
	\footnote{
		It is often assumed that Nature assigns positive probability to all of its available moves,
		but I am unaware of any good arguments to maintain this assumption.
		Consider for instance the following quote from \cite{KreWil}:
		``To keep matters simple, we henceforth assume that the players initial assessments
		[on Nature's choices] are strictly positive'', page 868.
		For further discussion see section \S\ref{disc:mistakes}.}
%%%
In order to define consistent beliefs, 
it is necessary to introduce new notation to denote players' beliefs about Nature's choices, other than $\chances$.
%%%
Let $\Chance$ and  $\Chance^+$ denote the sets of mixed strategies and strictly mixed strategies for Nature.

%%%%%%%%%%%%%%%%%%%%%%%%%%%%%%%%%%%%%%%%%%%%%%%%%%%%%%%%%%%%%%%%%%%%%%%%%%%%%%%%%%%%%%%%%%%%%%%%%%%%%%%%%%%%%%%%%%%%%%%%%%%%%%%%%%%%%%%%%
%%%%%%%%%%%%%%%%%%%%%%%%%%%%%%%%%%%%%%%%%%%%%%%%%%%%%%%%%%%%%%%%%%%%%%%%%%%%%%%%%%%%%%%%%%%%%%%%%%%%%%%%%%%%%%%%%%%%%%%%%%%%%%%%%%%%%%%%%
A conditional belief system for $i$ in an extensive form game $G$,
is a function $\psi_i$ mapping $i$'s information sets to distributions over nodes. 
%%%
$\psi_i(\decision|\h)$ is the probability that $i$ assigns in information set $\h$ to being in node $\decision$.%
	\footnote{See \S\ref{ssec:EFG} for the notation regarding extensive form games}
%%%
Let $\Psi_i$ denote the set of $i$'s conditional belief systems.
%%%
An assessment is a tuple $(\psi,\sigma)\in\Psi\times\Sigma$ that specifies both players interim and prior beliefs (or strategies).
%%%
An extended assessment is a tuple $(\psi,\sigma,\chance)\in\Psi\times\Sigma\times\Sigma_0$
that also specifies prior beliefs on Nature's choices.
%%%
Given an assessment $(\psi,\sigma)$, an information set $\h$ and an available move $\move$,
$\us_i(\move|\h)$ denotes $i$'s expected payoff from choosing $\move$ at $\h$.
%%%
The expectation is taken given his interim beliefs $\psi_i(\h)$ regarding the current state of the game,
and assuming that future choices will be made according to $\sigma$.

%%%%%%%%%%%%%%%%%%%%%%%%%%%%%%%%%%%%%%%%%%%%%%%%%%%%%%%%%%%%%%%%%%%%%%%%%%%%%%%%%%%%%%%%%%%%%%%%%%%%%%%%%%%%%%%%%%%%%%%%%%%%%%%%%%%%%%%%%
%%%%%%%%%%%%%%%%%%%%%%%%%%%%%%%%%%%%%%%%%%%%%%%%%%%%%%%%%%%%%%%%%%%%%%%%%%%%%%%%%%%%%%%%%%%%%%%%%%%%%%%%%%%%%%%%%%%%%%%%%%%%%%%%%%%%%%%%%
\begin{definition}[Quasi-sequential equilibrium] \label{defn:QSE} 
	An assessment $(\psi^*,\sigma^*)\in\Psi\times\Sigma$ is:
	\begin{itemize}
	\item \emph{Weakly consistent}\/ if and only if for every player 
		there exists a sequence of strictly mixed extended assessments  $(\psi^n,\sigma^n,\chance^n)$ 
		satisfying Bayes' rule and converging to $(\psi^*,\sigma^*,\chances)$.
	\item \emph{Sequentially rational}\/ if and only if $\us_i(\strat(\h)|\h)\geq \us_i(\move|\h)$ 
		for every player $i$, information set $\h\in\H_i$ and available move $\move\in\Moves(\h)$,
		and every strategy $s_i\in S_i$ such that $\sigma^*_i(s_i)>0$.
	\item A \emph{quasi-sequential equilibrium} (QSE) if it is both weakly consistent and sequentially rational.
	\end{itemize}
\end{definition} 

%%%%%%%%%%%%%%%%%%%%%%%%%%%%%%%%%%%%%%%%%%%%%%%%%%%%%%%%%%%%%%%%%%%%%%%%%%%%%%%%%%%%%%%%%%%%%%%%%%%%%%%%%%%%%%%%%%%%%%%%%%%%%%%%%%%%%%%%%
%%%%%%%%%%%%%%%%%%%%%%%%%%%%%%%%%%%%%%%%%%%%%%%%%%%%%%%%%%%%%%%%%%%%%%%%%%%%%%%%%%%%%%%%%%%%%%%%%%%%%%%%%%%%%%%%%%%%%%%%%%%%%%%%%%%%%%%%%
Sequential rationality requires that the choices that occur  off the equilibrium path should be optimal.  
%%%
This implies that players must always believe that the \emph{future} choices of their opponents will be rational,
and this fact is common knowledge. 
%%%
However, off the equilibrium path, QSE imposes no restrictions on beliefs about \emph{past} choices,
nor agreement of beliefs across different players. 
%%%
In that sense, the difference between QSE and Nash equilibrium can be thought of as a form of
\emph{future-looking} rationalizability off the equilibrium path.%
	\footnote{This idea closely resembles the notion of common belief in future rationality from \cite{perea13}.}
	
%%%%%%%%%%%%%%%%%%%%%%%%%%%%%%%%%%%%%%%%%%%%%%%%%%%%%%%%%%%%%%%%%%%%%%%%%%%%%%%%%%%%%%%%%%%%%%%%%%%%%%%%%%%%%%%%%%%%%%%%%%%%%%%%%%%%%%%%%
%%%%%%%%%%%%%%%%%%%%%%%%%%%%%%%%%%%%%%%%%%%%%%%%%%%%%%%%%%%%%%%%%%%%%%%%%%%%%%%%%%%%%%%%%%%%%%%%%%%%%%%%%%%%%%%%%%%%%%%%%%%%%%%%%%%%%%%%%
The only difference between QSE and sequential equilibrium, is that the former imposes a stronger notion of consistency. 
%%%
Namely, \emph{the same} sequence of strictly mixed assessments should work for all players. 
%%%
Loosely speaking, sequential equilibrium requires choices and beliefs to be in equilibrium,
not only along the equilibrium path, but also in every `subgame'.
%%%
In contrast, QSE requires equilibrium along the equilibrium path, but only imposes a form of rationalizability in null `subgames'.

%%%%%%%%%%%%%%%%%%%%%%%%%%%%%%%%%%%%%%%%%%%%%%%%%%%%%%%%%%%%%%%%%%%%%%%%%%%%%%%%%%%%%%%%%%%%%%%%%%%%%%%%%%%%%%%%%%%%%%%%%%%%%%%%%%%%%%%%%
%%%%%%%%%%%%%%%%%%%%%%%%%%%%%%%%%%%%%%%%%%%%%%%%%%%%%%%%%%%%%%%%%%%%%%%%%%%%%%%%%%%%%%%%%%%%%%%%%%%%%%%%%%%%%%%%%%%%%%%%%%%%%%%%%%%%%%%%%
%%%%%%%%%%%%%%%%%%%%%%%%%%%%%%%%%%%%%%%%%%%%%%%%%%%%%%%%%%%%%%%%%%%%%%%%%%%%%%%%%%%%%%%%%%%%%%%%%%%%%%%%%%%%%%%%%%%%%%%%%%%%%%%%%%%%%%%%%
%%%%%%%%%%%%%%%%%%%%%%%%%%%%%%%%%%%%%%%%%%%%%%%%%%%%%%%%%%%%%%%%%%%%%%%%%%%%%%%%%%%%%%%%%%%%%%%%%%%%%%%%%%%%%%%%%%%%%%%%%%%%%%%%%%%%%%%%%
%%%%%%%%%%%%%%%%%%%%%%%%%%%%%%%%%%%%%%%%%%%%%%%%%%%%%%%%%%%%%%%%%%%%%%%%%%%%%%%%%%%%%%%%%%%%%%%%%%%%%%%%%%%%%%%%%%%%%%%%%%%%%%%%%%%%%%%%%
%%%%%%%%%%%%%%%%%%%%%%%%%%%%%%%%%%%%%%%%%%%%%%%%%%%%%%%%%%%%%%%%%%%%%%%%%%%%%%%%%%%%%%%%%%%%%%%%%%%%%%%%%%%%%%%%%%%%%%%%%%%%%%%%%%%%%%%%%
%%%%%%%%%%%%%%%%%%%%%%%%%%%%%%%%%%%%%%%%%%%%%%%%%%%%%%%%%%%%%%%%%%%%%%%%%%%%%%%%%%%%%%%%%%%%%%%%%%%%%%%%%%%%%%%%%%%%%%%%%%%%%%%%%%%%%%%%%
%%%%%%%%%%%%%%%%%%%%%%%%%%%%%%%%%%%%%%%%%%%%%%%%%%%%%%%%%%%%%%%%%%%%%%%%%%%%%%%%%%%%%%%%%%%%%%%%%%%%%%%%%%%%%%%%%%%%%%%%%%%%%%%%%%%%%%%%%
%%%%%%%%%%%%%%%%%%%%%%%%%%%%%%%%%%%%%%%%%%%%%%%%%%%%%%%%%%%%%%%%%%%%%%%%%%%%%%%%%%%%%%%%%%%%%%%%%%%%%%%%%%%%%%%%%%%%%%%%%%%%%%%%%%%%%%%%%
%%%%%%%%%%%%%%%%%%%%%%%%%%%%%%%%%%%%%%%%%%%%%%%%%%%%%%%%%%%%%%%%%%%%%%%%%%%%%%%%%%%%%%%%%%%%%%%%%%%%%%%%%%%%%%%%%%%%%%%%%%%%%%% ssec: QSI
\subsection{Credible threats for quasi-sequential implementation} \label{ssec:QSI}

%%%%%%%%%%%%%%%%%%%%%%%%%%%%%%%%%%%%%%%%%%%%%%%%%%%%%%%%%%%%%%%%%%%%%%%%%%%%%%%%%%%%%%%%%%%%%%%%%%%%%%%%%%%%%%%%%%%%%%%%%%%%%%%%%%%%%%%%%
%%%%%%%%%%%%%%%%%%%%%%%%%%%%%%%%%%%%%%%%%%%%%%%%%%%%%%%%%%%%%%%%%%%%%%%%%%%%%%%%%%%%%%%%%%%%%%%%%%%%%%%%%%%%%%%%%%%%%%%%%%%%%%%%%%%%%%%%%
The preceding discussion suggests two kind of actions which can always be enforced as credible punishments for QS implementation.
%%%
C-rationalizable punishments are admissible because QSE implementation does not require agreement off the equilibrium path.
%%%
Hence, the player performing the punishment may very well have counterfactual beliefs which rationalize it.
%%%
Additionally, since QSE only imposes belief of rationality for future choices, beliefs about \emph{past} can be chosen freely.
%%%
Best responses to arbitrary \emph{degenerate} conjectures are thus also admissible. 
%%%
These two ideas are embodied in the notion of future-looking counterfactual rationalizablity.

%%%%%%%%%%%%%%%%%%%%%%%%%%%%%%%%%%%%%%%%%%%%%%%%%%%%%%%%%%%%%%%%%%%%%%%%%%%%%%%%%%%%%%%%%%%%%%%%%%%%%%%%%%%%%%%%%%%%%%%%%%%%%%%%%%%%%%%%%
%%%%%%%%%%%%%%%%%%%%%%%%%%%%%%%%%%%%%%%%%%%%%%%%%%%%%%%%%%%%%%%%%%%%%%%%%%%%%%%%%%%%%%%%%%%%%%%%%%%%%%%%%%%%%%%%%%%%%%%%%%%%%%%%%%%%%%%%%
\begin{definition}[Future-looking counterfactual rationalizability] {}\
	\begin{itemize}
	\item $a^*_i\in A_i$ is \emph{FC-rationalizable} with respect to $A'\in\A$ if and only if
		there exists a belief $\lambda_i^0\in\Delta(A_{-i})$, a counterfactual belief $\lambda_i^1\in\Lambda(A')$,
		and some $\mu\in[0,1]$ such that $a^*_i$ maximizes expected utility with respect to the counterfactual belief
		$\lambda_i = \mu\lambda_i^0 + (1-\mu)\lambda_i^1 \in \Lambda_i(A)$.
		Let $\FR_i(A')$ denote the set of profiles consisting of FC-rationalzable actions with respect to $A'$.
	\item $A'\in\A$ is \emph{self-FC-rationalizable} if and only if $A'\subseteq\FR(A')$.
	\item The set of \emph{FC-rationalizable} action profiles $\FLR\in\A$ is the largest self-FC-rationalizable set. 
\end{itemize}
\end{definition}

%%%%%%%%%%%%%%%%%%%%%%%%%%%%%%%%%%%%%%%%%%%%%%%%%%%%%%%%%%%%%%%%%%%%%%%%%%%%%%%%%%%%%%%%%%%%%%%%%%%%%%%%%%%%%%%%%%%%%%%%%%%%%%%%%%%%%%%%%
%%%%%%%%%%%%%%%%%%%%%%%%%%%%%%%%%%%%%%%%%%%%%%%%%%%%%%%%%%%%%%%%%%%%%%%%%%%%%%%%%%%%%%%%%%%%%%%%%%%%%%%%%%%%%%%%%%%%%%%%%%%%%%%%%%%%%%%%%
As before, $\FLR$ is guaranteed to exist because $\FR(\blank)$ is $\subseteq$-monotone,
and thus the union of all self-FC-rationalizable sets is self-FC-rationalizable. 
%%%
Also, it is non-empty because it always contains the set of C-rationalizable action profiles. 

%%%%%%%%%%%%%%%%%%%%%%%%%%%%%%%%%%%%%%%%%%%%%%%%%%%%%%%%%%%%%%%%%%%%%%%%%%%%%%%%%%%%%%%%%%%%%%%%%%%%%%%%%%%%%%%%%%%%%%%%%%%%%%%%%%%%%%%%%
%%%%%%%%%%%%%%%%%%%%%%%%%%%%%%%%%%%%%%%%%%%%%%%%%%%%%%%%%%%%%%%%%%%%%%%%%%%%%%%%%%%%%%%%%%%%%%%%%%%%%%%%%%%%%%%%%%%%%%%%%%%%%%%%%%%%%%%%%
Intuitively, one can think of $\lambda_i^0$  as the arbitrary beliefs (degenerate conjectures) over past deviations, 
and  think of $\lambda_i^1$ as the conjectures about future FC-rationalizable choices. 
%%%
With this interpretation, an action $a_i$ is FC-rationalizable with respect to an action space $A'$
if it is a best response to some conjecture $\lambda_i$ that assigns full probability to actions in $A'_{-i}$,
\emph{only for choices that occur in the future}.
%%%
$\lambda_i$ can assign positive probability to any action, provided that this probability is independent from $i$'s choice.
%%%
The set of FC-rationalizable actions is exactly the set of credible threats that characterizes QS implementation. 

%%%%%%%%%%%%%%%%%%%%%%%%%%%%%%%%%%%%%%%%%%%%%%%%%%%%%%%%%%%%%%%%%%%%%%%%%%%%%%%%%%%%%%%%%%%%%%%%%%%%%%%%%%%%%%%%%%%%%%%%%%%%%%%%%%%%%%%%%
%%%%%%%%%%%%%%%%%%%%%%%%%%%%%%%%%%%%%%%%%%%%%%%%%%%%%%%%%%%%%%%%%%%%%%%%%%%%%%%%%%%%%%%%%%%%%%%%%%%%%%%%%%%%%%%%%%%%%%%%%%%%%%%%%%%%%%%%%
\begin{theorem}\label{thm:QSI}
	A distribution over action profiles is quasi-sequentially implementable 
	if and only if it is an interdependent-choice equilibrium with respect to the set of 
	self-FC-rationalizable action profiles. 
\end{theorem}

%%%%%%%%%%%%%%%%%%%%%%%%%%%%%%%%%%%%%%%%%%%%%%%%%%%%%%%%%%%%%%%%%%%%%%%%%%%%%%%%%%%%%%%%%%%%%%%%%%%%%%%%%%%%%%%%%%%%%%%%%%%%%%%%%%%%%%%%%
%%%%%%%%%%%%%%%%%%%%%%%%%%%%%%%%%%%%%%%%%%%%%%%%%%%%%%%%%%%%%%%%%%%%%%%%%%%%%%%%%%%%%%%%%%%%%%%%%%%%%%%%%%%%%%%%%%%%%%%%%%%%%%%%%%%%%%%%%
There are two interesting corollaries of this result. 
%%%
First, since sequential implementability implies QS implementability, 
Theorem \ref{thm:QSI} can be interpreted as a necessary condition for sequential implementation in arbitrary environments. 

%%%%%%%%%%%%%%%%%%%%%%%%%%%%%%%%%%%%%%%%%%%%%%%%%%%%%%%%%%%%%%%%%%%%%%%%%%%%%%%%%%%%%%%%%%%%%%%%%%%%%%%%%%%%%%%%%%%%%%%%%%%%%%%%%%%%%%%%%
%%%%%%%%%%%%%%%%%%%%%%%%%%%%%%%%%%%%%%%%%%%%%%%%%%%%%%%%%%%%%%%%%%%%%%%%%%%%%%%%%%%%%%%%%%%%%%%%%%%%%%%%%%%%%%%%%%%%%%%%%%%%%%%%%%%%%%%%%
\begin{corollary}
	Every sequentially implementable distribution is an interdependent-choice equilibrium with respect to $\FLR$.
\end{corollary}

%%%%%%%%%%%%%%%%%%%%%%%%%%%%%%%%%%%%%%%%%%%%%%%%%%%%%%%%%%%%%%%%%%%%%%%%%%%%%%%%%%%%%%%%%%%%%%%%%%%%%%%%%%%%%%%%%%%%%%%%%%%%%%%%%%%%%%%%%
%%%%%%%%%%%%%%%%%%%%%%%%%%%%%%%%%%%%%%%%%%%%%%%%%%%%%%%%%%%%%%%%%%%%%%%%%%%%%%%%%%%%%%%%%%%%%%%%%%%%%%%%%%%%%%%%%%%%%%%%%%%%%%%%%%%%%%%%%
Second, since C-rationalziable actions are FC-rationalizable, 
in games with no absolute dominance a distribution is QS implementable if and only if it is an interdependent-choice equilibrium.
%%%
This means that requiring QSE instead of Nash equilibrium has a small impact,
because most games of interest have no absolutely dominated actions.

%%%%%%%%%%%%%%%%%%%%%%%%%%%%%%%%%%%%%%%%%%%%%%%%%%%%%%%%%%%%%%%%%%%%%%%%%%%%%%%%%%%%%%%%%%%%%%%%%%%%%%%%%%%%%%%%%%%%%%%%%%%%%%%%%%%%%%%%%
%%%%%%%%%%%%%%%%%%%%%%%%%%%%%%%%%%%%%%%%%%%%%%%%%%%%%%%%%%%%%%%%%%%%%%%%%%%%%%%%%%%%%%%%%%%%%%%%%%%%%%%%%%%%%%%%%%%%%%%%%%%%%%%%%%%%%%%%%
\begin{corollary}
	When there are no absolutely dominated actions, 
	a distribution is quasi-sequentially implementable if and only if it is an interdependent-choice equilibrium.
\end{corollary}

%%%%%%%%%%%%%%%%%%%%%%%%%%%%%%%%%%%%%%%%%%%%%%%%%%%%%%%%%%%%%%%%%%%%%%%%%%%%%%%%%%%%%%%%%%%%%%%%%%%%%%%%%%%%%%%%%%%%%%%%%%%%%%%%%%%%%%%%%
%%%%%%%%%%%%%%%%%%%%%%%%%%%%%%%%%%%%%%%%%%%%%%%%%%%%%%%%%%%%%%%%%%%%%%%%%%%%%%%%%%%%%%%%%%%%%%%%%%%%%%%%%%%%%%%%%%%%%%%%%%%%%%%%%%%%%%%%%
\begin{example}\label{eg:B}
	Consider the environment described by the payoff matrix in Figure \ref{fig:egA}, but now suppose that $k=5$.
	%%%
	Let $\Theta_i$ be the event that player $i$ is the first player to move.
	%%%
	Now, for player $1$ to choose $\aT$ he must assign probability at least $2/3$ to $\Theta_1$ and and player $2$ choosing $\aL$.
	%%%	
	Similarly, for player $2$ to choose $\aL$ he must assign probability at least $2/3$ to $\Theta_2$ and and player $1$ choosing $\aB$.
	%%%
	Moreover, for player $1$ to choose $\aB$ he must assign probability at least $2/3$ to $\Theta_1$ and and player $2$ choosing $\aR$.
	%%%
	Finally, for player $2$ to choose $\aR$ he must assign probability at least $2/3$ to $\Theta_2$ and and player $1$ choosing $\aT$.
	%%%
	Hence $\aT$ and $\aL$ can only be played if the players disagree
	about the order of play in a way that is not consistent with sequential implementation.
	%%%
	Therefore $\aT$ and $\aL$ cannot be used as credible threats for sequential implementation,
	and the only sequentially implementable outcome is $(\C,\C)$.
	%%%
	However, since there is no absolute dominance, $\aT$ and $\aL$ are credible threats for QS implementation,
	and thus $(\NC,\NC)$ is QS implementable.
\end{example}

%%%%%%%%%%%%%%%%%%%%%%%%%%%%%%%%%%%%%%%%%%%%%%%%%%%%%%%%%%%%%%%%%%%%%%%%%%%%%%%%%%%%%%%%%%%%%%%%%%%%%%%%%%%%%%%%%%%%%%%%%%%%%%%%%%%%%%%%%
%%%%%%%%%%%%%%%%%%%%%%%%%%%%%%%%%%%%%%%%%%%%%%%%%%%%%%%%%%%%%%%%%%%%%%%%%%%%%%%%%%%%%%%%%%%%%%%%%%%%%%%%%%%%%%%%%%%%%%%%%%%%%%%%%%%%%%%%%
This section concludes with a characterization of the operator $\FR$.
%%%
Loosely speaking, the following proposition shows that it is equivalent to the elimination
of strictly dominated actions in an auxiliary game. 
%%%
%Hence finding $\FLR$ is no more complicated than finding the set of rationalizable actions of a finite game.

%%%%%%%%%%%%%%%%%%%%%%%%%%%%%%%%%%%%%%%%%%%%%%%%%%%%%%%%%%%%%%%%%%%%%%%%%%%%%%%%%%%%%%%%%%%%%%%%%%%%%%%%%%%%%%%%%%%%%%%%%%%%%%%%%%%%%%%%%
%%%%%%%%%%%%%%%%%%%%%%%%%%%%%%%%%%%%%%%%%%%%%%%%%%%%%%%%%%%%%%%%%%%%%%%%%%%%%%%%%%%%%%%%%%%%%%%%%%%%%%%%%%%%%%%%%%%%%%%%%%%%%%%%%%%%%%%%%
\begin{proposition}\label{prop:FLR}
	An action $a_i\in A_i$ is FC-rationalizable with respect to 
	an action subspace $A'\in\A$ if and only if there is no $\alpha_i\in\Delta(A_i)$ such that:
	\begin{enumerate}	
		\item $\max\big\{ \ua_i(a_i,a_{-i}) \:\big|\: a_{-i}\in A_{-i}'\big\} 
			< \min\big\{ \uc_i(\alpha_i,a_{-i}) \:\big|\: a_{-i}\in A_{-i}'\big\}$
		\item $\ua_i(a_i,a_{-i}) < \uc_i(\alpha_i,a_{-i})$\: for all \:$a_{-i}\in A_{-i}\backslash A_{-i}'$
	\end{enumerate}
\end{proposition}

%%%%%%%%%%%%%%%%%%%%%%%%%%%%%%%%%%%%%%%%%%%%%%%%%%%%%%%%%%%%%%%%%%%%%%%%%%%%%%%%%%%%%%%%%%%%%%%%%%%%%%%%%%%%%%%%%%%%%%%%%%%%%%%%%%%%%%%%%
%%%%%%%%%%%%%%%%%%%%%%%%%%%%%%%%%%%%%%%%%%%%%%%%%%%%%%%%%%%%%%%%%%%%%%%%%%%%%%%%%%%%%%%%%%%%%%%%%%%%%%%%%%%%%%%%%%%%%%%%%%%%%%%%%%%%%%%%%
%%%%%%%%%%%%%%%%%%%%%%%%%%%%%%%%%%%%%%%%%%%%%%%%%%%%%%%%%%%%%%%%%%%%%%%%%%%%%%%%%%%%%%%%%%%%%%%%%%%%%%%%%%%%%%%%%%%%%%%%%%%%%%%%%%%%%%%%%
%%%%%%%%%%%%%%%%%%%%%%%%%%%%%%%%%%%%%%%%%%%%%%%%%%%%%%%%%%%%%%%%%%%%%%%%%%%%%%%%%%%%%%%%%%%%%%%%%%%%%%%%%%%%%%%%%%%%%%%%%%%%%%%%%%%%%%%%%
%%%%%%%%%%%%%%%%%%%%%%%%%%%%%%%%%%%%%%%%%%%%%%%%%%%%%%%%%%%%%%%%%%%%%%%%%%%%%%%%%%%%%%%%%%%%%%%%%%%%%%%%%%%%%%%%%%%%%%%%%%%%%%%%%%%%%%%%%
%%%%%%%%%%%%%%%%%%%%%%%%%%%%%%%%%%%%%%%%%%%%%%%%%%%%%%%%%%%%%%%%%%%%%%%%%%%%%%%%%%%%%%%%%%%%%%%%%%%%%%%%%%%%%%%%%%%%%%%%%%%%%%%%%%%%%%%%%
%%%%%%%%%%%%%%%%%%%%%%%%%%%%%%%%%%%%%%%%%%%%%%%%%%%%%%%%%%%%%%%%%%%%%%%%%%%%%%%%%%%%%%%%%%%%%%%%%%%%%%%%%%%%%%%%%%%%%%%%%%%%%%%%%%%%%%%%%
%%%%%%%%%%%%%%%%%%%%%%%%%%%%%%%%%%%%%%%%%%%%%%%%%%%%%%%%%%%%%%%%%%%%%%%%%%%%%%%%%%%%%%%%%%%%%%%%%%%%%%%%%%%%%%%%%%%%%%%%%%%%%%%%%%%%%%%%%
%%%%%%%%%%%%%%%%%%%%%%%%%%%%%%%%%%%%%%%%%%%%%%%%%%%%%%%%%%%%%%%%%%%%%%%%%%%%%%%%%%%%%%%%%%%%%%%%%%%%%%%%%%%%%%%%%%%%%%%%%%%%%%%%%%%%%%%%%
%%%%%%%%%%%%%%%%%%%%%%%%%%%%%%%%%%%%%%%%%%%%%%%%%%%%%%%%%%%%%%%%%%%%%%%%%%%%%%%%%%%%%%%%%%%%%%%%%%%%%%%%%%%%%%%%%%%%%%%%%% END OF SECTION

%% file: discussion.tex
%auto-ignore

%%%%%%%%%%%%%%%%%%%%%%%%%%%%%%%%%%%%%%%%%%%%%%%%%%%%%%%%%%%%%%%%%%%%%%%%%%%%%%%%%%%%%%%%%%%%%%%%%%%%%%%%%%%%%%%%%%%%%%%%%%%%%%%%%%%%%%%%%
%%%%%%%%%%%%%%%%%%%%%%%%%%%%%%%%%%%%%%%%%%%%%%%%%%%%%%%%%%%%%%%%%%%%%%%%%%%%%%%%%%%%%%%%%%%%%%%%%%%%%%%%%%%%%%%%%%%%%%%%%%%%%%%%%%%%%%%%%
% Section: Discussion
% Project: Coordination in Strategic Environments
%  Author: Bruno Salcedo
% Version: 7.1
%    Date: 09/08/13
%%%%%%%%%%%%%%%%%%%%%%%%%%%%%%%%%%%%%%%%%%%%%%%%%%%%%%%%%%%%%%%%%%%%%%%%%%%%%%%%%%%%%%%%%%%%%%%%%%%%%%%%%%%%%%%%%%%%%%%%%%%%%%%%%%%%%%%%%
%%%%%%%%%%%%%%%%%%%%%%%%%%%%%%%%%%%%%%%%%%%%%%%%%%%%%%%%%%%%%%%%%%%%%%%%%%%%%%%%%%%%%%%%%%%%%%%%%%%%%%%%%%%%%%%%%%%%%%%%%%%%%%%%%%%%%%%%%

%%%%%%%%%%%%%%%%%%%%%%%%%%%%%%%%%%%%%%%%%%%%%%%%%%%%%%%%%%%%%%%%%%%%%%%%%%%%%%%%%%%%%%%%%%%%%%%%%%%%%%%%%%%%%%%%%%%%%%%%%%%%%%%%%%%%%%%%%
%%%%%%%%%%%%%%%%%%%%%%%%%%%%%%%%%%%%%%%%%%%%%%%%%%%%%%%%%%%%%%%%%%%%%%%%%%%%%%%%%%%%%%%%%%%%%%%%%%%%%%%%%%%%%%%%%%%%%%%%%%%%%%%%%%%%%%%%%
The current paper analyzes choice-interdependence as a mechanism to generate incentives in moral hazard environments. 
%%%
It introduces a class of mediated games in which a mediator manages the game through private recommendations.
%%%
Two salient aspects of the model are that the recommendations are sequential and occur during the actual play of the game,
and that they can depend on previous choices. 
%%%
This enables reciprocal strategies that, for instance, may allow for cooperation in the prisoner's dilemma. 
%%%
ICE are defined as Nash equilibria of mediated games.
%%%
The set of ICE admits a canonical characterization  consisting of a finite set of affine inequalities.
%%%
Also, it characterizes all the outcomes that can be implemented in equilibrium  without repetition, commitment or side payments. 

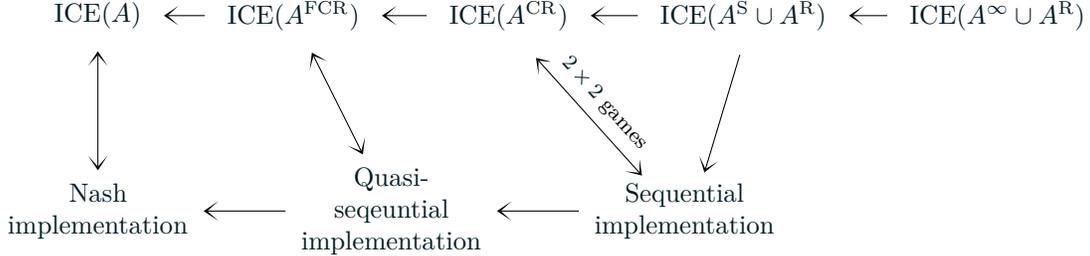
\begin{figure}[htb]
\centering
\psset{unit=13mm}
\begin{pspicture}(1,2.5)(12,5.5)
	%\psgrid%
	\footnotesize%
	% Equilibrium nodes 
	\newcommand{\ceH}{5}
	\newcommand{\imH}{3}
	% Equilibrium nodes 
	\psset{linestyle=none}
		\rput(2.0,\ceH){\rnode{Eall}{\psframebox{$\ICE(\Ac)$}}}
		\rput(4.0,\ceH){\rnode{Eflr}{\psframebox{$\ICE(\FLR)$}}}
		\rput(6.2,\ceH){\rnode{Ecrat}{\psframebox{$\ICE(\CRR)$}}}
		\rput(8.6,\ceH){\rnode{Epro}{\psframebox{$\ICE(\SEac\cup\R)$}}}
		\rput(11.2,\ceH){\rnode{Esuff}{\psframebox{$\ICE(\CEac\cup\R)$}}}
	% Implementation nodes
		\rput(2,\imH){\psframebox{\rnode{Inash}{\parbox{7em}{\centering Nash\\ implementation}}}}
		\rput(5,\imH){\psframebox{\rnode{Ipb}{\parbox{7em}{\centering Quasi-seqeuntial \\ implementation}}}}
		\rput(8,\imH){\psframebox{\rnode{Ipro}{\parbox{7em}{\centering Sequential\\ implementation}}}}
	\psreset%
	% Horizontal Implications
	\psset{fillcolor=white,doubleline=false,doublesep=1pt,linewidth=0.4pt,arrowinset=0.75,arrowsize=7.5pt,arrowlength=0.75,nodesep=5pt}
		\ncline{<->}{Eall}{Inash}
		\ncline{<->}{Eflr}{Ipb}
		\ncline[offset=-3pt]{<->}{Ecrat}{Ipro}
		\aput*[4pt]{:U}{\scriptsize $2\times 2$ games}
		\ncline[offset=3pt]{->}{Epro}{Ipro}
	% Vertical Implications
		\ncline{<-}{Eall}{Eflr}
		\ncline{<-}{Eflr}{Ecrat}
		\ncline{<-}{Ecrat}{Epro}
		\ncline{<-}{Epro}{Esuff}
		\ncline{<-}{Inash}{Ipb}
		\ncline{<-}{Ipb}{Ipro}
	\psreset%
\end{pspicture}
\caption{Summary of results}\label{fig:summ}
\end{figure}

%%%%%%%%%%%%%%%%%%%%%%%%%%%%%%%%%%%%%%%%%%%%%%%%%%%%%%%%%%%%%%%%%%%%%%%%%%%%%%%%%%%%%%%%%%%%%%%%%%%%%%%%%%%%%%%%%%%%%%%%%%%%%%%%%%%%%%%%
%%%%%%%%%%%%%%%%%%%%%%%%%%%%%%%%%%%%%%%%%%%%%%%%%%%%%%%%%%%%%%%%%%%%%%%%%%%%%%%%%%%%%%%%%%%%%%%%%%%%%%%%%%%%%%%%%%%%%%%%%%%%%%%%%%%%%%%%%
The paper also provides conditions for implementation according to different equilibrium refinements
requiring sequential rationality. 
%%%
The conditions restrict the set of credible threats that the mediator can recommend as credible threats off the equilibrium path.
%%%
Different sets of credible threats offer necessary and/or sufficient conditions for different solution concepts.
%%%
The implications are summarized in Figure \ref{fig:summ}.

%%%%%%%%%%%%%%%%%%%%%%%%%%%%%%%%%%%%%%%%%%%%%%%%%%%%%%%%%%%%%%%%%%%%%%%%%%%%%%%%%%%%%%%%%%%%%%%%%%%%%%%%%%%%%%%%%%%%%%%%%%%%%%%%%%%%%%%%%
%%%%%%%%%%%%%%%%%%%%%%%%%%%%%%%%%%%%%%%%%%%%%%%%%%%%%%%%%%%%%%%%%%%%%%%%%%%%%%%%%%%%%%%%%%%%%%%%%%%%%%%%%%%%%%%%%%%%%%%%%%%%%%%%%%%%%%%%%

%%%%%%%%%%%%%%%%%%%%%%%%%%%%%%%%%%%%%%%%%%%%%%%%%%%%%%%%%%%%%%%%%%%%%%%%%%%%%%%%%%%%%%%%%%%%%%%%%%%%%%%%%%%%%%%%%%%%%%%%%%%%%%%%%%%%%%%%%
%%%%%%%%%%%%%%%%%%%%%%%%%%%%%%%%%%%%%%%%%%%%%%%%%%%%%%%%%%%%%%%%%%%%%%%%%%%%%%%%%%%%%%%%%%%%%%%%%%%%%%%%%%%%%%%%%%%%%%%%%%%%%%%%%%%%%%%%%
%%%%%%%%%%%%%%%%%%%%%%%%%%%%%%%%%%%%%%%%%%%%%%%%%%%%%%%%%%%%%%%%%%%%%%%%%%%%%%%%%%%%%%%%%%%%%%%%%%%%%%%%%%%%%%%%%%%%%%%%%%%%%%%%%%%%%%%%%
%%%%%%%%%%%%%%%%%%%%%%%%%%%%%%%%%%%%%%%%%%%%%%%%%%%%%%%%%%%%%%%%%%%%%%%%%%%%%%%%%%%%%%%%%%%%%%%%%%%%%%%%%%%%%%%%%%%%%%%%%%%%%%%%%%%%%%%%%
%%%%%%%%%%%%%%%%%%%%%%%%%%%%%%%%%%%%%%%%%%%%%%%%%%%%%%%%%%%%%%%%%%%%%%%%%%%%%%%%%%%%%%%%%%%%%%%%%%%%%%%%%%%%%%%%%%%%%%%%%%%%%%%%%%%%%%%%%
%%%%%%%%%%%%%%%%%%%%%%%%%%%%%%%%%%%%%%%%%%%%%%%%%%%%%%%%%%%%%%%%%%%%%%%%%%%%%%%%%%%%%%%%%%%%%%%%%%%%%%%%%%%%%%%%%%%%%%%%%%%%%%%%%%%%%%%%%
%%%%%%%%%%%%%%%%%%%%%%%%%%%%%%%%%%%%%%%%%%%%%%%%%%%%%%%%%%%%%%%%%%%%%%%%%%%%%%%%%%%%%%%%%%%%%%%%%%%%%%%%%%%%%%%%%%%%%%%%%%%%%%%%%%%%%%%%%
%%%%%%%%%%%%%%%%%%%%%%%%%%%%%%%%%%%%%%%%%%%%%%%%%%%%%%%%%%%%%%%%%%%%%%%%%%%%%%%%%%%%%%%%%%%%%%%%%%%%%%%%%%%%%%%%%%%%%%%%%%%%%%%%%%%%%%%%%
%%%%%%%%%%%%%%%%%%%%%%%%%%%%%%%%%%%%%%%%%%%%%%%%%%%%%%%%%%%%%%%%%%%%%%%%%%%%%%%%%%%%%%%%%%%%%%%%%%%%%%%%%%%%%%%%%%%%%%%%%%%%%%%%%%%%%%%%%
%%%%%%%%%%%%%%%%%%%%%%%%%%%%%%%%%%%%%%%%%%%%%%%%%%%%%%%%%%%%%%%%%%%%%%%%%%%%%%%%%%%%%%%%%%%%%%%%%%%%%%%%%%%%%%%%%%%%%%%% ssec: discussion
\subsection{Discussion} \label{ssec:discussion}

%%%%%%%%%%%%%%%%%%%%%%%%%%%%%%%%%%%%%%%%%%%%%%%%%%%%%%%%%%%%%%%%%%%%%%%%%%%%%%%%%%%%%%%%%%%%%%%%%%%%%%%%%%%%%%%%%%%%%%%%%%%%%%%%%%%%%%%%%
%%%%%%%%%%%%%%%%%%%%%%%%%%%%%%%%%%%%%%%%%%%%%%%%%%%%%%%%%%%%%%%%%%%%%%%%%%%%%%%%%%%%%%%%%%%%%%%%%%%%%%%%%%%%%%%%%%%%%%%%%%%%%%%%%%%%%%%%%
%%%%%%%%%%%%%%%%%%%%%%%%%%%%%%%%%%%%%%%%%%%%%%%%%%%%%%%%%%%%%%%%%%%%%%%%%%%%%%%%%%%%%%%%%%%%%%%%%%%%%%%%%%%%%%%%%%%%%%%%%%%%%%%%%%%%%%%%%
%%%%%%%%%%%%%%%%%%%%%%%%%%%%%%%%%%%%%%%%%%%%%%%%%%%%%%%%%%%%%%%%%%%%%%%%%%%%%%%%%%%%%%%%%%%%%%%%%%%%%%%%%%%%%%%%%%%%%%%%%%%%%%%%%%%%%%%%%
%%%%%%%%%%%%%%%%%%%%%%%%%%%%%%%%%%%%%%%%%%%%%%%%%%%%%%%%%%%%%%%%%%%%%%%%%%%%%%%%%%%%%%%%%%%%%%%%%%%%%%%%%%%%%%%%%%%%%%%%%%%% Alternatives
\subsubsection{Additional restrictions}\label{disc:alternative}
The formulation of ICE allows to capture additional restrictions by adjusting the worst punishment functions $\uw$. 
%%%
For example, the assumption that all deviations from equilibrium are publicly observed can be captured 
by replacing the worst punishments function $\uw_i$ 
with $\uw_i'(a_i') = \min_{a_{-i}\mathrm{BR}_{-i}(a_i')}\ua_i(a_i',a_{-i})$,
where $\mathrm{BR}_{-i}$ is ${-i}$'s best response correspondence.
%%%
The resulting solution concept would lie somewhere between correlated equilibrium and ICE. 

%%%%%%%%%%%%%%%%%%%%%%%%%%%%%%%%%%%%%%%%%%%%%%%%%%%%%%%%%%%%%%%%%%%%%%%%%%%%%%%%%%%%%%%%%%%%%%%%%%%%%%%%%%%%%%%%%%%%%%%%%%%%%%%%%%%%%%%%%
%%%%%%%%%%%%%%%%%%%%%%%%%%%%%%%%%%%%%%%%%%%%%%%%%%%%%%%%%%%%%%%%%%%%%%%%%%%%%%%%%%%%%%%%%%%%%%%%%%%%%%%%%%%%%%%%%%%%%%%%%%%%%%%%%%%%%%%%%
Alternatively, instead of assuming that the mediator controls the order of choices,
suppose that she can control the order of her recommendations but players can choose to act before or after they encounter her. 
%%%
In such cases, the mediator could not recommend action-specific punishments.
%%%
A player who intended to deviate would make his choice after the mediator has left,
and thus the mediator could no longer observe the specific deviation. 
%%%
The set of implementable outcomes under these conditions could be characterized by replacing  $\uw_i$
with the constant minimax punishment  $\uw_i'(a'_i) = \min_{\alpha_{-i}\in\Delta(B^*_{-i})} \max_{a_i\in A_i} \uc_i(a_i,\alpha_{-i})$.
%%%
Notice that this makes no difference when each player has at most two actions.

%%%%%%%%%%%%%%%%%%%%%%%%%%%%%%%%%%%%%%%%%%%%%%%%%%%%%%%%%%%%%%%%%%%%%%%%%%%%%%%%%%%%%%%%%%%%%%%%%%%%%%%%%%%%%%%%%%%%%%%%%%%%%%%%%%%%%%%%%
%%%%%%%%%%%%%%%%%%%%%%%%%%%%%%%%%%%%%%%%%%%%%%%%%%%%%%%%%%%%%%%%%%%%%%%%%%%%%%%%%%%%%%%%%%%%%%%%%%%%%%%%%%%%%%%%%%%%%%%%%%%%%%%%%%%%%%%%%
%%%%%%%%%%%%%%%%%%%%%%%%%%%%%%%%%%%%%%%%%%%%%%%%%%%%%%%%%%%%%%%%%%%%%%%%%%%%%%%%%%%%%%%%%%%%%%%%%%%%%%%%%%%%%%%%%%%%%%%%%%%%%%%%%%%%%%%%%
%%%%%%%%%%%%%%%%%%%%%%%%%%%%%%%%%%%%%%%%%%%%%%%%%%%%%%%%%%%%%%%%%%%%%%%%%%%%%%%%%%%%%%%%%%%%%%%%%%%%%%%%%%%%%%%%%%%%%%%%%%%%%%%%%%%%%%%%%
%%%%%%%%%%%%%%%%%%%%%%%%%%%%%%%%%%%%%%%%%%%%%%%%%%%%%%%%%%%%%%%%%%%%%%%%%%%%%%%%%%%%%%%%%%%%%%%%%%%%%%%%%%%%%%%%%%%%%%%%%%%%%%%%%%%% Many
\subsubsection{Many players} \label{disc:many}
The definitions and results can be extended to $n$-player environments, but the notation becomes cumbersome. 
%%%
For one thing, when the mediator chooses an ordering of the players she is no longer choosing the player who moves first
but an entire enumeration $\ordn$ of $I$.
%%%
Hence, $\theta$ must specify distributions over such enumerations, and the incentive constraints become:
%%%
\begin{align*}
	\sum_{a_{-i}\in A_{-i}} \alpha(a)\ua_i(a) 
	\geq \sum_{a_{-i}\in A_{-i}} \sum_{\ordn} \alpha(a)\theta(\ordn|a) 
		\cdot \min\left\{ \ua_i\big(a_i',a'_{\post},a_{\prev}\big) \:\Big|\: a'_{\post}\in B^*_{\post} \right\}
\end{align*}
%%%
where $\post = \left\{j\in I\:\big|\: \ordn(j)>\ordn(i)\right\}$  and $\prev = \left\{j\in I\:\big|\: \ordn(j)<\ordn(i)\right\}$ 
are the set of players that move before and after $i$ according to $\ordn$. 
%%%
Of course, assuming of costless and perfect monitoring becomes less appealing as the number of players increases.

%%%%%%%%%%%%%%%%%%%%%%%%%%%%%%%%%%%%%%%%%%%%%%%%%%%%%%%%%%%%%%%%%%%%%%%%%%%%%%%%%%%%%%%%%%%%%%%%%%%%%%%%%%%%%%%%%%%%%%%%%%%%%%%%%%%%%%%%%
%%%%%%%%%%%%%%%%%%%%%%%%%%%%%%%%%%%%%%%%%%%%%%%%%%%%%%%%%%%%%%%%%%%%%%%%%%%%%%%%%%%%%%%%%%%%%%%%%%%%%%%%%%%%%%%%%%%%%%%%%%%%%%%%%%%%%%%%%
%%%%%%%%%%%%%%%%%%%%%%%%%%%%%%%%%%%%%%%%%%%%%%%%%%%%%%%%%%%%%%%%%%%%%%%%%%%%%%%%%%%%%%%%%%%%%%%%%%%%%%%%%%%%%%%%%%%%%%%%%%%%%%%%%%%%%%%%%
%%%%%%%%%%%%%%%%%%%%%%%%%%%%%%%%%%%%%%%%%%%%%%%%%%%%%%%%%%%%%%%%%%%%%%%%%%%%%%%%%%%%%%%%%%%%%%%%%%%%%%%%%%%%%%%%%%%%%%%%%%%%%%%%%%%%%%%%%
%%%%%%%%%%%%%%%%%%%%%%%%%%%%%%%%%%%%%%%%%%%%%%%%%%%%%%%%%%%%%%%%%%%%%%%%%%%%%%%%%%%%%%%%%%%%%%%%%%%%%%%%%%%%%%%%%%%%%%%%%%%%%%%%%% Nature
\subsubsection{Mistakes by nature} \label{disc:mistakes}
I allow players to attribute deviations from the equilibrium pathto unexpected moves by Nature (or a mediator). 
%%%
In equilibrium, when an agent finds himself in the equilibrium path he may believe that is is because Nature made a mistake
and not because his opponent deviated.
%%%
Since this is not a common feature in other models, it deserves some justification.

%%%%%%%%%%%%%%%%%%%%%%%%%%%%%%%%%%%%%%%%%%%%%%%%%%%%%%%%%%%%%%%%%%%%%%%%%%%%%%%%%%%%%%%%%%%%%%%%%%%%%%%%%%%%%%%%%%%%%%%%%%%%%%%%%%%%%%%%%
%%%%%%%%%%%%%%%%%%%%%%%%%%%%%%%%%%%%%%%%%%%%%%%%%%%%%%%%%%%%%%%%%%%%%%%%%%%%%%%%%%%%%%%%%%%%%%%%%%%%%%%%%%%%%%%%%%%%%%%%%%%%%%%%%%%%%%%%%
Consider for instance the hypothetical situation of a loving marriage
after the wife finds unfamiliar lingerie mixed in the laundry.
%%%
A plausible explanation is that the husband deviated from the marital arrangement by involving in an extramarital relationship, 
and made the mistake of bringing home evidence of his deviation.
%%%
However, more often than not, a \emph{trusing} wife is likely  to ignore this story and instead recur to intricate explanations
involving unexpected chance events. 
%%%
Back to our abstract environment, each player $i$ knows that his opponents cannot gain from deviating,
as long as he sticks to his equilibrium strategy.
%%%
Hence, he has no reason to be suspicious about them, and attributing deviations to Nature may be reasonable. 

%%%%%%%%%%%%%%%%%%%%%%%%%%%%%%%%%%%%%%%%%%%%%%%%%%%%%%%%%%%%%%%%%%%%%%%%%%%%%%%%%%%%%%%%%%%%%%%%%%%%%%%%%%%%%%%%%%%%%%%%%%%%%%%%%%%%%%%%%
%%%%%%%%%%%%%%%%%%%%%%%%%%%%%%%%%%%%%%%%%%%%%%%%%%%%%%%%%%%%%%%%%%%%%%%%%%%%%%%%%%%%%%%%%%%%%%%%%%%%%%%%%%%%%%%%%%%%%%%%%%%%%%%%%%%%%%%%%
A key element in this previous example is the trusting nature of the relationship. 
%%%
This line of thought might find less favor in situations in which the agents have reasons to be suspicious about each other. 
%%%
A extramarital affair is bound to be the favored explanation  by a suspicious wife who expects to be cheated.
%%%
The sense in which allowing for null chance moves is sensible might depend on the level of trust or suspicion among the agents. 
%%%
In any case, Nature's mistakes play a crucial role for Theorem \ref{thm:QSI}, but not for Theorem \ref{thm:SE}. 

%%%%%%%%%%%%%%%%%%%%%%%%%%%%%%%%%%%%%%%%%%%%%%%%%%%%%%%%%%%%%%%%%%%%%%%%%%%%%%%%%%%%%%%%%%%%%%%%%%%%%%%%%%%%%%%%%%%%%%%%%%%%%%%%%%%%%%%%%
%%%%%%%%%%%%%%%%%%%%%%%%%%%%%%%%%%%%%%%%%%%%%%%%%%%%%%%%%%%%%%%%%%%%%%%%%%%%%%%%%%%%%%%%%%%%%%%%%%%%%%%%%%%%%%%%%%%%%%%%%%%%%%%%%%%%%%%%%
%%%%%%%%%%%%%%%%%%%%%%%%%%%%%%%%%%%%%%%%%%%%%%%%%%%%%%%%%%%%%%%%%%%%%%%%%%%%%%%%%%%%%%%%%%%%%%%%%%%%%%%%%%%%%%%%%%%%%%%%%%%%%%%%%%%%%%%%%
%%%%%%%%%%%%%%%%%%%%%%%%%%%%%%%%%%%%%%%%%%%%%%%%%%%%%%%%%%%%%%%%%%%%%%%%%%%%%%%%%%%%%%%%%%%%%%%%%%%%%%%%%%%%%%%%%%%%%%%%%%%%%%%%%%%%%%%%%
%%%%%%%%%%%%%%%%%%%%%%%%%%%%%%%%%%%%%%%%%%%%%%%%%%%%%%%%%%%%%%%%%%%%%%%%%%%%%%%%%%%%%%%%%%%%%%%%%%%%%%%%%%%%%%%%%%%%%%%%%%%%%%%%%%%%% QSE
\subsubsection{A rationale for QSE} \label{disc:QSE}
The focus on QSE is partially motivated by the fact that it is the finer refinement
for which I can provide a complete characterization.
%%%
However, there may be situations for which it is more appealing than sequential equilibrium. 
%%%
In general, equilibrium is not a straightforward consequence of rational behavior. 
%%%
In order to guarantee equilibrium one must assume mutual or common knowledge of choices or conjectures \citep{aumann95},
%%%
which may be hard to justify off the equilibrium path. 

%%%%%%%%%%%%%%%%%%%%%%%%%%%%%%%%%%%%%%%%%%%%%%%%%%%%%%%%%%%%%%%%%%%%%%%%%%%%%%%%%%%%%%%%%%%%%%%%%%%%%%%%%%%%%%%%%%%%%%%%%%%%%%%%%%%%%%%%%
%%%%%%%%%%%%%%%%%%%%%%%%%%%%%%%%%%%%%%%%%%%%%%%%%%%%%%%%%%%%%%%%%%%%%%%%%%%%%%%%%%%%%%%%%%%%%%%%%%%%%%%%%%%%%%%%%%%%%%%%%%%%%%%%%%%%%%%%%
In this respect, 
focal point arguments may be questioned because of the complexity of determining whether an equilibrium is sequential. 
%%%
Communication can be questioned along similar lines,
because planning for all possible contingencies or agreeing on their likelihood may be too complex. 
%%%
Finally, precedence may provide a justification for equilibrium,
but repetition provides no experience about events which only happen off the equilibrium path \citep{FudLev}.
%%%
Hence there might be situations in which 
(\emph{i}) it makes sense to assume agreement \emph{exclusively} along the equilibrium path;
and yet (\emph{ii}) rationality and common certainty of rationality may also be defended in every subgame.

%%%%%%%%%%%%%%%%%%%%%%%%%%%%%%%%%%%%%%%%%%%%%%%%%%%%%%%%%%%%%%%%%%%%%%%%%%%%%%%%%%%%%%%%%%%%%%%%%%%%%%%%%%%%%%%%%%%%%%%%%%%%%%%%%%%%%%%%%
%%%%%%%%%%%%%%%%%%%%%%%%%%%%%%%%%%%%%%%%%%%%%%%%%%%%%%%%%%%%%%%%%%%%%%%%%%%%%%%%%%%%%%%%%%%%%%%%%%%%%%%%%%%%%%%%%%%%%%%%%%%%%%%%%%%%%%%%%
%%%%%%%%%%%%%%%%%%%%%%%%%%%%%%%%%%%%%%%%%%%%%%%%%%%%%%%%%%%%%%%%%%%%%%%%%%%%%%%%%%%%%%%%%%%%%%%%%%%%%%%%%%%%%%%%%%%%%%%%%%%%%%%%%%%%%%%%%
%%%%%%%%%%%%%%%%%%%%%%%%%%%%%%%%%%%%%%%%%%%%%%%%%%%%%%%%%%%%%%%%%%%%%%%%%%%%%%%%%%%%%%%%%%%%%%%%%%%%%%%%%%%%%%%%%%%%%%%%%%%%%%%%%%%%%%%%%
%%%%%%%%%%%%%%%%%%%%%%%%%%%%%%%%%%%%%%%%%%%%%%%%%%%%%%%%%%%%%%%%%%%%%%%%%%%%%%%%%%%%%%%%%%%%%%%%%%%%%%%%%%%%%%%%%%%%%%%%%%%%%% Extensions
\subsubsection{Extensions} \label{disc:ext}
The current paper leaves a number of open questions. 
%%%
The definition of mediated games assumes that choices are instantaneous and that players don't know the order in which choices are made.
%%%
It is not clear whether this is possible or what outcomes remain to be implementable 
once the \emph{temporal} dimension is taken into consideration.  
%%%
Other aspects yet to consider are incomplete information, imperfect or costly monitoring, and bargaining. 
%%%
Also, some degree of commitment is assumed from part of the mediator. 
%%%
An interesting question is whether the mediator can be replaced by cheap-talk,
or whether the required information structures can be generated transparently \citep{izmalkov05}.

%%%%%%%%%%%%%%%%%%%%%%%%%%%%%%%%%%%%%%%%%%%%%%%%%%%%%%%%%%%%%%%%%%%%%%%%%%%%%%%%%%%%%%%%%%%%%%%%%%%%%%%%%%%%%%%%%%%%%%%%%%%%%%%%%%%%%%%%%
%%%%%%%%%%%%%%%%%%%%%%%%%%%%%%%%%%%%%%%%%%%%%%%%%%%%%%%%%%%%%%%%%%%%%%%%%%%%%%%%%%%%%%%%%%%%%%%%%%%%%%%%%%%%%%%%%%%%%%%%%%%%%%%%%%%%%%%%%
%%%%%%%%%%%%%%%%%%%%%%%%%%%%%%%%%%%%%%%%%%%%%%%%%%%%%%%%%%%%%%%%%%%%%%%%%%%%%%%%%%%%%%%%%%%%%%%%%%%%%%%%%%%%%%%%%%%%%%%%%%%%%%%%%%%%%%%%%
%%%%%%%%%%%%%%%%%%%%%%%%%%%%%%%%%%%%%%%%%%%%%%%%%%%%%%%%%%%%%%%%%%%%%%%%%%%%%%%%%%%%%%%%%%%%%%%%%%%%%%%%%%%%%%%%%%%%%%%%%%%%%%%%%%%%%%%%%
%%%%%%%%%%%%%%%%%%%%%%%%%%%%%%%%%%%%%%%%%%%%%%%%%%%%%%%%%%%%%%%%%%%%%%%%%%%%%%%%%%%%%%%%%%%%%%%%%%%%%%%%%%%%%%%%%%%%%%%%%%%%%%%%%%%%%%%%%
%%%%%%%%%%%%%%%%%%%%%%%%%%%%%%%%%%%%%%%%%%%%%%%%%%%%%%%%%%%%%%%%%%%%%%%%%%%%%%%%%%%%%%%%%%%%%%%%%%%%%%%%%%%%%%%%%%%%%%%%%%%%%%%%%%%%%%%%%
%%%%%%%%%%%%%%%%%%%%%%%%%%%%%%%%%%%%%%%%%%%%%%%%%%%%%%%%%%%%%%%%%%%%%%%%%%%%%%%%%%%%%%%%%%%%%%%%%%%%%%%%%%%%%%%%%%%%%%%%%%%%%%%%%%%%%%%%%
%%%%%%%%%%%%%%%%%%%%%%%%%%%%%%%%%%%%%%%%%%%%%%%%%%%%%%%%%%%%%%%%%%%%%%%%%%%%%%%%%%%%%%%%%%%%%%%%%%%%%%%%%%%%%%%%%%%%%%%%%%%%%%%%%%%%%%%%%
%%%%%%%%%%%%%%%%%%%%%%%%%%%%%%%%%%%%%%%%%%%%%%%%%%%%%%%%%%%%%%%%%%%%%%%%%%%%%%%%%%%%%%%%%%%%%%%%%%%%%%%%%%%%%%%%%%%%%%%%%%%%%%%%%%%%%%%%%
%%%%%%%%%%%%%%%%%%%%%%%%%%%%%%%%%%%%%%%%%%%%%%%%%%%%%%%%%%%%%%%%%%%%%%%%%%%%%%%%%%%%%%%%%%%%%%%%%%%%%%%%%%%%%%%%%%%%%%%%%% END OF SECTION

%% file: EFM.tex
\subsection{Extensive form games}\label{ssec:EFG}

%%%%%%%%%%%%%%%%%%%%%%%%%%%%%%%%%%%%%%%%%%%%%%%%%%%%%%%%%%%%%%%%%%%%%%%%%%%%%%%%%%%%%%%%%%%%%%%%%%%%%%%%%%%%%%%%%%%%%%%%%%%%%%%%%%%%%%%%%
%%%%%%%%%%%%%%%%%%%%%%%%%%%%%%%%%%%%%%%%%%%%%%%%%%%%%%%%%%%%%%%%%%%%%%%%%%%%%%%%%%%%%%%%%%%%%%%%%%%%%%%%%%%%%%%%%%%%%%%%%%%%%%%%%%%%%%%%%
Extensive form games are defined as in \cite{OsbRub}, with some differences in notation.
%%%
An extensive form game is a tuple $G=\big(\Moves,\Nodes,\mover,\H,\chances,\uz\big)$.
%%%
$\Moves$ denotes a set of moves.
%%%
$\Nodes\subseteq\cup_{t\in\Natural}\Moves^t$  denotes a finite set of histories or nodes.
%%%
$\pre$ denotes precedence among nodes.
%%%
$\Moves(\node) = \{\move\in\Moves\:|\:(\node,\move)\in\Nodes\}$ is the set of moves available at $\node$.
%%%
$\mover(\node)\in I \cup\{\nat\}$ is the agent moving at $\node$, where  $\nat$ represents Nature (or a mediator).
%%%
$\Terminal$ and $\Decision_i$ are the sets of terminal nodes and $i$'s decision nodes respectively.
%%%
$\Terminal(\node) = \{\terminal\in\Terminal\:|\:\node\pre\terminal\}$ 
is the set of terminal nodes that can be reached after $\node$.
%%%
$\H_i$ partitions $i$'s decision nodes into information sets and satisfies perfect recall.
%%%
$\chances$ specifies  the players' common prior beliefs about Natures' choices.
%%%
Finally, $\uz_i:\Terminal\rightarrow\Real$ represents $i$'s preferences over terminal nodes.
%%%
Notice that attention is restricted to \emph{finite} games with \emph{perfect recall}.

%%%%%%%%%%%%%%%%%%%%%%%%%%%%%%%%%%%%%%%%%%%%%%%%%%%%%%%%%%%%%%%%%%%%%%%%%%%%%%%%%%%%%%%%%%%%%%%%%%%%%%%%%%%%%%%%%%%%%%%%%%%%%%%%%%%%%%%%%
%%%%%%%%%%%%%%%%%%%%%%%%%%%%%%%%%%%%%%%%%%%%%%%%%%%%%%%%%%%%%%%%%%%%%%%%%%%%%%%%%%%%%%%%%%%%%%%%%%%%%%%%%%%%%%%%%%%%%%%%%%%%%%%%%%%%%%%%%
A pure strategy for player $i$ is a function $s_i:\H\rightarrow\Moves$, 
with  $a_i(\h)\in\Moves(\h)$ for every $\h\in\H$.
%%%
A mixed strategy for $i$ is a distribution $\sigma_i$ over his pure strategies, it is strictly mixed if it has full support.
%%%
$S_i$, $\Sigma_i$ and $\Sigma^+_i$ denote the sets of $i$'s pure, mixed and strictly mixed strategies respectively. 
%%%
Given that Nature chooses according to $\chances$, a strategy profile $\mixed$ induces a distribution over nodes
$\disZ(\blank|\sigma,\chances)\in\Delta(\Nodes)$.
%%%
$\disZ(\node|\mixed,\chances)$ is the probability that the game will reach $\node$
if players choose according to $\mixed$ and Nature chooses according to $\chances$.
%%%
When there is no ambiguity, I omit the reference to $\mixed$ and $\chances$  and simply write $\disZ(\node)$.
%%%
Expected payoffs $\us:\Sigma\rightarrow\Real$ are defined in the obvious way. 
%%%
A Nash equilibrium (NE) is a strategy profile $\sigma^*$ such that 
$\us_i(\sigma^*)\geq \us_i(\sigma'_i,\sigma^*_{-i})$ for every $i$ and $\sigma_i'$.

%%%%%%%%%%%%%%%%%%%%%%%%%%%%%%%%%%%%%%%%%%%%%%%%%%%%%%%%%%%%%%%%%%%%%%%%%%%%%%%%%%%%%%%%%%%%%%%%%%%%%%%%%%%%%%%%%%%%%%%%%%%%%%%%%%%%%%%%%
%%%%%%%%%%%%%%%%%%%%%%%%%%%%%%%%%%%%%%%%%%%%%%%%%%%%%%%%%%%%%%%%%%%%%%%%%%%%%%%%%%%%%%%%%%%%%%%%%%%%%%%%%%%%%%%%%%%%%%%%%%%%%%%%%%%%%%%%%
%%%%%%%%%%%%%%%%%%%%%%%%%%%%%%%%%%%%%%%%%%%%%%%%%%%%%%%%%%%%%%%%%%%%%%%%%%%%%%%%%%%%%%%%%%%%%%%%%%%%%%%%%%%%%%%%%%%%%%%%%%%%%%%%%%%%%%%%%
%%%%%%%%%%%%%%%%%%%%%%%%%%%%%%%%%%%%%%%%%%%%%%%%%%%%%%%%%%%%%%%%%%%%%%%%%%%%%%%%%%%%%%%%%%%%%%%%%%%%%%%%%%%%%%%%%%%%%%%%%%%%%%%%%%%%%%%%%
%%%%%%%%%%%%%%%%%%%%%%%%%%%%%%%%%%%%%%%%%%%%%%%%%%%%%%%%%%%%%%%%%%%%%%%%%%%%%%%%%%%%%%%%%%%%%%%%%%%%%%%%%%%%%%%%%%%%%%%%%%%%%%%%%%%%%%%%%
%%%%%%%%%%%%%%%%%%%%%%%%%%%%%%%%%%%%%%%%%%%%%%%%%%%%%%%%%%%%%%%%%%%%%%%%%%%%%%%%%%%%%%%%%%%%%%%%%%%%%%%%%%%%%%%%%%%%%%%%%%%%%%%%%%%%%%%%%
%%%%%%%%%%%%%%%%%%%%%%%%%%%%%%%%%%%%%%%%%%%%%%%%%%%%%%%%%%%%%%%%%%%%%%%%%%%%%%%%%%%%%%%%%%%%%%%%%%%%%%%%%%%%%%%%%%%%%%%%%%%%%%%%%%%%%%%%%
%%%%%%%%%%%%%%%%%%%%%%%%%%%%%%%%%%%%%%%%%%%%%%%%%%%%%%%%%%%%%%%%%%%%%%%%%%%%%%%%%%%%%%%%%%%%%%%%%%%%%%%%%%%%%%%%%%%%%%%%%%%%%%%%%%%%%%%%%
%%%%%%%%%%%%%%%%%%%%%%%%%%%%%%%%%%%%%%%%%%%%%%%%%%%%%%%%%%%%%%%%%%%%%%%%%%%%%%%%%%%%%%%%%%%%%%%%%%%%%%%%%%%%%%%%%%%%%%%%%%%%%%%%%%%%%%%%%
%%%%%%%%%%%%%%%%%%%%%%%%%%%%%%%%%%%%%%%%%%%%%%%%%%%%%%%%%%%%%%%%%%%%%%%%%%%%%%%%%%%%%%%%%%%%%%%%%%%%%%%%%%%%%%%%%%%%%%%%%%%%%%% ssec: EFM
\subsection{Extensive form mechanisms}\label{ssec:EFM}

%%%%%%%%%%%%%%%%%%%%%%%%%%%%%%%%%%%%%%%%%%%%%%%%%%%%%%%%%%%%%%%%%%%%%%%%%%%%%%%%%%%%%%%%%%%%%%%%%%%%%%%%%%%%%%%%%%%%%%%%%%%%%%%%%%%%%%%%%
%%%%%%%%%%%%%%%%%%%%%%%%%%%%%%%%%%%%%%%%%%%%%%%%%%%%%%%%%%%%%%%%%%%%%%%%%%%%%%%%%%%%%%%%%%%%%%%%%%%%%%%%%%%%%%%%%%%%%%%%%%%%%%%%%%%%%%%%%
This section formalizes the three requirements characterizing extensive form mechanisms 
according to definition \ref{defn:EFMinformal}.
%%%
The first requirement for an extensive form game to be an extensive form mechanism
is that it must preserve the outcome and preference structure of the environment.
%%%
That is, there must be a preference-preserving map from terminal nodes (outcomes of the game)
to action profiles (outcomes of the environment).

%%%%%%%%%%%%%%%%%%%%%%%%%%%%%%%%%%%%%%%%%%%%%%%%%%%%%%%%%%%%%%%%%%%%%%%%%%%%%%%%%%%%%%%%%%%%%%%%%%%%%%%%%%%%%%%%%%%%%%%%%%%%%%%%%%%%%%%%%
%%%%%%%%%%%%%%%%%%%%%%%%%%%%%%%%%%%%%%%%%%%%%%%%%%%%%%%%%%%%%%%%%%%%%%%%%%%%%%%%%%%%%%%%%%%%%%%%%%%%%%%%%%%%%%%%%%%%%%%%%%%%%%%%%%%%%%%%%
\begin{definition}\label{def:tra}
	An \emph{outcome homomorphism} is a function $\tra$ from terminal nodes \emph{onto} action profiles preserving preferences,
	i.e.\ such that $\uz(\terminal)=\ua(\tra(\terminal))$ for every terminal node $\terminal$.
	$G$ is \emph{outcome equivalent} to $E$ if it admits an outcome homomorphism.
\end{definition}

%%%%%%%%%%%%%%%%%%%%%%%%%%%%%%%%%%%%%%%%%%%%%%%%%%%%%%%%%%%%%%%%%%%%%%%%%%%%%%%%%%%%%%%%%%%%%%%%%%%%%%%%%%%%%%%%%%%%%%%%%%%%%%%%%%%%%%%%%
%%%%%%%%%%%%%%%%%%%%%%%%%%%%%%%%%%%%%%%%%%%%%%%%%%%%%%%%%%%%%%%%%%%%%%%%%%%%%%%%%%%%%%%%%%%%%%%%%%%%%%%%%%%%%%%%%%%%%%%%%%%%%%%%%%%%%%%%%
The second and third conditions from definition \ref{defn:EFMinformal} require that,
each player should \emph{freely} choose his own action at some point in the game.
%%%
Formalizing them requires a form of identifying moves (choices in the game) with actions (choices in the environment).
%%%
For the remainder of this section,  let $G$ be outcome equivalent to $E$ and fix an outcome homeomorphism $\tra$.
%%%
For every player $i$ and every corresponding decision node $\decision$, $\tra$ induces a \emph{representation} relationship $\rep$
from the set of moves available at $\decision$ in the game to the set of $i$'s actions in the environment.
%%%
A move $\move$ represents action $a_i$ at $\decision$, if and only if choosing $\move$ at $\decision$ in the game has the same effect 
in (payoff-relevant) outcomes as choosing $a_i$ in the environment.
%%%
This idea is formalized by the following definition.

%%%%%%%%%%%%%%%%%%%%%%%%%%%%%%%%%%%%%%%%%%%%%%%%%%%%%%%%%%%%%%%%%%%%%%%%%%%%%%%%%%%%%%%%%%%%%%%%%%%%%%%%%%%%%%%%%%%%%%%%%%%%%%%%%%%%%%%%%
%%%%%%%%%%%%%%%%%%%%%%%%%%%%%%%%%%%%%%%%%%%%%%%%%%%%%%%%%%%%%%%%%%%%%%%%%%%%%%%%%%%%%%%%%%%%%%%%%%%%%%%%%%%%%%%%%%%%%%%%%%%%%%%%%%%%%%%%%
\begin{definition}\label{def:rep}
	Given a player $i\in I$ and a decision node $\decision\in\Decision_i$, a move $\move\in\Moves(\decision)$
	\emph{represents} an action $a_i\in A_i$ at $\decision$ if and only if:
	\begin{enumerate}
		\item \label{rep:A} $\tra_i(\terminal)=a_i$ for every $\terminal\in\Terminal(\decision,\move)$
		\item \label{rep:B} There exist $\move'\in\Moves(\decision)$ and
			$\terminal\in\Terminal(\decision,\move')$ such that $\tra_i(\terminal)\neq a_i$
	\end{enumerate}
	%%%
	The representation relationship is denoted by $\move \rep a_i$, 
	and $\Moves^{a_i}(\decision)$ denotes the set of moves that represent $a_i$ at $\decision$.
	%%%
	A move is \emph{pivotal} at $\decision$ if and only if it represents some action.
\end{definition}

%%%%%%%%%%%%%%%%%%%%%%%%%%%%%%%%%%%%%%%%%%%%%%%%%%%%%%%%%%%%%%%%%%%%%%%%%%%%%%%%%%%%%%%%%%%%%%%%%%%%%%%%%%%%%%%%%%%%%%%%%%%%%%%%%%%%%%%%%
%%%%%%%%%%%%%%%%%%%%%%%%%%%%%%%%%%%%%%%%%%%%%%%%%%%%%%%%%%%%%%%%%%%%%%%%%%%%%%%%%%%%%%%%%%%%%%%%%%%%%%%%%%%%%%%%%%%%%%%%%%%%%%%%%%%%%%%%%
The first requirement for $\move\rep a_i$ is that, if $i$ chooses $\move$ at $\decision$,
then the game will end at a terminal node  which is equivalent to $\ac_i$ according to $\tra_i$.
%%%
This is regardless of any previous or future moves by either $i$ or his opponents.
%%%
The second requirement is that, after the game reaches $\decision$,  $i$ could still choose a different move $\move'$
after which the game remains open to the possibility of ending at a terminal node that is \emph{not} equivalent to $a_i$.

%%%%%%%%%%%%%%%%%%%%%%%%%%%%%%%%%%%%%%%%%%%%%%%%%%%%%%%%%%%%%%%%%%%%%%%%%%%%%%%%%%%%%%%%%%%%%%%%%%%%%%%%%%%%%%%%%%%%%%%%%%%%%%%%%%%%%%%%%
%%%%%%%%%%%%%%%%%%%%%%%%%%%%%%%%%%%%%%%%%%%%%%%%%%%%%%%%%%%%%%%%%%%%%%%%%%%%%%%%%%%%%%%%%%%%%%%%%%%%%%%%%%%%%%%%%%%%%%%%%%%%%%%%%%%%%%%%%
\begin{definition}\label{def:pivotal}
	A decision node $\decision\in\Decision_i$ is \emph{pivotal} for player $i\in I$
	if and only if $\Moves^{a_i}(\decision)\neq\emptyset$ for every $a_i\in A_i$.
	%%%
	$\Pivotal_i\subseteq\Decision$ denotes the set of pivotal nodes for $i$. 
\end{definition}

%%%%%%%%%%%%%%%%%%%%%%%%%%%%%%%%%%%%%%%%%%%%%%%%%%%%%%%%%%%%%%%%%%%%%%%%%%%%%%%%%%%%%%%%%%%%%%%%%%%%%%%%%%%%%%%%%%%%%%%%%%%%%%%%%%%%%%%%%
%%%%%%%%%%%%%%%%%%%%%%%%%%%%%%%%%%%%%%%%%%%%%%%%%%%%%%%%%%%%%%%%%%%%%%%%%%%%%%%%%%%%%%%%%%%%%%%%%%%%%%%%%%%%%%%%%%%%%%%%%%%%%%%%%%%%%%%%%
In words, a decision node $\decision$ is pivotal for player $i$ if for every action $a_i\in A_i$
there exists a pivotal move which represents it at $\decision$.
%%%
Using this language, the second and third conditions of definition \ref{defn:EFMinformal}
require that \emph{every player makes a pivotal move at a pivotal node} along every possible play of the game. 
%%%
A final technical condition is that a player should always know when he is making a pivotal move
representing some action. 

%%%%%%%%%%%%%%%%%%%%%%%%%%%%%%%%%%%%%%%%%%%%%%%%%%%%%%%%%%%%%%%%%%%%%%%%%%%%%%%%%%%%%%%%%%%%%%%%%%%%%%%%%%%%%%%%%%%%%%%%%%%%%%%%%%%%%%%%%
%%%%%%%%%%%%%%%%%%%%%%%%%%%%%%%%%%%%%%%%%%%%%%%%%%%%%%%%%%%%%%%%%%%%%%%%%%%%%%%%%%%%%%%%%%%%%%%%%%%%%%%%%%%%%%%%%%%%%%%%%%%%%%%%%%%%%%%%%
\begin{definition}\label{def:fullDisclosure}
	$(G,\tra)$ satisfies \emph{full disclosure of consequences} if and only if
	$\rep  = \rep[\decision']$ whenever $\decision$ and $\decision'$ belong to the same information set.% 
%		\footnote{$\rep = \rep[\decision']$ means that $\move\rep a_i$
%			if and only if $\move\rep[\decision']a_i$
%			for all $\move\in\Moves(\decision)=\Moves(\decision')$ and all $a_i\in A_i$.}
\end{definition}

%%%%%%%%%%%%%%%%%%%%%%%%%%%%%%%%%%%%%%%%%%%%%%%%%%%%%%%%%%%%%%%%%%%%%%%%%%%%%%%%%%%%%%%%%%%%%%%%%%%%%%%%%%%%%%%%%%%%%%%%%%%%%%%%%%%%%%%%%
%%%%%%%%%%%%%%%%%%%%%%%%%%%%%%%%%%%%%%%%%%%%%%%%%%%%%%%%%%%%%%%%%%%%%%%%%%%%%%%%%%%%%%%%%%%%%%%%%%%%%%%%%%%%%%%%%%%%%%%%%%%%%%%%%%%%%%%%%
%When $(\efg,\tra)$ satisfies full disclosure of consequences,
%the previous definitions can be extended to talk about information sets instead of nodes.
%In particular one can say that $\move$ represents $\ac_\pl$ at $\h$, or that $\h$ is pivotal,
%and $\Moves^{\ac_\pl}(\h)$ can be defined in the obvious way.

%%%%%%%%%%%%%%%%%%%%%%%%%%%%%%%%%%%%%%%%%%%%%%%%%%%%%%%%%%%%%%%%%%%%%%%%%%%%%%%%%%%%%%%%%%%%%%%%%%%%%%%%%%%%%%%%%%%%%%%%%%%%%%%%%%%%%%%%%
%%%%%%%%%%%%%%%%%%%%%%%%%%%%%%%%%%%%%%%%%%%%%%%%%%%%%%%%%%%%%%%%%%%%%%%%%%%%%%%%%%%%%%%%%%%%%%%%%%%%%%%%%%%%%%%%%%%%%%%%%%%%%%%%%%%%%%%%%
Finally, definitions \ref{defn:EFMinformal} and \ref{defn:IMPinformal} can be formally restated as follows:

%%%%%%%%%%%%%%%%%%%%%%%%%%%%%%%%%%%%%%%%%%%%%%%%%%%%%%%%%%%%%%%%%%%%%%%%%%%%%%%%%%%%%%%%%%%%%%%%%%%%%%%%%%%%%%%%%%%%%%%%%%%%%%%%%%%%%%%%%
%%%%%%%%%%%%%%%%%%%%%%%%%%%%%%%%%%%%%%%%%%%%%%%%%%%%%%%%%%%%%%%%%%%%%%%%%%%%%%%%%%%%%%%%%%%%%%%%%%%%%%%%%%%%%%%%%%%%%%%%%%%%%%%%%%%%%%%%%
\newcounter{auxi}
\setcounter{definition}{\value{EFM}}
\renewcommand{\thedefinition}{\arabic{definition}'}
\begin{definition}
	A \emph{extensive form mechanism} is a tuple $\big(G,\tra\big)$
	%consisting of an extensive form game $G$ and an outcome homomorphism $\tra$,
	satisfying full disclosure of consequences and such that
	for every terminal node $\terminal$ and every player $i$,
	there exists a pivotal node $\decision\in\Pivotal_i$
	and a pivotal move $\move\in\Moves^{\tra_i(\terminal)}$
	such that $\terminal\in\Terminal(\decision,\move)$.
\end{definition}

%%%%%%%%%%%%%%%%%%%%%%%%%%%%%%%%%%%%%%%%%%%%%%%%%%%%%%%%%%%%%%%%%%%%%%%%%%%%%%%%%%%%%%%%%%%%%%%%%%%%%%%%%%%%%%%%%%%%%%%%%%%%%%%%%%%%%%%%%
%%%%%%%%%%%%%%%%%%%%%%%%%%%%%%%%%%%%%%%%%%%%%%%%%%%%%%%%%%%%%%%%%%%%%%%%%%%%%%%%%%%%%%%%%%%%%%%%%%%%%%%%%%%%%%%%%%%%%%%%%%%%%%%%%%%%%%%%%
\setcounter{definition}{\value{IMP}}
\begin{definition}
	$\alpha\in\Delta(A)$ is (Nash, sequentially, \ldots)
	\emph{implementable} if and only if it there exist a mechanism $\big(G,\tra\big)$ 
	and a (Nash, sequential, \ldots) equilibrium $\sigma^*\in\Sigma$ such that for every $a\in A$:
	\begin{align*}
	\mac(\ac) = \disZ^*\left(\trainv(\ac)\right)
	= \sum_{\terminal\in\Terminal} \disZ(\terminal,\mixeds,\chances) \cdot \Char\big(\tra(\node)=\ac\big)
	\end{align*}
\end{definition}
\renewcommand{\thedefinition}{\arabic{definition}}
\setcounter{definition}{\value{auxi}}

%%%%%%%%%%%%%%%%%%%%%%%%%%%%%%%%%%%%%%%%%%%%%%%%%%%%%%%%%%%%%%%%%%%%%%%%%%%%%%%%%%%%%%%%%%%%%%%%%%%%%%%%%%%%%%%%%%%%%%%%%%%%%%%%%%%%%%%%%
%%%%%%%%%%%%%%%%%%%%%%%%%%%%%%%%%%%%%%%%%%%%%%%%%%%%%%%%%%%%%%%%%%%%%%%%%%%%%%%%%%%%%%%%%%%%%%%%%%%%%%%%%%%%%%%%%%%%%%%%%%%%%%%%%%%%%%%%%
%%%%%%%%%%%%%%%%%%%%%%%%%%%%%%%%%%%%%%%%%%%%%%%%%%%%%%%%%%%%%%%%%%%%%%%%%%%%%%%%%%%%%%%%%%%%%%%%%%%%%%%%%%%%%%%%%%%%%%%%%%%%%%%%%%%%%%%%%
%%%%%%%%%%%%%%%%%%%%%%%%%%%%%%%%%%%%%%%%%%%%%%%%%%%%%%%%%%%%%%%%%%%%%%%%%%%%%%%%%%%%%%%%%%%%%%%%%%%%%%%%%%%%%%%%%%%%%%%%%%%%%%%%%%%%%%%%%
%%%%%%%%%%%%%%%%%%%%%%%%%%%%%%%%%%%%%%%%%%%%%%%%%%%%%%%%%%%%%%%%%%%%%%%%%%%%%%%%%%%%%%%%%%%%%%%%%%%%%%%%%%%%%%%%%%%%%%%%%%%%%%%%%%%%%%%%%
%%%%%%%%%%%%%%%%%%%%%%%%%%%%%%%%%%%%%%%%%%%%%%%%%%%%%%%%%%%%%%%%%%%%%%%%%%%%%%%%%%%%%%%%%%%%%%%%%%%%%%%%%%%%%%%%%%%%%%%%%%%%%%%%%%%%%%%%%
%%%%%%%%%%%%%%%%%%%%%%%%%%%%%%%%%%%%%%%%%%%%%%%%%%%%%%%%%%%%%%%%%%%%%%%%%%%%%%%%%%%%%%%%%%%%%%%%%%%%%%%%%%%%%%%%%%%%%%%%%%%%%%%%%%%%%%%%%
%%%%%%%%%%%%%%%%%%%%%%%%%%%%%%%%%%%%%%%%%%%%%%%%%%%%%%%%%%%%%%%%%%%%%%%%%%%%%%%%%%%%%%%%%%%%%%%%%%%%%%%%%%%%%%%%%%%%%%%%%%%%%%%%%%%%%%%%%
%%%%%%%%%%%%%%%%%%%%%%%%%%%%%%%%%%%%%%%%%%%%%%%%%%%%%%%%%%%%%%%%%%%%%%%%%%%%%%%%%%%%%%%%%%%%%%%%%%%%%%%%%%%%%%%%%%%%%%%%%%%%%%%%%%%%%%%%%
%%%%%%%%%%%%%%%%%%%%%%%%%%%%%%%%%%%%%%%%%%%%%%%%%%%%%%%%%%%%%%%%%%%%%%%%%%%%%%%%%%%%%%%%%%%%%%%%%%%%%%%%%%%%%%%%%%%%%%%%%% END OF SECTION

%% file: proofs.tex
\subsection{Nash implementation}

%%%%%%%%%%%%%%%%%%%%%%%%%%%%%%%%%%%%%%%%%%%%%%%%%%%%%%%%%%%%%%%%%%%%%%%%%%%%%%%%%%%%%%%%%%%%%%%%%%%%%%%%%%%%%%%%%%%%%%%%%%%%%%%%%%%%%%%%%
%%%%%%%%%%%%%%%%%%%%%%%%%%%%%%%%%%%%%%%%%%%%%%%%%%%%%%%%%%%%%%%%%%%%%%%%%%%%%%%%%%%%%%%%%%%%%%%%%%%%%%%%%%%%%%%%%%%%%%%%%%%%%%%%%%%%%%%%%
\begin{BSproof*}[Proof of Theorem \ref{thm:Nash}]
	For sufficiency, a mediated game is an EFM and ICE result from NE of mediated games.
	%%%
	For necessity, consider a mechanism $(G,\tra)$, a NE $\sigma^*$ and let $\alpha$ be the induced distribution. 
	%%%
	I will show that $\alpha\in\ICE$.

%%%%%%%%%%%%%%%%%%%%%%%%%%%%%%%%%%%%%%%%%%%%%%%%%%%%%%%%%%%%%%%%%%%%%%%%%%%%%%%%%%%%%%%%%%%%%%%%%%%%%%%%%%%%%%%%%%%%%%%%%%%%%%%%%%%%%%%%%
%%%%%%%%%%%%%%%%%%%%%%%%%%%%%%%%%%%%%%%%%%%%%%%%%%%%%%%%%%%%%%%%%%%%%%%%%%%%%%%%%%%%%%%%%%%%%%%%%%%%%%%%%%%%%%%%%%%%%%%%%%%%%%%%%%%%%%%%%
	Fix any two of actions $a^*_i,a_i'\in A_i$ with $\alpha_i(a_i^*)>0$ and $a_i'\neq a_i^*$.
	%%%
	For each information set $\h\in\H_i$, let $\Moves^*(\h)$ be the set of moves that represent $a_i^*$ at $\h$ 
	\emph{and} are chosen with positive probability.
	%%%
	Also, let $\H_i^*$ be the set of information sets \emph{along the equilibrium path}
	in which $i$ chooses a move representing $a_i^*$ with positive probability according to $\sigma^*$.
	%%%
%	i.e.:
%	\begin{align*}
%		\Moves^*(\h) &= \left\{
%			\move\in\Moves^{a_i^*}(\h) \:\big|\quad 
%			\big(\exists s_i\in S_i\big) \big(\sigma_i^*(s_i)>0 \,\wedge\, s_i(\h)=\move \big)
%			\right\}\\[1ex]
%		%%%
%		\H^* &= \left\{
%			\h\in\H \:\big|\quad \disZ^*(\h)>0 \:\wedge\:\Moves^*(\h)\neq\emptyset
%			\right\}
%	\end{align*}
	%%%
	Finally, let $\disZ^*$ be distribution over nodes induced by $\sigma^*$.
	%%%
	All the expectations and conditional distributions in this proof are with respect to $\disZ^*$. 

%%%%%%%%%%%%%%%%%%%%%%%%%%%%%%%%%%%%%%%%%%%%%%%%%%%%%%%%%%%%%%%%%%%%%%%%%%%%%%%%%%%%%%%%%%%%%%%%%%%%%%%%%%%%%%%%%%%%%%%%%%%%%%%%%%%%%%%%%
%%%%%%%%%%%%%%%%%%%%%%%%%%%%%%%%%%%%%%%%%%%%%%%%%%%%%%%%%%%%%%%%%%%%%%%%%%%%%%%%%%%%%%%%%%%%%%%%%%%%%%%%%%%%%%%%%%%%%%%%%%%%%%%%%%%%%%%%%
	Every $\h\in\H_i^*$ must be pivotal, and thus admits a move  $\move'\in\Moves^{a_i'}(\h)$ representing $a_i'$.
	%%%
	Since $\sigma^*$ is a NE, and $\h$ is along the equilibrium path, for each $\move^*\in\Moves^*(\h)$:
	%%%
	\begin{align}\label{eqn:A}
		\Exp{\ua_i(a_i^*,a_{-i}) \,\big|\, \h,\move^* }
		\geq
		\Exp{\ua_i(a_i',a_{-i}) \,\big|\, \h,\move' }
	\end{align}
	%%%
	where $\h,\move$ denotes the set of nodes $\h\times\{\move\}$ for $\move\in\{\move^*,\move'\}$.

%%%%%%%%%%%%%%%%%%%%%%%%%%%%%%%%%%%%%%%%%%%%%%%%%%%%%%%%%%%%%%%%%%%%%%%%%%%%%%%%%%%%%%%%%%%%%%%%%%%%%%%%%%%%%%%%%%%%%%%%%%%%%%%%%%%%%%%%%
%%%%%%%%%%%%%%%%%%%%%%%%%%%%%%%%%%%%%%%%%%%%%%%%%%%%%%%%%%%%%%%%%%%%%%%%%%%%%%%%%%%%%%%%%%%%%%%%%%%%%%%%%%%%%%%%%%%%%%%%%%%%%%%%%%%%%%%%%
	Let $\Phi^{\h}\subseteq\h$ denote the event that $\tra_{-i}$ is already determined at $\h$, i.e.:
	%%%
	\begin{align}\label{eqn:Phi}
		\Phi^{\h} = \left\{\decision \in\h \:\Big|\quad
			\big(\forall \terminal,\terminal'\in\Terminal(\decision)\big)
			\big(\tra_{-i}(\terminal) =  \tra_{-i}(\terminal')\big)\right\}
	\end{align}
	%%%
	and let $\bar{\Phi}^{\h} = \h \backslash \Phi^{\h}$ be its complement.
	%%%
	Notice that the probability of $\Phi^{\h}$ and the distribution of $\tra^{-1}_{-i}(a_{-i})$ conditional on $\Phi^{\h}$, 
	are independent from $i$'s choice at $\h$.
	%%%
	Hence, by Bayes' rule:
	%%%
	\begin{align*} 
		\Exp{ \ua_i(a_i',a_{-i}) \,\big|\, \h,\move'}
			&= \disZ^*\left( \Phi^{\h} \,\big|\, \h,\move' \right)
					\Exp{ \ua_i(a_i',a_{-i}) \,\big|\, \h,\move',\Phi^{\h}} \\
				\nonumber & \qquad\qquad  + \disZ^*\left( \bar\Phi^{\h} \,\big|\, \h,\move'\right)
					\Exp{ \ua_i(a_i',a_{-i}) \,\big|\, \h,\move',\bar\Phi^{\h}} \\
			 &= 
				\disZ^*\left( \Phi^{\h} \,\big|\, \h,\move^* \right)
					\Exp{ \ua_i(a_i',a_{-i}) \,\big|\, \h,\move^*} \\
				\nonumber &\qquad\qquad  + \disZ^*\left(\bar\Phi^{\h} \,\big|\, \h,\move^*\right)
					\Exp{ \ua_i(a_i',a_{-i}) \,\big|\, \h,\move',\bar\Phi^{\h}} \\
			&\geq
				\disZ^*\left(\Phi^{\h} \,\big|\, \h,\move^*\right) \Exp{\ua_i(a_i',a_{-i}) \,\big|\, \h,\move^*}  
				+ \disZ^*\left(\bar\Phi^{\h} \,\big|\, \h,\move^*\right) \uw_i(a_i')
			%\label{eqn:B}
	\end{align*}
	%%%
	Together with (\ref{eqn:A}), this yields the following inequality which \emph{does not depend on} $\move'$:
	%%%
	\begin{align*}
		\Exp{\ua_i(a_i^*,a_{-i}) \,\big|\, \h,\move^*}
			\geq \disZ^*\left(\Phi^{\h}\,\big|\,\h,\move^*\right) \Exp{\ua_i(a_i',a_{-i})\,\big|\,\h,\move^*}
			+ \disZ^*\left(\bar\Phi^{\h}\,\big|\,\h,\move^*\right) \uw_i(a_i') 
	\end{align*}
	
%%%%%%%%%%%%%%%%%%%%%%%%%%%%%%%%%%%%%%%%%%%%%%%%%%%%%%%%%%%%%%%%%%%%%%%%%%%%%%%%%%%%%%%%%%%%%%%%%%%%%%%%%%%%%%%%%%%%%%%%%%%%%%%%%%%%%%%%%
%%%%%%%%%%%%%%%%%%%%%%%%%%%%%%%%%%%%%%%%%%%%%%%%%%%%%%%%%%%%%%%%%%%%%%%%%%%%%%%%%%%%%%%%%%%%%%%%%%%%%%%%%%%%%%%%%%%%%%%%%%%%%%%%%%%%%%%%%
	The last inequality holds for for each point in the game where $i$ chooses $a_i^*$ with positive probability.
	%%%
	Integrating over them yields:
	%%%
	\begin{align*}
		\sum_{a_{-i}\in A_{-i}} \!\! \disZ^*(a_i^*,a_{-i}) \ua_i(a_i^*,a_{-i})
			\geq \sum_{a_{-i}\in A_{-i}} \!\! \Big[
				\disZ^*(-i,a_i^*,a_{-i}) \ua_i(a_i',a_{-i}) 
				+ \disZ^*(i,a_i^*,a_{-i}) \uw_i(a_i') \Big]
	\end{align*}
	%%%
	Which correspond to the alternative characterization if ICE from equation (\ref{eqn:ICEalt}).
	%%%
	Since $i$, $a_i^*$ and $a_i'$ were arbitrary, it follows that $\alpha$ is an ICE.
\end{BSproof*}
\subsection{C-rationalizability and FC-rationalizability}

%%%%%%%%%%%%%%%%%%%%%%%%%%%%%%%%%%%%%%%%%%%%%%%%%%%%%%%%%%%%%%%%%%%%%%%%%%%%%%%%%%%%%%%%%%%%%%%%%%%%%%%%%%%%%%%%%%%%%%%%%%%%%%%%%%%%%%%%%
%%%%%%%%%%%%%%%%%%%%%%%%%%%%%%%%%%%%%%%%%%%%%%%%%%%%%%%%%%%%%%%%%%%%%%%%%%%%%%%%%%%%%%%%%%%%%%%%%%%%%%%%%%%%%%%%%%%%%%%%%%%%%%%%%%%%%%%%%
%%%%%%%%%%%%%%%%%%%%%%%%%%%%%%%%%%%%%%%%%%%%%%%%%%%%%%%%%%%%%%%%%%%%%%%%%%%%%%%%%%%%%%%%%%%%%%%%%%%%%%%%%%%%%%%%%%%%%%%%%%%%%%%%%%%%%%%%%
%%%%%%%%%%%%%%%%%%%%%%%%%%%%%%%%%%%%%%%%%%%%%%%%%%%%%%%%%%%%%%%%%%%%%%%%%%%%%%%%%%%%%%%%%%%%%%%%%%%%%%%%%%%%%%%%%%%%%%%%%%%%%%%%%%%%%%%%%
%%%%%%%%%%%%%%%%%%%%%%%%%%%%%%%%%%%%%%%%%%%%%%%%%%%%%%%%%%%%%%%%%%%%%%%%%%%%%%%%%%%%%%%%%%%%%%%%%%%%%%%%%%%%%%%%%%%%%%%%%%%%%%%%%%% C-rat
\begin{BSproof*}[Proof of Proposition \ref{prop:Crat}]
	C-rationalizable actions are clearly not absolutely dominated.
	%%%
	For the opposite direction, fix an action $a_i^*\in A_i$ that is not absolutely dominated in $A'$.
	%%%
	Let $a_{-i}^*\in\argmax_{a_{-i}\in A_{-i}} \ua_i(a_i^*,a_{-i})$.
	%%%
	Since $a_i^*$ is not dominated in $A'$,
	for every $a_i'\in A_i$ there exists some $a_{-i}(a_i')\in A'_{-i}$ 
	such that  $\ua_i(a^*)\geq\ua_i\big(a_i',a_{-i}(a_i')\big)$.
	%%%
	Hence $a_i^*$ is a best response to $\lambda_i\in\Lambda_i(A')$,
	with $\lambda_i(a_{-i}^*|a_i^*)=1$ and $\lambda_i\big(a_{-i}(a_i')\big)=1$ for $a_i'\neq a_i^*$.
	
%%%%%%%%%%%%%%%%%%%%%%%%%%%%%%%%%%%%%%%%%%%%%%%%%%%%%%%%%%%%%%%%%%%%%%%%%%%%%%%%%%%%%%%%%%%%%%%%%%%%%%%%%%%%%%%%%%%%%%%%%%%%%%%%%%%%%%%%%
%%%%%%%%%%%%%%%%%%%%%%%%%%%%%%%%%%%%%%%%%%%%%%%%%%%%%%%%%%%%%%%%%%%%%%%%%%%%%%%%%%%%%%%%%%%%%%%%%%%%%%%%%%%%%%%%%%%%%%%%%%%%%%%%%%%%%%%%%
	An elimination procedure can be described by a function $\keep:\A\rightarrow\A$,
	describing kept actions, such that for $A' \in\A$: 
	%%%
		(\emph{i}) never adds new actions, i.e.\ $\keep(A')\subseteq A'$;
		%%%
		(\emph{ii}) never eliminates undominated actions, i.e.\ $\CR(A')\subseteq \keep(A')$;
		%%%
		and  (\emph{iii}) if there are dominated actions then it always eliminates at least one,
			i.e.\ $\CR(A')\neq A'$ implies $\keep(A')\neq A'$.
		
%%%%%%%%%%%%%%%%%%%%%%%%%%%%%%%%%%%%%%%%%%%%%%%%%%%%%%%%%%%%%%%%%%%%%%%%%%%%%%%%%%%%%%%%%%%%%%%%%%%%%%%%%%%%%%%%%%%%%%%%%%%%%%%%%%%%%%%%%
%%%%%%%%%%%%%%%%%%%%%%%%%%%%%%%%%%%%%%%%%%%%%%%%%%%%%%%%%%%%%%%%%%%%%%%%%%%%%%%%%%%%%%%%%%%%%%%%%%%%%%%%%%%%%%%%%%%%%%%%%%%%%%%%%%%%%%%%%
	Now consider the corresponding sequence of surviving actions $(A^n)\in\A^\Natural$
	defined recursively by $A^1=A$ and $A^{n+1} = \keep(A^n)$.
	%%%
	For $n\in\Natural$ with $A^n\neq \emptyset$, there exists some action profile $a^0\in A^n$.
	%%%
	Ans, since the game is finite, for each player $i$ there exists a best response $a_i^*$ to $a_{-i}^0$.
	%%%
	By (\emph{ii}) this implies that $a^*\in \CR(A^n) \subseteq A^{n+1}$.
	%%%
	Thus, by induction, $(A^n)$ is weakly decreasing sequence of \emph{nonempty} sets.
	%%%
	Therefore, since $\A$ is finite, $(A^n)$ converges in finite iterations to a nonempty limit $A^*$.
	%%%
	Since $\CRR \subseteq\CR(\CRR)$ and $\CR(\blank)$ is $\subseteq$-monotone,
	(\emph{ii}) implies that $\CRR\subseteq A^n$ for all $n\in\Natural$, and thus $\CRR\subseteq A^*$.
	%%%
	Finally, (\emph{iii}) implies that $A^{*} \subseteq \CR(A^{*})$ and thus $A^{*} \subseteq \CRR$.
\end{BSproof*}
%%%%%%%%%%%%%%%%%%%%%%%%%%%%%%%%%%%%%%%%%%%%%%%%%%%%%%%%%%%%%%%%%%%%%%%%%%%%%%%%%%%%%%%%%%%%%%%%%%%%%%%%%%%%%%%%%%%%%%%%%%%%%%%%%%%%% QED

%%%%%%%%%%%%%%%%%%%%%%%%%%%%%%%%%%%%%%%%%%%%%%%%%%%%%%%%%%%%%%%%%%%%%%%%%%%%%%%%%%%%%%%%%%%%%%%%%%%%%%%%%%%%%%%%%%%%%%%%%%%%%%%%%%%%%%%%%
%%%%%%%%%%%%%%%%%%%%%%%%%%%%%%%%%%%%%%%%%%%%%%%%%%%%%%%%%%%%%%%%%%%%%%%%%%%%%%%%%%%%%%%%%%%%%%%%%%%%%%%%%%%%%%%%%%%%%%%%%%%%%%%%%%%%%%%%%
%%%%%%%%%%%%%%%%%%%%%%%%%%%%%%%%%%%%%%%%%%%%%%%%%%%%%%%%%%%%%%%%%%%%%%%%%%%%%%%%%%%%%%%%%%%%%%%%%%%%%%%%%%%%%%%%%%%%%%%%%%%%%%%%%%%%%%%%%
%%%%%%%%%%%%%%%%%%%%%%%%%%%%%%%%%%%%%%%%%%%%%%%%%%%%%%%%%%%%%%%%%%%%%%%%%%%%%%%%%%%%%%%%%%%%%%%%%%%%%%%%%%%%%%%%%%%%%%%%%%%%%%%%%%%%%%%%%
%%%%%%%%%%%%%%%%%%%%%%%%%%%%%%%%%%%%%%%%%%%%%%%%%%%%%%%%%%%%%%%%%%%%%%%%%%%%%%%%%%%%%%%%%%%%%%%%%%%%%%%%%%%%%%%%%%%%%%%%%%%%%%%%%% FC-rat
\begin{BSproof*}[Proof of Proposition \ref{prop:FLR}]
	$a^*_i\in\FR_i(A')$ if and only if it is a best response to \emph{some} $\con_i=\mu\con_i^0+(1-\mu)\con_i^1$,
	with  $\con_i^0\in\Delta(A_{-i}\backslash A_{-i}')$, $\con^1_i\in\Con(A'_{-i})$ and $\mu\in[0,1]$.
	%%%
	Which holds if and only if it is a best response to those beliefs which are more favorable for $a_i^*$,
	i.e. beliefs with:
	%%%
	\begin{align*}
		\con_i^1\left( \argmax_{a_{-i}\in A_{-i}} \Big\{\ua_i(a_i^*,a_{-i})\Big\} \:\Big|\, a_i^* \right)=1
		\quad\text{\BSCblack and }\quad
		\con_i^1\left( \argmin_{a_{-i}\in A_{-i}} \Big\{\ua_i(a_i,a_{-i})\Big\} \:\Big|\, a_i \neq a_i^* \right)=1
	\end{align*}		
	%%%
	Hence, after some simple algebra, $a_i^*\in\FR_i(A')$ if and only if  for every $a_i'\in A_i$:
	%%%
%	\begin{align*}
%	& (1-\mu) \max_{a_{-i}\in A_{-i}'} \Big\{\ua_i(a_i^*,a_{-i})\Big\} 
%		+  \sum_{a_{-i}\not\in A_{-i}'} \mu\con_i^0(a_{-i}) \ua_i(a_i^*,a_{-i}) \\
%	& \qquad \geq (1-\mu) \min_{a_{-i}\in A_{-i}'} \Big\{\ua_i(a_i',a_{-i})\Big\} 
%		+ \sum_{a_{-i}\not\in A_{-i}'} \mu\con_i^0(a_{-i}) \ua_i(a_i',a_{-i})
%	\end{align*}	
	\begin{align*}
	& (1-\mu) \Big[ \bar{w}_i(a_i^*) - \uw_i (a_i') \Big] 
		+ \sum_{a_{-i}\not\in A_{-i}'} \mu\con_i^0(a_{-i}) \Big[ \ua_i(a_i^*,a_{-i}) - \ua_i(a_i',a_{-i}) \Big]  \geq 0
	\end{align*}
	%%%
	where $\bar{w}_i(a_i^*,A')\equiv \max_{a_{-i}\in A_{-i}'} \big\{\ua_i(a_i^*,a_{-i})\big\}$
	%%%
	That is, if and only if it is a best response to some (non-counterfactual) belief in the auxiliary
	strategic form game $(I,\tilde{A},\tilde{\ua})$ with $\tilde{A}_i = A_i$, 
	$\tilde{A}_{-i}= \big(A_{-i}\backslash A_{-i}\big) \cup \{a^0_{-i}\}$,
	and $\tilde{\ua}_i:\tilde{A}\rightarrow\Real$ given by:
	%%%
	\begin{align*}
		\tilde{\ua}_i(a_i,a_{-i}) = \left\{\begin{array}{l@{\quad\text{if}\quad}l}
				\ua_i(a_i,a_{-i})   & a_{-i} \not\in A_{-i}' \\[1ex]
				\bar{w}_i(a_i^*,A') & a_{-i} \in A_{-i}' \,\wedge\, a_i=a_i^*\\[1ex]
				\uw_i(a_i',A')      & a_{-i} \in A_{-i}' \,\wedge\, a_i\neq a_i^*
		\end{array}\right.
	\end{align*}
	%%%
	The result then follows from the well known equivalence between never best responses and dominated actions,
	cf.\ Lemma 3 in \cite{pearce}.
\end{BSproof*}
%%%%%%%%%%%%%%%%%%%%%%%%%%%%%%%%%%%%%%%%%%%%%%%%%%%%%%%%%%%%%%%%%%%%%%%%%%%%%%%%%%%%%%%%%%%%%%%%%%%%%%%%%%%%%%%%%%%%%%%%%%%%%%%%%%%%% QED

%%%%%%%%%%%%%%%%%%%%%%%%%%%%%%%%%%%%%%%%%%%%%%%%%%%%%%%%%%%%%%%%%%%%%%%%%%%%%%%%%%%%%%%%%%%%%%%%%%%%%%%%%%%%%%%%%%%%%%%%%%%%%%%%%%%%%%%%%
%%%%%%%%%%%%%%%%%%%%%%%%%%%%%%%%%%%%%%%%%%%%%%%%%%%%%%%%%%%%%%%%%%%%%%%%%%%%%%%%%%%%%%%%%%%%%%%%%%%%%%%%%%%%%%%%%%%%%%%%%%%%%%%%%%%%%%%%%
%%%%%%%%%%%%%%%%%%%%%%%%%%%%%%%%%%%%%%%%%%%%%%%%%%%%%%%%%%%%%%%%%%%%%%%%%%%%%%%%%%%%%%%%%%%%%%%%%%%%%%%%%%%%%%%%%%%%%%%%%%%%%%%%%%%%%%%%%
%%%%%%%%%%%%%%%%%%%%%%%%%%%%%%%%%%%%%%%%%%%%%%%%%%%%%%%%%%%%%%%%%%%%%%%%%%%%%%%%%%%%%%%%%%%%%%%%%%%%%%%%%%%%%%%%%%%%%%%%%%%%%%%%%%%%%%%%%
%%%%%%%%%%%%%%%%%%%%%%%%%%%%%%%%%%%%%%%%%%%%%%%%%%%%%%%%%%%%%%%%%%%%%%%%%%%%%%%%%%%%%%%%%%%%%%%%%%%%%%%%%%%%%%%%%%%%%%%%%%%%%%%%%%%%%%%%%
%%%%%%%%%%%%%%%%%%%%%%%%%%%%%%%%%%%%%%%%%%%%%%%%%%%%%%%%%%%%%%%%%%%%%%%%%%%%%%%%%%%%%%%%%%%%%%%%%%%%%%%%%%%%%%%%%%%%%%%%%%%%%%%%%%%%%%%%%
%%%%%%%%%%%%%%%%%%%%%%%%%%%%%%%%%%%%%%%%%%%%%%%%%%%%%%%%%%%%%%%%%%%%%%%%%%%%%%%%%%%%%%%%%%%%%%%%%%%%%%%%%%%%%%%%%%%%%%%%%%%%%%%%%%%%%%%%%
%%%%%%%%%%%%%%%%%%%%%%%%%%%%%%%%%%%%%%%%%%%%%%%%%%%%%%%%%%%%%%%%%%%%%%%%%%%%%%%%%%%%%%%%%%%%%%%%%%%%%%%%%%%%%%%%%%%%%%%%%%%%%%%%%%%%%%%%%
%%%%%%%%%%%%%%%%%%%%%%%%%%%%%%%%%%%%%%%%%%%%%%%%%%%%%%%%%%%%%%%%%%%%%%%%%%%%%%%%%%%%%%%%%%%%%%%%%%%%%%%%%%%%%%%%%%%%%%%%%%%%%%%%%%%%%%%%%
%%%%%%%%%%%%%%%%%%%%%%%%%%%%%%%%%%%%%%%%%%%%%%%%%%%%%%%%%%%%%%%%%%%%%%%%%%%%%%%%%%%%%%%%%%%%%%%%%%%%%%%%%%%%%%%%%%%%%%%%%%%%%% Sequential
\subsection{Sequential implementation}\label{app:sequential}

%%%%%%%%%%%%%%%%%%%%%%%%%%%%%%%%%%%%%%%%%%%%%%%%%%%%%%%%%%%%%%%%%%%%%%%%%%%%%%%%%%%%%%%%%%%%%%%%%%%%%%%%%%%%%%%%%%%%%%%%%%%%%%%%%%%%%%%%%
%%%%%%%%%%%%%%%%%%%%%%%%%%%%%%%%%%%%%%%%%%%%%%%%%%%%%%%%%%%%%%%%%%%%%%%%%%%%%%%%%%%%%%%%%%%%%%%%%%%%%%%%%%%%%%%%%%%%%%%%%%%%%%%%%%%%%%%%%
%%%%%%%%%%%%%%%%%%%%%%%%%%%%%%%%%%%%%%%%%%%%%%%%%%%%%%%%%%%%%%%%%%%%%%%%%%%%%%%%%%%%%%%%%%%%%%%%%%%%%%%%%%%%%%%%%%%%%%%%%%%%%%%%%%%%%%%%%
%%%%%%%%%%%%%%%%%%%%%%%%%%%%%%%%%%%%%%%%%%%%%%%%%%%%%%%%%%%%%%%%%%%%%%%%%%%%%%%%%%%%%%%%%%%%%%%%%%%%%%%%%%%%%%%%%%%%%%%%%%%%%%%%%%%%%%%%%
%%%%%%%%%%%%%%%%%%%%%%%%%%%%%%%%%%%%%%%%%%%%%%%%%%%%%%%%%%%%%%%%%%%%%%%%%%%%%%%%%%%%%%%%%%%%%%%%%%%%%%%%%%%%%%%%%%%%%%%%%%%%%%%%%%%%% APS
\begin{BSproof*}[Proof of Proposition \ref{prop:APS}]
	By definition $A^2\subseteq A^1=A$.
	%%%
	Now suppose that $\Ac^{n+1}\subseteq \Ac^{n}$ for some $n\in\Natural$.
	%%%
	Since $\ICE(\blank)$ is a monotone correspondence, then so is $\B(\blank)$.
	%%%
	Hence, $A^{n+2} = \B(A^{n+1}) \subseteq \B(A^n) = A^{n+1}$.
	%%%
	Therefore, by induction, $A^{n+1}\subseteq A^n$ for all $n\in\Natural$.
	%%%
	Since $\A$ is finite,
	this implies that $(A^n)$ converges \emph{after finite iterations} to a limit $\CEac$. 

%%%%%%%%%%%%%%%%%%%%%%%%%%%%%%%%%%%%%%%%%%%%%%%%%%%%%%%%%%%%%%%%%%%%%%%%%%%%%%%%%%%%%%%%%%%%%%%%%%%%%%%%%%%%%%%%%%%%%%%%%%%%%%%%%%%%%%%%%
%%%%%%%%%%%%%%%%%%%%%%%%%%%%%%%%%%%%%%%%%%%%%%%%%%%%%%%%%%%%%%%%%%%%%%%%%%%%%%%%%%%%%%%%%%%%%%%%%%%%%%%%%%%%%%%%%%%%%%%%%%%%%%%%%%%%%%%%%
	It follows that there exists some $m\in\Natural$ such that $\CEac = A^m = \B(A^m)$,
	and thus $\CEac=\B(\CEac)$.
	%%%
	Furthermore, since $\ICE(\empty)\subseteq \ICE(A^m)$ contains the set if correlated equilibria, 
	it follows that $A^m = \B(A^m) \supseteq \B(\emptyset) \neq \emptyset$.
	%%%
	Finally, let $A'\in\A$ be such that $A'\subseteq\B(A')$.
	%%%
	By definition $\Aci\subseteq\Ac = \Ac^1$.
	%%%
	By monotonicity of $\B$, if $A'\subseteq A^n$ then  $A' \subseteq \B(A') \subseteq \B(A^n) = A^{n+1}$.
	%%%
	Hence, by the induction principle, $A' \subseteq \Ac^n$ for all $n\in\Natural$,
	which implies that $A'\subseteq \CEac$.
\end{BSproof*}
%%%%%%%%%%%%%%%%%%%%%%%%%%%%%%%%%%%%%%%%%%%%%%%%%%%%%%%%%%%%%%%%%%%%%%%%%%%%%%%%%%%%%%%%%%%%%%%%%%%%%%%%%%%%%%%%%%%%%%%%%%%%%%%%%%%%% QED

%%%%%%%%%%%%%%%%%%%%%%%%%%%%%%%%%%%%%%%%%%%%%%%%%%%%%%%%%%%%%%%%%%%%%%%%%%%%%%%%%%%%%%%%%%%%%%%%%%%%%%%%%%%%%%%%%%%%%%%%%%%%%%%%%%%%%%%%%
%%%%%%%%%%%%%%%%%%%%%%%%%%%%%%%%%%%%%%%%%%%%%%%%%%%%%%%%%%%%%%%%%%%%%%%%%%%%%%%%%%%%%%%%%%%%%%%%%%%%%%%%%%%%%%%%%%%%%%%%%%%%%%%%%%%%%%%%%
%%%%%%%%%%%%%%%%%%%%%%%%%%%%%%%%%%%%%%%%%%%%%%%%%%%%%%%%%%%%%%%%%%%%%%%%%%%%%%%%%%%%%%%%%%%%%%%%%%%%%%%%%%%%%%%%%%%%%%%%%%%%%%%%%%%%%%%%%
%%%%%%%%%%%%%%%%%%%%%%%%%%%%%%%%%%%%%%%%%%%%%%%%%%%%%%%%%%%%%%%%%%%%%%%%%%%%%%%%%%%%%%%%%%%%%%%%%%%%%%%%%%%%%%%%%%%%%%%%%%%%%%%%%%%%%%%%%
%%%%%%%%%%%%%%%%%%%%%%%%%%%%%%%%%%%%%%%%%%%%%%%%%%%%%%%%%%%%%%%%%%%%%%%%%%%%%%%%%%%%%%%%%%%%%%%%%%%%%%%%%%%%%%%%%%%%%%%%%%%%%% sufficient
\begin{BSproof*}[Proof of Theorem \ref{thm:SE}]
	$\CEac=\B(\CEac)$ by Proposition \ref{prop:APS}.
	%%%
	$\CEac\subseteq\B(\CEac)$ implies that $\supp(\alpha)\subseteq\CEac\subseteq \CEac\cup\R$ for $\alpha\in\ICE(\CEac\cup\R)$.
	%%%
	Hence, $\ICE(\CEac\cup\R)$ is characterized by a \emph{finite} set of \emph{affine} inequalities, and is thus convex.
	%%% 
	$\CEac\supseteq\B(\CEac)$ then implies that there exists some $\alpha^*\in\ICE(\CEac\cup\R)$
	such that $\alpha^*_i(a_i)>0$ for every $i$ and $a_i\in\CEac_i$.

%%%%%%%%%%%%%%%%%%%%%%%%%%%%%%%%%%%%%%%%%%%%%%%%%%%%%%%%%%%%%%%%%%%%%%%%%%%%%%%%%%%%%%%%%%%%%%%%%%%%%%%%%%%%%%%%%%%%%%%%%%%%%%%%%%%%%%%%%
%%%%%%%%%%%%%%%%%%%%%%%%%%%%%%%%%%%%%%%%%%%%%%%%%%%%%%%%%%%%%%%%%%%%%%%%%%%%%%%%%%%%%%%%%%%%%%%%%%%%%%%%%%%%%%%%%%%%%%%%%%%%%%%%%%%%%%%%%
	Let $G^*$ be the mediated game which implements $\alpha^*$ as an ICE with respect to $\CEac\cup\R$,
	and consider an action $a_i^0\in \CEac\cup\R$ used as a threat.
	%%%
	If $a_i^0\in\CEac$, then the \emph{unique} information set in which $i$ is asked to choose $a_i^0$ is along the equilibrium path.
	%%%
	If $a_i^0\in\R_i\backslash\CEac_i$ then, by definition, 
	$a_i^0$ is a best response to a (non-counterfactual) belief $\con\in\Delta(A_{-i})$.
	%%%
	Every information set in which $i$ is asked to use $a_i^0$ is off the equilibrium path,
	and thus $i$ may believe that the most likely tremble leading to it corresponds to $-i$ choosing according to $\con$.
	%%%
	In either case, choosing $a^0_i$ is indeed sequentially rational.
	%%%
	Following recommendations thus constitutes a sequential equilibrium.

%%%%%%%%%%%%%%%%%%%%%%%%%%%%%%%%%%%%%%%%%%%%%%%%%%%%%%%%%%%%%%%%%%%%%%%%%%%%%%%%%%%%%%%%%%%%%%%%%%%%%%%%%%%%%%%%%%%%%%%%%%%%%%%%%%%%%%%%%
%%%%%%%%%%%%%%%%%%%%%%%%%%%%%%%%%%%%%%%%%%%%%%%%%%%%%%%%%%%%%%%%%%%%%%%%%%%%%%%%%%%%%%%%%%%%%%%%%%%%%%%%%%%%%%%%%%%%%%%%%%%%%%%%%%%%%%%%%
	Now consider any $\alpha\in\ICE(\CEac\cup\R)$ and the corresponding mediated game $G$. 
	%%%
	Let $\hat{G}$ be the extensive form mechanism in which 
		%%%
		(\emph{i}) the mediator randomizes  between $G$ and $G^*$ with probabilities $(1-\epsilon)$ and $\epsilon$ respectively;
		%%%
		and (\emph{ii}) players are only informed about recommendations, in particular they cannot distinguish between $\G$ or $\G^*$.
	%%%		
	Now suppose players agree that trembles in $G^*$ are more likely than trembles in $G$,
	so that whenever they are asked to perform an action in $\CEac\cup\R$, they will believe 
	that they are either along the equilibrium path, or in $G^*$.
	%%%
	From the previous analysis it follows that complying remains to be sequentially optimal. 
	%%%	
	And hence the distribution $\hat{\alpha} = (1-\epsilon)\alpha + \epsilon \alpha^*$ is sequentially implementable. 
	%%%
	Of course $\hat\alpha$ approximates $\alpha$ as $\epsilon$ approaches $0$.
	%%%
	If one allows for $\epsilon=0$ (meaning that the mediator can make mistakes), then $\mac$ is sequentially implementable. 
\end{BSproof*}
%%%%%%%%%%%%%%%%%%%%%%%%%%%%%%%%%%%%%%%%%%%%%%%%%%%%%%%%%%%%%%%%%%%%%%%%%%%%%%%%%%%%%%%%%%%%%%%%%%%%%%%%%%%%%%%%%%%%%%%%%%%%%%%%%%%%% QED

%%%%%%%%%%%%%%%%%%%%%%%%%%%%%%%%%%%%%%%%%%%%%%%%%%%%%%%%%%%%%%%%%%%%%%%%%%%%%%%%%%%%%%%%%%%%%%%%%%%%%%%%%%%%%%%%%%%%%%%%%%%%%%%%%%%%%%%%%
%%%%%%%%%%%%%%%%%%%%%%%%%%%%%%%%%%%%%%%%%%%%%%%%%%%%%%%%%%%%%%%%%%%%%%%%%%%%%%%%%%%%%%%%%%%%%%%%%%%%%%%%%%%%%%%%%%%%%%%%%%%%%%%%%%%%%%%%%
%%%%%%%%%%%%%%%%%%%%%%%%%%%%%%%%%%%%%%%%%%%%%%%%%%%%%%%%%%%%%%%%%%%%%%%%%%%%%%%%%%%%%%%%%%%%%%%%%%%%%%%%%%%%%%%%%%%%%%%%%%%%%%%%%%%%%%%%%
%%%%%%%%%%%%%%%%%%%%%%%%%%%%%%%%%%%%%%%%%%%%%%%%%%%%%%%%%%%%%%%%%%%%%%%%%%%%%%%%%%%%%%%%%%%%%%%%%%%%%%%%%%%%%%%%%%%%%%%%%%%%%%%%%%%%%%%%%
%%%%%%%%%%%%%%%%%%%%%%%%%%%%%%%%%%%%%%%%%%%%%%%%%%%%%%%%%%%%%%%%%%%%%%%%%%%%%%%%%%%%%%%%%%%%%%%%%%%%%%%%%%%%%%%%%%%%%%%%%%%%%%%%%%%%% 2x2
\begin{BSproof*}[Proof of Proposition \ref{prop:2by2}]
	Let $A_i = \{a_i,b_i\}$ for $i=1,2$, and suppose that there are no repeated payoffs.
	%%%
	If some player $i$ has an absolutely dominated strategy, say $a_i$, 
	and let $a_{-i}$ be $-i$'s \emph{unique} best response to $a_i$.
	%%%
	In this case $(a_i,a_{-i})$ is the \emph{unique} ICE with respect to $\CRR$,
	and it is a sequential equilibrium of the simultaneous move game.
	%%% 
	In every other case, I will show that every action can be played 
	with positive probability in a sequentially implementable distribution (i.e.\ $A^\SE = A$), 
	and hence the result follows from Theorem \ref{thm:SE}.
	%%%
	This is straightforward if there are no strictly dominated strategies,
	because then there exists a completely mixed (sequential) equilibrium. 

%%%%%%%%%%%%%%%%%%%%%%%%%%%%%%%%%%%%%%%%%%%%%%%%%%%%%%%%%%%%%%%%%%%%%%%%%%%%%%%%%%%%%%%%%%%%%%%%%%%%%%%%%%%%%%%%%%%%%%%%%%%%%%%%%%%%%%%%%
%%%%%%%%%%%%%%%%%%%%%%%%%%%%%%%%%%%%%%%%%%%%%%%%%%%%%%%%%%%%%%%%%%%%%%%%%%%%%%%%%%%%%%%%%%%%%%%%%%%%%%%%%%%%%%%%%%%%%%%%%%%%%%%%%%%%%%%%%
	The interesting cases are when there are no absolutely dominated strategies, 
	but at least one player has a strictly dominated strategy.
	%%%
	Let $\con_i,\con_i'\in\Con_i$ denote the counterfactual beliefs:
	\begin{align}
		 \con_i(a_{-i}|a_i) = 1 \:\:\wedge\:\:  \con_i(b_{-i}|b_{-i}) = 1
		 \qquad \text{\BSCblack and } \qquad  
		\con'_i(b_{-i}|a_i) = 1 \:\:\wedge\:\: \con_i'(a_{-i}|b_{-i}) = 1
		\label{eqn:2x2}
	\end{align}
	%%%
	If $b_i$ is not absolutely dominated but it is strictly dominated by $a_i$,
	then it must be a best response to either $\con_i$ or $\con_i'$.
	%%%
	Furthermore, since there are no repeated payoffs, it must be a \emph{strict} best response.
	%%%	
	There are two cases to consider depending on whether one or two players have dominated strategies. 

%%%%%%%%%%%%%%%%%%%%%%%%%%%%%%%%%%%%%%%%%%%%%%%%%%%%%%%%%%%%%%%%%%%%%%%%%%%%%%%%%%%%%%%%%%%%%%%%%%%%%%%%%%%%%%%%%%%%%%%%%%%%%%%%%%%%%%%%%
%%%%%%%%%%%%%%%%%%%%%%%%%%%%%%%%%%%%%%%%%%%%%%%%%%%%%%%%%%%%%%%%%%%%%%%%%%%%%%%%%%%%%%%%%%%%%%%%%%%%%%%%%%%%%%%%%%%%%%%%%%%%%%%%%%%%%%%%%
	\begin{figure}[htb]
	\centering
	\psset{unit=7mm}
	\begin{pspicture}(0,-4.5)(14,6.5)
		%\psgrid%
		\footnotesize%
		% Equilibrium path 
		\psset{linewidth=1.2pt,linecolor=red,arrows=->}	
			\psline(04,03)(02,05) \psline(02,01)(00,02) \psline(02,05)(00,06)
			\psline(07,00)(05,-2) \psline(05,-2)(04,-4) \psline(09,-2)(10,-4)
			\psline(10,03)(12,01) \psline(12,05)(14,06) \psline(12,01)(14,02)
		\psreset%
		% Branches
			\psline(4,3)(7,3)(10,3)(7,3)(7,0)
			\psline(00,06)(02,05)(00,04) \psline(00,02)(02,01)(00,00) \psline(02,05)(04,03)(02,01)
			\psline(14,06)(12,05)(14,04) \psline(14,02)(12,01)(14,00) \psline(12,05)(10,03)(12,01)
			\psline(04,-4)(05,-2)(06,-4) \psline(08,-4)(09,-2)(10,-4) \psline(09,-2)(07,00)(05,-2)
		% Information sets
		\psset{linestyle=dashed,linecolor=BSCdark}
			\psline(02,05)(02,01)(05,-2)
			\psline(12,05)(12,01)(09,-2)
			\psline(07,00)(10,03)
		\psreset%
		% Nodes
		\psset{fillcolor=white}
			\psdots[dotstyle=o](7,3)
			\psdots(00,06)(00,04)(00,02)(00,00)(02,05)(02,01)(04,03)
			\psdots(10,03)(12,05)(12,01)(14,06)(14,04)(14,02)(14,00)
			\psdots(07,00)(09,-2)(05,-2)(04,-4)(06,-4)(08,-4)(10,-4)
		\psreset%
		% Labels
			\rput(7,3.5){$0$}
			\rput[b](4,3.3){$1$} \rput[b](10,3.3){$1$} \rput[r](6.7,0){$1$}
			\rput[b](2,5.3){$2$} \rput[t](2,0.7){$2$} \rput[t](12,0.7){$2$}
			\rput[b](12,5.3){$2$} \rput[r](4.7,-2){$2$} \rput[l](9.3,-2){$2$}	
			\rput[t](5.5,2.9){$\big[1-\epsilon-\epsilon^2\big]$}
			\rput[t](8.5,2.9){$\big[\epsilon^2\big]$}
			\rput[r](6.9,1.5){$\big[\epsilon\big]$}
			\rput[lb](3.1,4.1){$a_1$} \rput[br](10.9,4.1){$a_1$} \rput[bl](8.1,-0.9){$a_1$}
			\rput[tl](3.1,1.9){$b_1$} \rput[tr](10.9,1.9){$b_1$} \rput[br](5.9,-0.9){$b_1$}
			\rput[tl](1.1,4.4){$b_2$} \rput[br](12.9,1.6){$b_2$} \rput[br](12.9,5.6){$b_2$}
			\rput[tl](1.1,0.4){$b_2$} \rput[bl](5.6,-2.9){$b_2$} \rput[bl](9.6,-2.9){$b_2$}
			\rput[lb](1.1,5.6){$a_2$} \rput[br](4.4,-2.9){$a_2$} \rput[br](8.4,-2.9){$a_2$}
			\rput[lb](1.1,1.6){$a_2$} \rput[tr](12.9,4.4){$a_2$} \rput[tr](12.9,0.4){$a_2$}
	\end{pspicture}
	\caption{Implementation of $b_1$ when it is the only dominated action.}
	\label{fig:2x2A}
	\end{figure}
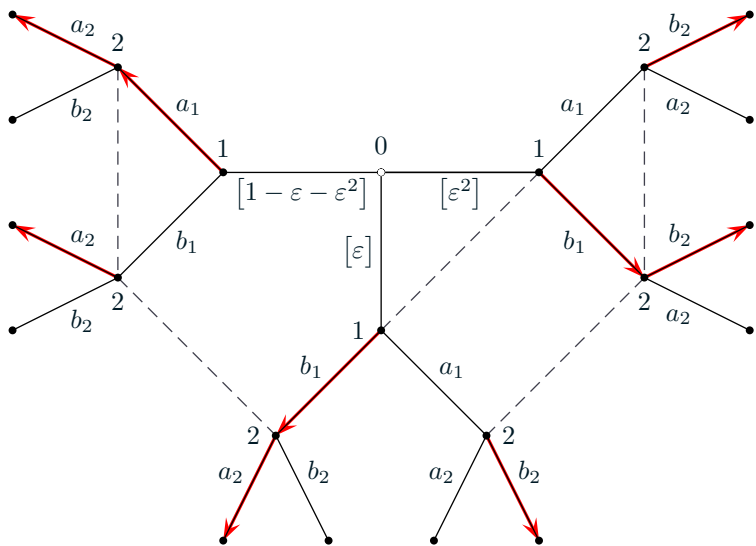

%%%%%%%%%%%%%%%%%%%%%%%%%%%%%%%%%%%%%%%%%%%%%%%%%%%%%%%%%%%%%%%%%%%%%%%%%%%%%%%%%%%%%%%%%%%%%%%%%%%%%%%%%%%%%%%%%%%%%%%%%%%%%%%%%%%%%%%%%
%%%%%%%%%%%%%%%%%%%%%%%%%%%%%%%%%%%%%%%%%%%%%%%%%%%%%%%%%%%%%%%%%%%%%%%%%%%%%%%%%%%%%%%%%%%%%%%%%%%%%%%%%%%%%%%%%%%%%%%%%%%%%%%%%%%%%%%%%
	Fist suppose that player $2$ has no dominated strategies but $b_1$ is dominated by $a_1$.
	%%%
	Further assume (without loss of generality) that $a_2$ is a best response to $a_1$.
	%%%
	This implies that $b_2$ is the unique best response to $b_1$,
	and that $(a_1,a_2)$ is a \emph{strict} NE of the simultaneous move game.
	%%%
	If $b_1$ is a best response to $\con_1$, then it suffices to have player $1$ move first and make his choice public. 
	%%%
	By backward induction, in the unique SPNE, player $2$ will choose $a_2$ if he chooses $a_1$ and $b_2$ if he chooses $b_1$.
	%%%
	Hence, $1$'s counterfactual beliefs are $\con_1$ and $b_1$ is the unique best response.

%%%%%%%%%%%%%%%%%%%%%%%%%%%%%%%%%%%%%%%%%%%%%%%%%%%%%%%%%%%%%%%%%%%%%%%%%%%%%%%%%%%%%%%%%%%%%%%%%%%%%%%%%%%%%%%%%%%%%%%%%%%%%%%%%%%%%%%%%
%%%%%%%%%%%%%%%%%%%%%%%%%%%%%%%%%%%%%%%%%%%%%%%%%%%%%%%%%%%%%%%%%%%%%%%%%%%%%%%%%%%%%%%%%%%%%%%%%%%%%%%%%%%%%%%%%%%%%%%%%%%%%%%%%%%%%%%%%
	Otherwise, if $\aci_1$ is a best response to $\con_1'$,
	then it can be implemented as an equilibrium of the mechanism in Figure (\ref{fig:2x2A}), with $\epsilon>0$ small enough.
	%%%
	The equilibrium strategies are represented with arrows. 
	%%%
	Player's are willing to choose $a_i$ because $(a_1,a_2)$ is a strict Nash equilibrium.
	%%%
	Player $2$ is willing to choose $b_2$ because it is a best response to $b_1$.
	%%%
	Player $1$ is willing to choose $b_1$ because his conjectures at that moment are close enough to $\con_1'$.
	%%%
	Since all the information sets are on the equilibrium path, the equilibrium is sequential.

%%%%%%%%%%%%%%%%%%%%%%%%%%%%%%%%%%%%%%%%%%%%%%%%%%%%%%%%%%%%%%%%%%%%%%%%%%%%%%%%%%%%%%%%%%%%%%%%%%%%%%%%%%%%%%%%%%%%%%%%%%%%%%%%%%%%%%%%%
%%%%%%%%%%%%%%%%%%%%%%%%%%%%%%%%%%%%%%%%%%%%%%%%%%%%%%%%%%%%%%%%%%%%%%%%%%%%%%%%%%%%%%%%%%%%%%%%%%%%%%%%%%%%%%%%%%%%%%%%%%%%%%%%%%%%%%%%%
	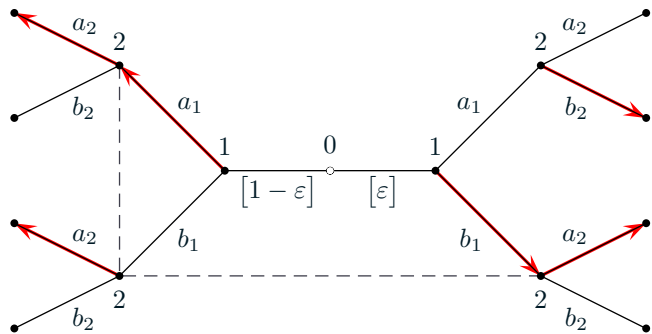
\begin{figure}[htb]
	\centering
	\psset{unit=7mm}
	\begin{pspicture}(0,-1)(14,7)
		%\psgrid%
		\footnotesize%
		% Equilibrium path 
		\psset{linewidth=1.2pt,linecolor=red,arrows=->}	
			\psline(05,03)(03,05) \psline(03,01)(01,02) \psline(03,05)(01,06)
			\psline(09,03)(11,01) \psline(11,05)(13,04) \psline(11,01)(13,02)
		\psreset%
		% Branches
			\psline(05,03)(09,03)
			\psline(03,05)(05,03)(03,01) \psline(01,06)(03,05)(01,04) \psline(01,02)(03,01)(01,00)
			\psline(11,05)(09,03)(11,01) \psline(13,06)(11,05)(13,04) \psline(13,02)(11,01)(13,00)
		% Information sets
		\psset{linestyle=dashed,linecolor=BSCdark}
			\psline(3,5)(3,1)(11,1)
		\psreset%
		% Nodes
		\psset{fillcolor=white}
			\psdots[dotstyle=o](7,3)
			\psdots(1,6)(1,4)(1,2)(1,0)(3,5)(3,1)(5,3)
			\psdots(9,3)(11,5)(11,1)(13,6)(13,4)(13,2)(13,0)
		\psreset%
		% Labels
			\rput(7,3.5){$0$}
			\rput[t](6,2.9){$\big[1-\epsilon\big]$}
			\rput[t](8,2.9){$\big[\epsilon\big]$}
			\rput[b](5,3.3){$1$} \rput[b](9,3.3){$1$}
			\rput[b](3,5.3){$2$} \rput[t](3,0.7){$2$} \rput[t](11,0.7){$2$} \rput[b](11,5.3){$2$}
			\rput[lb](4.1,4.1){$a_1$} \rput[br](9.9,4.1){$a_1$}
			\rput[tl](4.1,1.9){$b_1$} \rput[tr](9.9,1.9){$b_1$}
			\rput[lb](2.1,5.6){$a_2$} \rput[lb](2.1,1.6){$a_2$} \rput[br](11.9,1.6){$a_2$} \rput[br](11.9,5.6){$a_2$}
			\rput[tl](2.1,4.4){$b_2$} \rput[tl](2.1,0.4){$b_2$} \rput[tr](11.9,4.4){$b_2$} \rput[tr](11.9,0.4){$b_2$}
	\end{pspicture}
	\caption{Implementation of $b_1$ when $b_2$ is also dominated.}
	\label{fig:2x2B}
	\end{figure}

%%%%%%%%%%%%%%%%%%%%%%%%%%%%%%%%%%%%%%%%%%%%%%%%%%%%%%%%%%%%%%%%%%%%%%%%%%%%%%%%%%%%%%%%%%%%%%%%%%%%%%%%%%%%%%%%%%%%%%%%%%%%%%%%%%%%%%%%%
%%%%%%%%%%%%%%%%%%%%%%%%%%%%%%%%%%%%%%%%%%%%%%%%%%%%%%%%%%%%%%%%%%%%%%%%%%%%%%%%%%%%%%%%%%%%%%%%%%%%%%%%%%%%%%%%%%%%%%%%%%%%%%%%%%%%%%%%%
	Finally, suppose that both players have strictly dominated strategies, say $b_1$ and $b_2$.
	%%%
	In this case $(a_1,a_2)$ is a \emph{strict} NE. 
	%%%
	If $b_i$ is a best response to $\con_i$, then it can be implemented as a NE
	of the mechanism where $i$ moves first and $-i$ chooses $b_{-i}$ along the equilibrium path
	and punishes deviations with $a_{-i}$.
	%%%
	Otherwise, if $b_i$ is a best response to $\con_i'$, then it can be played with positive probability in a NE
	of the mechanism depicted in figure \ref{fig:2x2B}, with $\epsilon>0$ small enough. 
	%%%
	Hence there always exists EFMs $G^1$ and $G^2$ with NE in which $b_1$ and $b_2$ are played with positive probability.

%%%%%%%%%%%%%%%%%%%%%%%%%%%%%%%%%%%%%%%%%%%%%%%%%%%%%%%%%%%%%%%%%%%%%%%%%%%%%%%%%%%%%%%%%%%%%%%%%%%%%%%%%%%%%%%%%%%%%%%%%%%%%%%%%%%%%%%%%
%%%%%%%%%%%%%%%%%%%%%%%%%%%%%%%%%%%%%%%%%%%%%%%%%%%%%%%%%%%%%%%%%%%%%%%%%%%%%%%%%%%%%%%%%%%%%%%%%%%%%%%%%%%%%%%%%%%%%%%%%%%%%%%%%%%%%%%%%
	The proof is not complete because the equilibria are not subgame perfect.
	%%%
	For that purpose, one can construct a third mechanism in which nature randomizes between $G^1$ and $G^2$ and the simultaneous move game, 
	and every action is played with positive probability along the equilibrium path. 
	%%% 
	Information sets can be connected so that,  whenever a player is supposed to choose $b_i$ he believes that he is in $G^i$.
	%%%
	Doing so guarantees that the equilibrium is sequential.
\end{BSproof*}
%%%%%%%%%%%%%%%%%%%%%%%%%%%%%%%%%%%%%%%%%%%%%%%%%%%%%%%%%%%%%%%%%%%%%%%%%%%%%%%%%%%%%%%%%%%%%%%%%%%%%%%%%%%%%%%%%%%%%%%%%%%%%%%%%%%%% QED

%%%%%%%%%%%%%%%%%%%%%%%%%%%%%%%%%%%%%%%%%%%%%%%%%%%%%%%%%%%%%%%%%%%%%%%%%%%%%%%%%%%%%%%%%%%%%%%%%%%%%%%%%%%%%%%%%%%%%%%%%%%%%%%%%%%%%%%%%
%%%%%%%%%%%%%%%%%%%%%%%%%%%%%%%%%%%%%%%%%%%%%%%%%%%%%%%%%%%%%%%%%%%%%%%%%%%%%%%%%%%%%%%%%%%%%%%%%%%%%%%%%%%%%%%%%%%%%%%%%%%%%%%%%%%%%%%%%
%%%%%%%%%%%%%%%%%%%%%%%%%%%%%%%%%%%%%%%%%%%%%%%%%%%%%%%%%%%%%%%%%%%%%%%%%%%%%%%%%%%%%%%%%%%%%%%%%%%%%%%%%%%%%%%%%%%%%%%%%%%%%%%%%%%%%%%%%
%%%%%%%%%%%%%%%%%%%%%%%%%%%%%%%%%%%%%%%%%%%%%%%%%%%%%%%%%%%%%%%%%%%%%%%%%%%%%%%%%%%%%%%%%%%%%%%%%%%%%%%%%%%%%%%%%%%%%%%%%%%%%%%%%%%%%%%%%
%%%%%%%%%%%%%%%%%%%%%%%%%%%%%%%%%%%%%%%%%%%%%%%%%%%%%%%%%%%%%%%%%%%%%%%%%%%%%%%%%%%%%%%%%%%%%%%%%%%%%%%%%%%%%%%%%%%%%%%%%%%%%%%%%%%%%%%%%
%%%%%%%%%%%%%%%%%%%%%%%%%%%%%%%%%%%%%%%%%%%%%%%%%%%%%%%%%%%%%%%%%%%%%%%%%%%%%%%%%%%%%%%%%%%%%%%%%%%%%%%%%%%%%%%%%%%%%%%%%%%%%%%%%%%%%%%%%
%%%%%%%%%%%%%%%%%%%%%%%%%%%%%%%%%%%%%%%%%%%%%%%%%%%%%%%%%%%%%%%%%%%%%%%%%%%%%%%%%%%%%%%%%%%%%%%%%%%%%%%%%%%%%%%%%%%%%%%%%%%%%%%%%%%%%%%%%
%%%%%%%%%%%%%%%%%%%%%%%%%%%%%%%%%%%%%%%%%%%%%%%%%%%%%%%%%%%%%%%%%%%%%%%%%%%%%%%%%%%%%%%%%%%%%%%%%%%%%%%%%%%%%%%%%%%%%%%%%%%%%%%%%%%%%%%%%
%%%%%%%%%%%%%%%%%%%%%%%%%%%%%%%%%%%%%%%%%%%%%%%%%%%%%%%%%%%%%%%%%%%%%%%%%%%%%%%%%%%%%%%%%%%%%%%%%%%%%%%%%%%%%%%%%%%%%%%%%%%%%%%%%%%%%%%%%
%%%%%%%%%%%%%%%%%%%%%%%%%%%%%%%%%%%%%%%%%%%%%%%%%%%%%%%%%%%%%%%%%%%%%%%%%%%%%%%%%%%%%%%%%%%%%%%%%%%%%%%%%%%%%%%%%%%%%%%%%%%%%%%%%%%%% QSI
\subsection{QS implementation}

%%%%%%%%%%%%%%%%%%%%%%%%%%%%%%%%%%%%%%%%%%%%%%%%%%%%%%%%%%%%%%%%%%%%%%%%%%%%%%%%%%%%%%%%%%%%%%%%%%%%%%%%%%%%%%%%%%%%%%%%%%%%%%%%%%%%%%%%%
%%%%%%%%%%%%%%%%%%%%%%%%%%%%%%%%%%%%%%%%%%%%%%%%%%%%%%%%%%%%%%%%%%%%%%%%%%%%%%%%%%%%%%%%%%%%%%%%%%%%%%%%%%%%%%%%%%%%%%%%%%%%%%%%%%%%%%%%%
The proof of theorem theorem \ref{thm:QSI} is divided in two parts regarding necessity and sufficiency. 
%%%
To establish necessity it suffices to show that given a QSE of an EFM,  every action played with positive probability 
(on or off the equilibrium path) is in $\FLR$.
%%%
Then the proof of Theorem \ref{thm:Nash} applies simply replacing $\uw_i(a_i',A)$ with $\uw_i(a_i,\FLR)$.
%%%
This fact is established in Lemma \ref{lem:QSnec}. 
%%%
Given an EFM and an QSE $(\sigma^*,\psi^*)$, let $A_i^*\subseteq A_i$ denote the set of actions that 
$i$ plays with positive probability is some information set, i.e.:
%%%
\begin{align*}
	A_i^* = \left\{a_i\in A_i \:\Big|\quad 
		\Big(\exists\h\in\H\Big)\Big(\exists s_i\in S_i\Big)
		\Big(\sigma^*_i(s_i)>0 \:\wedge\: s_i\big(\h\big)\in\Moves^{a_i}(\h)\right\}
\end{align*}

%%%%%%%%%%%%%%%%%%%%%%%%%%%%%%%%%%%%%%%%%%%%%%%%%%%%%%%%%%%%%%%%%%%%%%%%%%%%%%%%%%%%%%%%%%%%%%%%%%%%%%%%%%%%%%%%%%%%%%%%%%%%%%%%%%%%%%%%%
%%%%%%%%%%%%%%%%%%%%%%%%%%%%%%%%%%%%%%%%%%%%%%%%%%%%%%%%%%%%%%%%%%%%%%%%%%%%%%%%%%%%%%%%%%%%%%%%%%%%%%%%%%%%%%%%%%%%%%%%%%%%%%%%%%%%%%%%%
%%%%%%%%%%%%%%%%%%%%%%%%%%%%%%%%%%%%%%%%%%%%%%%%%%%%%%%%%%%%%%%%%%%%%%%%%%%%%%%%%%%%%%%%%%%%%%%%%%%%%%%%%%%%%%%%%%%%%%%%%%%%%%%%%%%%%%%%%
%%%%%%%%%%%%%%%%%%%%%%%%%%%%%%%%%%%%%%%%%%%%%%%%%%%%%%%%%%%%%%%%%%%%%%%%%%%%%%%%%%%%%%%%%%%%%%%%%%%%%%%%%%%%%%%%%%%%%%%%%%%%%%%%%%%%%%%%%
%%%%%%%%%%%%%%%%%%%%%%%%%%%%%%%%%%%%%%%%%%%%%%%%%%%%%%%%%%%%%%%%%%%%%%%%%%%%%%%%%%%%%%%%%%%%%%%%%%%%%%%%%%%%%%%%%%%%%%%%%%%%%%% necessity
\begin{lemma}\label{lem:QSnec}
	Every quasi-sequential equilibrium $\sigma^*$ of an extensive form mechanism satisfies $A^*\subseteq\FLR$.
\end{lemma}
%%%%%%%%%%%%%%%%%%%%%%%%%%%%%%%%%%%%%%%%%%%%%%%%%%%%%%%%%%%%%%%%%%%%%%%%%%%%%%%%%%%%%%%%%%%%%%%%%%%%%%%%%%%%%%%%%%%%%%%%%%%%%%%%%%%%%%%%%
%%%%%%%%%%%%%%%%%%%%%%%%%%%%%%%%%%%%%%%%%%%%%%%%%%%%%%%%%%%%%%%%%%%%%%%%%%%%%%%%%%%%%%%%%%%%%%%%%%%%%%%%%%%%%%%%%%%%%%%%%%%%%%%%%%%%%%%%%
\begin{BSproof}
	Fix some $a_i^*\in A_i^*$ chosen with positive probability in some  $\h\in\H_i$,
	and a move  $\move^{a_i^*}\in\Moves^{a_i^*}(\h)$ that represents $a_i^*$ and is chosen with positive probability.
	%%%
	For each other action $a_i' \neq a_i^*$, pick a move $\move^{a_i'}\in\Moves^{a_i}(\h)$ representing $a_i'$ at $\h$.
	%%%
	Now let $\mu = \psi_i^*\big(\Phi^{\h}\,\big|\,\h\big)\in[0,1]$,
	where $\Phi^{\h}$ is the event that  $\tra_{-i}$ is already determined at $\h$, as defined in (\ref{eqn:Phi}).
	%%%
	Finally, let $\con_i^0\in\Delta(A_{-i})$ and $\con^1_i\in\Con_i(A^*)$ be the given by:
	%%%	
	\begin{align*}
		\con^0_i(a_{-i}) = \disZ_i^*\left(\trainv_{-i}(a_{-i}) \,\big|\, \h,\Phi^{\h}\right)
		\qquad\wedge\qquad
		\con^1_i(a_{-i}|a_i) = \disZ^*_i\left(\trainv_{-i}(a_{-i}) \,\big|\, \h,\move^{a_i},\bar\Phi^{\h}\right)
	\end{align*}
	%%%
	and let $\con_i=\mu\con_i^0 + (1-\mu)\con_i^1$.
	
%%%%%%%%%%%%%%%%%%%%%%%%%%%%%%%%%%%%%%%%%%%%%%%%%%%%%%%%%%%%%%%%%%%%%%%%%%%%%%%%%%%%%%%%%%%%%%%%%%%%%%%%%%%%%%%%%%%%%%%%%%%%%%%%%%%%%%%%%
%%%%%%%%%%%%%%%%%%%%%%%%%%%%%%%%%%%%%%%%%%%%%%%%%%%%%%%%%%%%%%%%%%%%%%%%%%%%%%%%%%%%%%%%%%%%%%%%%%%%%%%%%%%%%%%%%%%%%%%%%%%%%%%%%%%%%%%%%
	Sequential rationality together with the fact that $\disZ_i^*\big(\Phi^{\h}\,\big|\,\h,\move\big)$
	and $\disZ_i^* \big(\tra^{-1}_{-i}(a_{-i})|\h,\move,\Phi^{\h}\big)$
	are independent from $\move$, imply that for every deviation $a_i'$:
	%%%
	\begin{align*}
		\sum_{a_{-i}\in A_{-i}} \con_i(a_{-i} \,|\, a_i^*) \ua_i(a_i^*,a_{-i})
			   =& \sum_{a_{-i}\in A_{-i}} \disZ^*_i\left(\trainv_{-i}(a_{-i}) \,|\, \h,\move^{\acs_\pl}\right)\ua_i(a_i^*,a_{-i}) \\
			\geq& \sum_{a_{-i}\in A_{-i}} \disZ^*_i\left(\trainv_{-i}(a_{-i}) \,|\, \h,\move^{\aci_\pl}\right)\ua_i(a_i',a_{-i})\\
			   =& \sum_{a_{-i}\in A_{-i}} \con_i(a_{-i} \,|\, a_i')\ua_i(a_i',a_{-i})
	\end{align*}
	%%%
	Hence $a_i^*$ is a best response to $\con_i^*\in\Lambda_i(A^*)$, and thus $a^*_i\in\FR(\Acs)$.
	%%%
	This holds for all $i$ and $a_i^*\in A_i^*$.
	%%%
	Hence, $A^*\subseteq\FR(A^*)$ and thus $A^*\subseteq \FLR$.
\end{BSproof}
%%%%%%%%%%%%%%%%%%%%%%%%%%%%%%%%%%%%%%%%%%%%%%%%%%%%%%%%%%%%%%%%%%%%%%%%%%%%%%%%%%%%%%%%%%%%%%%%%%%%%%%%%%%%%%%%%%%%%%%%%%%%%%%%%%%%% QED

%%%%%%%%%%%%%%%%%%%%%%%%%%%%%%%%%%%%%%%%%%%%%%%%%%%%%%%%%%%%%%%%%%%%%%%%%%%%%%%%%%%%%%%%%%%%%%%%%%%%%%%%%%%%%%%%%%%%%%%%%%%%%%%%%%%%%%%%%
%%%%%%%%%%%%%%%%%%%%%%%%%%%%%%%%%%%%%%%%%%%%%%%%%%%%%%%%%%%%%%%%%%%%%%%%%%%%%%%%%%%%%%%%%%%%%%%%%%%%%%%%%%%%%%%%%%%%%%%%%%%%%%%%%%%%%%%%%
%%%%%%%%%%%%%%%%%%%%%%%%%%%%%%%%%%%%%%%%%%%%%%%%%%%%%%%%%%%%%%%%%%%%%%%%%%%%%%%%%%%%%%%%%%%%%%%%%%%%%%%%%%%%%%%%%%%%%%%%%%%%%%%%%%%%%%%%%
%%%%%%%%%%%%%%%%%%%%%%%%%%%%%%%%%%%%%%%%%%%%%%%%%%%%%%%%%%%%%%%%%%%%%%%%%%%%%%%%%%%%%%%%%%%%%%%%%%%%%%%%%%%%%%%%%%%%%%%%%%%%%%%%%%%%%%%%%
%%%%%%%%%%%%%%%%%%%%%%%%%%%%%%%%%%%%%%%%%%%%%%%%%%%%%%%%%%%%%%%%%%%%%%%%%%%%%%%%%%%%%%%%%%%%%%%%%%%%%%%%%%%%%%%%%%%%%%%%%%%%% sufficiency
The sufficiency proof is constructive, and the mechanics behind the construction are as follows.
%%% 
Every action $a_i^0 \in \FLR_i$ can be rationalized by some beliefs about future choices in $\FLR_{-i}$
and about arbitrary equilibrium or arbitrary past choices.
%%%
Off path beliefs are assigned in such a way that, whenever $i$ is asked to choose $a_i^0$,
he naively believes that doing so is in his best interest.
%%%
Since weak consistency does not imply any consistency requirements \emph{across} players,
this can always be done even if it implies that $i$ \emph{must be certain that his opponent is or will be mistaken}.

%%%%%%%%%%%%%%%%%%%%%%%%%%%%%%%%%%%%%%%%%%%%%%%%%%%%%%%%%%%%%%%%%%%%%%%%%%%%%%%%%%%%%%%%%%%%%%%%%%%%%%%%%%%%%%%%%%%%%%%%%%%%%%%%%%%%%%%%%
%%%%%%%%%%%%%%%%%%%%%%%%%%%%%%%%%%%%%%%%%%%%%%%%%%%%%%%%%%%%%%%%%%%%%%%%%%%%%%%%%%%%%%%%%%%%%%%%%%%%%%%%%%%%%%%%%%%%%%%%%%%%%%%%%%%%%%%%%
\begin{BSproof*}[Proof of sufficiency for Theorem \ref{thm:QSI}]
	Fix an equilibrium $\alpha\in\ICE(\FLR)$.
	%%%
	I will construct an extensive form mechanism $(G,\tra)$ and a QSE $(\sigma^*,\psi)$ implementing it.
	%%%
	As an intermediate step, let $G^0$ denote the mediated game which implements $\alpha$
	as an ICE with respect to $\FLR$.
	%%%
	I will add additional off-path histories to guarantee that the equilibrium becomes QS. 
	%%%
	Since equilibrium path remains unchaged, 
	it is sufficient to ensure that sequential rationality off the equilibrium path, 
	and that the off-path beliefs are weakly consistent. 

%%%%%%%%%%%%%%%%%%%%%%%%%%%%%%%%%%%%%%%%%%%%%%%%%%%%%%%%%%%%%%%%%%%%%%%%%%%%%%%%%%%%%%%%%%%%%%%%%%%%%%%%%%%%%%%%%%%%%%%%%%%%%%%%%%%%%%%%%
%%%%%%%%%%%%%%%%%%%%%%%%%%%%%%%%%%%%%%%%%%%%%%%%%%%%%%%%%%%%%%%%%%%%%%%%%%%%%%%%%%%%%%%%%%%%%%%%%%%%%%%%%%%%%%%%%%%%%%%%%%%%%%%%%%%%%%%%%
	In the construction, all the players' information sets are pivotal 
	and have a unique pivotal move representing each action, and all the moves in each pivotal information set are pivotal, i.e.\
	$\Moves(\h) = \cup_{a_i \in A_i} \Moves^{a_i}(\h)$ and $\# \Moves(\h)= \# A_i$ for $\h\in\H_i$.
	%%%
	Furthermore, the only information that a player has at the moment of making his choice
	is the action that he is supposed to choose.
	%%%
	Hence, it is possible to specify equilibrium strategies by labelling each information set
	with the distribution of actions that the corresponding player is supposed to follow. 
	%%%	
	For instance $\h^{a_i}$ represents a pivotal information set in which, according to $\sigma^*$, 
	$i$ chooses the only move which represents $a_i$ in $\h^{a_i}$.

%%%%%%%%%%%%%%%%%%%%%%%%%%%%%%%%%%%%%%%%%%%%%%%%%%%%%%%%%%%%%%%%%%%%%%%%%%%%%%%%%%%%%%%%%%%%%%%%%%%%%%%%%%%%%%%%%%%%%%%%%%%%%%%%%%%%%%%%%
%%%%%%%%%%%%%%%%%%%%%%%%%%%%%%%%%%%%%%%%%%%%%%%%%%%%%%%%%%%%%%%%%%%%%%%%%%%%%%%%%%%%%%%%%%%%%%%%%%%%%%%%%%%%%%%%%%%%%%%%%%%%%%%%%%%%%%%%%
	Fix a player $i$ and some action $a_i^0 \in \FLR_i \backslash \supp(\alpha_i)$.
	%%%
	Since $\FLR$ is self-FC-rationalizable,
	$a_i^0$ is a best response to some counterfactual belief $\con_i = (1-\mu)\con_i^0 + \mu \con_i^3$,
	with $\mu\in[0,1]$, $\con_i^0 \in \Delta(A_{-i})$ and $\con_i^3 \in \Con_i(\FLR)$.
	%%%
	$(1-\mu)\con_i^0$ can be further decomposed as $(1-\mu)\con_i^0 = \gamma \con^1_i + \eta \con^2_i$
	with  $\gamma,\eta\in[0,1]$, $\con_i^1 \in \Delta(A_{-i}\backslash \FLR_{-i})$ and $\con^2_i \in \Delta(\FLR_{-i})$.
	%%%
	Assume without loss of generality that $\con_i^3(\bar{a}_{-i}|a_i^0) = 1$ 
	and $\con^3_i(\ushort{a}_{-i}(a_i')|a_i')=1$ for every $a_i'\neq a_i^0$,
	where $\bar{a}_{-i} \in \argmax_{a_{-i}\in \FLR_{-i}}\{ \ua_i(a_i^0,a_{-i})\}$ 
	and $\ushort{a}_{-i}(a_{-i}) \in \argmin_{a_{-i}\in \FLR_{-i}}\{ \ua_i(a_i',a_{-i})\}$.

%%%%%%%%%%%%%%%%%%%%%%%%%%%%%%%%%%%%%%%%%%%%%%%%%%%%%%%%%%%%%%%%%%%%%%%%%%%%%%%%%%%%%%%%%%%%%%%%%%%%%%%%%%%%%%%%%%%%%%%%%%%%%%%%%%%%%%%%%
%%%%%%%%%%%%%%%%%%%%%%%%%%%%%%%%%%%%%%%%%%%%%%%%%%%%%%%%%%%%%%%%%%%%%%%%%%%%%%%%%%%%%%%%%%%%%%%%%%%%%%%%%%%%%%%%%%%%%%%%%%%%%%%%%%%%%%%%%
	The entire mechanism starts from an initial node where Nature chooses between $G^0$ and other additional paths. 
	%%%
	For each action $a^0_i\in \FR_i^\infty \backslash \supp(\alpha_i)$,
	$G^{a_i^0}$ denotes a set of paths on which player $i$ is willing to choose $a_i$
	and believe that the future choices of his opponents will be restricted to $\FR_{-i}^\infty$.
	%%%
	The set of paths $G^{a_i^0}$ is depicted in Figure \ref{fig:QSI}.
	%%%
	The nodes are labelled with circled numbers, 
	and the player moving at each node can be inferred from the subindexes of the information sets. 

%%%%%%%%%%%%%%%%%%%%%%%%%%%%%%%%%%%%%%%%%%%%%%%%%%%%%%%%%%%%%%%%%%%%%%%%%%%%%%%%%%%%%%%%%%%%%%%%%%%%%%%%%%%%%%%%%%%%%%%%%%%%%%%%%%%%%%%%%
%%%%%%%%%%%%%%%%%%%%%%%%%%%%%%%%%%%%%%%%%%%%%%%%%%%%%%%%%%%%%%%%%%%%%%%%%%%%%%%%%%%%%%%%%%%%%%%%%%%%%%%%%%%%%%%%%%%%%%%%%%%%%%%%%%%%%%%%%
	\begin{figure}[htb]
	\centering
	\psset{unit=9.5mm}%,arrowsize=7pt,linewidth=0.8pt}
	\begin{pspicture}(0,0)(15.5,9)
		%\psgrid%
		\scriptsize%
		% Nodes
		\psset{dotsize=4pt}
			\psdots(14,0)(12,0)(14,2)(12,2)(10,2)(8,2)(6,2)(13,4)(10,4)(8,4)(6,4)(10,6)(7,6)(2,6)(10,8)
		\psreset%
		% Branches
			\psline(13,4)(13,6)(10,8)(10,6)(10,8)(7,6)(10,8)(2,6)
			\psline(6,4)(7,6)(8,4) \psline(10,6)(10,4)
			\psline(10,4)(10,2) \psline(8,4)(8,2) \psline(6,4)(6,2) \psline(12,2)(13,4)(14,2)
			\psline(12,2)(12,0) \psline(14,2)(14,0)
		% Information sets
		\psset{linestyle=dashed}
			\psline[linearc=1](2,5)(6,6)(7,6)(10,6)
			\psline(8,4)(10,4)(13,4)(14,4)
			\psline(14,2)(15,2)
			\psline(12,2)(11,2)
			\psline(6,4)(5,4)
		\psreset%
		% Solid objects
		\psset{fillcolor=white,fillstyle=solid}
			\pswedge(2,6){4}{240}{300}
		\psset{linestyle=dashed}
			\pscircle(6,6){0.5} \pscircle(14,4){0.5} \pscircle(15,2){0.5} \pscircle(11,2){0.5} \pscircle(5,4){0.5}
		\psreset%
		% Node Labels
		\psset{fillstyle=solid,fillcolor=white}
			\cput(07,6){1} \cput(10,6){2}
			\cput(06,4){3} \cput(08,4){4} \cput(10,4){5} \cput(13,4){6}
			\cput(12,2){7} \cput(14,2){8}
		\psreset%
		% Block labels
			\rput(5,4){$\h_i^{a_i^{\mathrm{BR}}}$}
			\rput(6,6){$\h_{-i}^{\con_i^2}$}
			\rput(14,4){$\h_i^{a_i^0}$}
			\rput(15,2){$\h^{\ushort{a}_{-i}}_\plo$}
			\rput(11,2){$\h^{\bar{a}_{-i}}_\plo$}
			\rput(2,3){\parbox{30mm}{\centering Equilibrium path\\ and other deviations}}
		% Action labels
			\rput[br](9.9,8.1){$0$}
			\rput[br](12.3,2.7){$a_i^0$} \rput[bl](13.7,2.7){$a_i'\neq a_i^0$}
			\rput[br](5.9,7.1){$\big[1 - \epsilon_n\gamma/N_i - \epsilon_n^2\eta/N_i - \epsilon_n^2\mu/N_i \big]$}
			\rput[br](7.7,6.6){$\big[ \epsilon_n \gamma/N_i \big]$}
			\rput[br](9.9,6.6){$\big[ \epsilon^2_n \eta/N_i \big]$}
			\rput[bl](12.3,6.6){$\big[ \epsilon^2_n \mu/N_i \big]$}
			\rput[br](6.3,4.7){$\big[ 1 - \epsilon_n \gamma/N_i \big]$}
			\rput[bl](7.7,4.7){$\big[ \con^1 \epsilon_n \gamma/N_i \big]$}
	\end{pspicture}
	\caption{Incentives for $a_i^0 \in \FLR_i \backslash A^*_i$}\label{fig:QSI}
	\end{figure}

%%%%%%%%%%%%%%%%%%%%%%%%%%%%%%%%%%%%%%%%%%%%%%%%%%%%%%%%%%%%%%%%%%%%%%%%%%%%%%%%%%%%%%%%%%%%%%%%%%%%%%%%%%%%%%%%%%%%%%%%%%%%%%%%%%%%%%%%%
%%%%%%%%%%%%%%%%%%%%%%%%%%%%%%%%%%%%%%%%%%%%%%%%%%%%%%%%%%%%%%%%%%%%%%%%%%%%%%%%%%%%%%%%%%%%%%%%%%%%%%%%%%%%%%%%%%%%%%%%%%%%%%%%%%%%%%%%%
	The numbers within brackets, specify the sequence of mixed strategies that converges to the equilibrium assessment.
	%%%
	$(\epsilon_n)$ denotes an arbitrary sequence of sufficiently small positive numbers converging to $0$,
	%%%
	and $N_i = \# \FLR_i$ is the number of FC-rationalizable actions.
	%%%
	The sequence is not strictly mixed, but reach all the relevant information sets with positive probability.%
		\footnote{One could use a strictly mixed sequences by assigning probabilities or order $\epsilon_n^3$ or less to other strategies, 
			but this would only complicate the exposition unnecessarily.}
	%%%
	The limit of this sequence generates weakly consistent beliefs.
	%%%
	Hence, it only remains to verify the incentive constraints:
%%%%%%%%%%%%%%%%%%%%%%%%%%%%%%%%%%%%%%%%%%%%%%%%%%%%%%%%%%%%%%%%%%%%%%%%%%%%%%%%%%%%%%%%%%%%%%%%%%%%%%%%%%%%%%%%%%%%%%%%%%%%%%%%%%%%%%%%%
%%%%%%%%%%%%%%%%%%%%%%%%%%%%%%%%%%%%%%%%%%%%%%%%%%%%%%%%%%%%%%%%%%%%%%%%%%%%%%%%%%%%%%%%%%%%%%%%%%%%%%%%%%%%%%%%%%%%%%%%%%%%%%%%%%%%%%%%%
\begin{itemize}
	%%%%%%%%%%%%%%%%%%%%%%%%%%%%%%%%%%%%%%%%%%%%%%%%%%%%%%%%%%%%%%%%%%%%%%%%%%%%%%%%%%%%%%%%%%%%%%%%%%%%%%%%%%%%%%%%%%%%%%%%%%%%%%%%%%%%%
	\item At nodes (1) and (2), player $-i$ is willing to make choices according to $\con_i^2$ 
		because he believes that he is on the equilibrium path. 
	%%%%%%%%%%%%%%%%%%%%%%%%%%%%%%%%%%%%%%%%%%%%%%%%%%%%%%%%%%%%%%%%%%%%%%%%%%%%%%%%%%%%%%%%%%%%%%%%%%%%%%%%%%%%%%%%%%%%%%%%%%%%%%%%%%%%%
	\item At nodes (7) and (8), $\bar{a}_{-i}$ and $\ushort{a}_{-i}$ may not be best responses to $a_i^0$ or $a_i'$.
		%%%
		However, they are in $\FR_{-i}^\infty$ and thus $-i$ is willing to play them either along the equilibrium path,
		or on $G^{\bar{a}_{-i}}$ and $G^{\ushort{a}_{-i}}$.
		%%%
		Since $-i$ will consider the deviations to and in $G^{a_i^0}$ to be unlikely (of order at most $\epsilon^3$),
		the incentives for these actions are independent from what happens in this figure. 
	%%%%%%%%%%%%%%%%%%%%%%%%%%%%%%%%%%%%%%%%%%%%%%%%%%%%%%%%%%%%%%%%%%%%%%%%%%%%%%%%%%%%%%%%%%%%%%%%%%%%%%%%%%%%%%%%%%%%%%%%%%%%%%%%%%%%%
	\item First suppose that the information sets for $i$ are fully contained in the figure:
		\begin{itemize}
		\item At (3) player $i$ is supposed to choose an action which is a best response to $\con_i^2$.
			And therefore his choice is trivially incentive compatible. 
		\item It is straightforward to see that equilibrium beliefs for player $i$ would generate a conjecture $\con_i$ at nodes (4)--(6),
		 	and thus he would be willing to choose $a_i^0$.
		\end{itemize}
	%%%%%%%%%%%%%%%%%%%%%%%%%%%%%%%%%%%%%%%%%%%%%%%%%%%%%%%%%%%%%%%%%%%%%%%%%%%%%%%%%%%%%%%%%%%%%%%%%%%%%%%%%%%%%%%%%%%%%%%%%%%%%%%%%%%%%
	\item Now suppose that either $\h^{a_i^0}$ or $\h^{a_i^\mathrm{BR}}$ appear in other parts of the game. 
		There are only two possibilities:
		\begin{itemize}
		\item They could appear as punishments in the position analogous to (7) or (8) in some $G^{a_{-i}^0}$.
			%%%
			From $i$'s perspective, this has probability of order $\epsilon^3$ or lower, and hence it is irrelevant for $i$.
		\item They could appear in the equilibrium path, or in some $G^{a_{-i}^0}$ in the positions of (3) - (8).
			In such cases, it will also be a best response to the conditional beliefs and thus to the average beliefs.
		\end{itemize}
	\end{itemize}
\end{BSproof*}
%%%%%%%%%%%%%%%%%%%%%%%%%%%%%%%%%%%%%%%%%%%%%%%%%%%%%%%%%%%%%%%%%%%%%%%%%%%%%%%%%%%%%%%%%%%%%%%%%%%%%%%%%%%%%%%%%%%%%%%%%%%%%%%%%%%%% QED

%%%%%%%%%%%%%%%%%%%%%%%%%%%%%%%%%%%%%%%%%%%%%%%%%%%%%%%%%%%%%%%%%%%%%%%%%%%%%%%%%%%%%%%%%%%%%%%%%%%%%%%%%%%%%%%%%%%%%%%%%%%%%%%%%%%%%%%%%
%%%%%%%%%%%%%%%%%%%%%%%%%%%%%%%%%%%%%%%%%%%%%%%%%%%%%%%%%%%%%%%%%%%%%%%%%%%%%%%%%%%%%%%%%%%%%%%%%%%%%%%%%%%%%%%%%%%%%%%%%%%%%%%%%%%%%%%%%
%%%%%%%%%%%%%%%%%%%%%%%%%%%%%%%%%%%%%%%%%%%%%%%%%%%%%%%%%%%%%%%%%%%%%%%%%%%%%%%%%%%%%%%%%%%%%%%%%%%%%%%%%%%%%%%%%%%%%%%%%%%%%%%%%%%%%%%%%
%%%%%%%%%%%%%%%%%%%%%%%%%%%%%%%%%%%%%%%%%%%%%%%%%%%%%%%%%%%%%%%%%%%%%%%%%%%%%%%%%%%%%%%%%%%%%%%%%%%%%%%%%%%%%%%%%%%%%%%%%%%%%%%%%%%%%%%%%
%%%%%%%%%%%%%%%%%%%%%%%%%%%%%%%%%%%%%%%%%%%%%%%%%%%%%%%%%%%%%%%%%%%%%%%%%%%%%%%%%%%%%%%%%%%%%%%%%%%%%%%%%%%%%%%%%%%%%%%%%%%%%%%%%%%%%%%%%
%%%%%%%%%%%%%%%%%%%%%%%%%%%%%%%%%%%%%%%%%%%%%%%%%%%%%%%%%%%%%%%%%%%%%%%%%%%%%%%%%%%%%%%%%%%%%%%%%%%%%%%%%%%%%%%%%%%%%%%%%%%%%%%%%%%%%%%%%
%%%%%%%%%%%%%%%%%%%%%%%%%%%%%%%%%%%%%%%%%%%%%%%%%%%%%%%%%%%%%%%%%%%%%%%%%%%%%%%%%%%%%%%%%%%%%%%%%%%%%%%%%%%%%%%%%%%%%%%%%%%%%%%%%%%%%%%%%
%%%%%%%%%%%%%%%%%%%%%%%%%%%%%%%%%%%%%%%%%%%%%%%%%%%%%%%%%%%%%%%%%%%%%%%%%%%%%%%%%%%%%%%%%%%%%%%%%%%%%%%%%%%%%%%%%%%%%%%%%%%%%%%%%%%%%%%%%
%%%%%%%%%%%%%%%%%%%%%%%%%%%%%%%%%%%%%%%%%%%%%%%%%%%%%%%%%%%%%%%%%%%%%%%%%%%%%%%%%%%%%%%%%%%%%%%%%%%%%%%%%%%%%%%%%%%%%%%%%%%%%%%%%%%%%%%%%
%%%%%%%%%%%%%%%%%%%%%%%%%%%%%%%%%%%%%%%%%%%%%%%%%%%%%%%%%%%%%%%%%%%%%%%%%%%%%%%%%%%%%%%%%%%%%%%%%%%%%%%%%%%%%%%%%%%%%%%%%% END OF SECTION

%% file: Salcedo.bbl
\begin{thebibliography}{}

\bibitem[\protect\astroncite{Abreu et~al.}{1990}]{APS}
Abreu, D., Pearce, D., and Stacchetti, E. (1990).
\newblock Towards a theory of discounted repeated games with imperfect
  monitoring.
\newblock {\em Econometrica}, 58(5):1041--1063.

\bibitem[\protect\astroncite{Aumann and Brandenburger}{1995}]{aumann95}
Aumann, R. and Brandenburger, A. (1995).
\newblock Epistemic conditions for nash equilibrium.
\newblock {\em Econometrica}, 63(5):1161--1180.

\bibitem[\protect\astroncite{Aumann}{1987}]{aumann87}
Aumann, R.~J. (1987).
\newblock Correlated equilibrium as an expression of {B}ayesian rationality.
\newblock {\em Econometrica}, 55(12):1--18.

\bibitem[\protect\astroncite{Bade et~al.}{2009}]{bade09}
Bade, S., Guillaume, H., and Renou, L. (2009).
\newblock Bilateral commitment.
\newblock {\em Journal of Economic Theory}, 144(4):1817--1831.

\bibitem[\protect\astroncite{Brams}{1993}]{brams93}
Brams, S.~J. (1993).
\newblock {\em Theory of moves}.
\newblock Cambridge University Press.

\bibitem[\protect\astroncite{Calcagno et~al.}{2012}]{calcagno12}
Calcagno, R., Kamada, Y., Lovo, S., and Sugaya, T. (2012).
\newblock Asynchronicity and coordination in common and opposing interest
  games.
\newblock mimeo.

\bibitem[\protect\astroncite{Eisert et~al.}{1999}]{eisert99}
Eisert, J., Wilkens, M., and Lewenstein, M. (1999).
\newblock Quantum games and quantum strategies.
\newblock {\em Physical Review Letters}, 83(15):3077--3080.

\bibitem[\protect\astroncite{Forges}{1986}]{forges86}
Forges, F. (1986).
\newblock An approach to communication equilibria.
\newblock {\em Econometrica}, 54(6):1375--1385.

\bibitem[\protect\astroncite{Forg\'{o}}{2010}]{forgo10}
Forg\'{o}, F. (2010).
\newblock A generalization of correlated equilibrium: a new protocol.
\newblock {\em {M}athematical {S}ocial {S}ciences}, 60(1):186--190.

\bibitem[\protect\astroncite{Fudenberg and Levine}{1993}]{FudLev}
Fudenberg, D. and Levine, D.~K. (1993).
\newblock Self-confirming equilibrium.
\newblock {\em Econometrica}, 61(3):523--545.

\bibitem[\protect\astroncite{Gibbard and Harper}{1980}]{GibHar}
Gibbard, A. and Harper, W.~L. (1980).
\newblock Counterfactuals and two kinds of expected utility.
\newblock In Harper, W.~L., Stalnaker, R., and Pearce, G., editors, {\em Ifs,
  conditionals, beliefs, decision, chance and time}. Springer.

\bibitem[\protect\astroncite{Halpern and Pass}{2012}]{halpern12}
Halpern, J.~Y. and Pass, R. (2012).
\newblock Game theory with translucent players.
\newblock Cornell Univeristy, mimeo.

\bibitem[\protect\astroncite{Halpern and Rong}{2010}]{halpern10}
Halpern, J.~Y. and Rong, N. (2010).
\newblock Cooperative equilibrium.
\newblock In {\em Proceedings of the 9th International Conference on Autonomous
  Agents and Multiagent Systems}.

\bibitem[\protect\astroncite{Hofstadter}{1996}]{hofstadter96}
Hofstadter, D. (1996).
\newblock {\em Metamagical themas: Questing for the essence of mind and
  pattern}.
\newblock Basic books.

\bibitem[\protect\astroncite{Howard}{1971}]{howard71}
Howard, N. (1971).
\newblock {\em Paradoxes of rationality: theory of metagames and political
  behavior}.
\newblock MIT Press.

\bibitem[\protect\astroncite{Izmalkov et~al.}{2005}]{izmalkov05}
Izmalkov, S., Micali, S., and Lepinski, M. (2005).
\newblock Rational secure computation and ideal mechanism design.
\newblock In {\em Proceedings of the 46th Annual IEEE Symposium on Foundations
  of Computer Science}, pages 585--594.

\bibitem[\protect\astroncite{Jackson and Wilkie}{2005}]{jackson05}
Jackson, M.~O. and Wilkie, S. (2005).
\newblock Endogenous games and mechanisms: side payments among players.
\newblock {\em The Review of Economic Studies}, 72(2):543--566.

\bibitem[\protect\astroncite{Kalai et~al.}{2010}]{kalai10}
Kalai, A.~T., Kalai, E., Lehrer, E., and Samet, D. (2010).
\newblock A commitment folk theorem.
\newblock {\em Games and Economic Behavior}, 69(1):123--137.

\bibitem[\protect\astroncite{Kalai}{1981}]{kalai81}
Kalai, E. (1981).
\newblock Preplay negotiations and the prisoner's dilemma.
\newblock {\em {M}athematical {S}social {S}ciences}, 1(4):375--379.

\bibitem[\protect\astroncite{Kalai and Stanford}{1985}]{kalSta}
Kalai, E. and Stanford, W.~G. (1985).
\newblock Conjectural variations strategies in accelerated cournot games.
\newblock {\em International Journal of Industrial Organization},
  3(2):133--152.

\bibitem[\protect\astroncite{Kamada and Kandori}{2009}]{KamKan09}
Kamada, Y. and Kandori, M. (2009).
\newblock Revision games.
\newblock Harvard University, mimeo.

\bibitem[\protect\astroncite{Kamada and Kandori}{2012}]{KamKan12}
Kamada, Y. and Kandori, M. (2012).
\newblock Asynchronous revision games.
\newblock Harvard University, mimeo.

\bibitem[\protect\astroncite{Kreps and Wilson}{1982}]{KreWil}
Kreps, D.~M. and Wilson, R. (1982).
\newblock Sequential equilibria.
\newblock {\em Econometrica}, 50(4):863--894.

\bibitem[\protect\astroncite{Levine and Pesendorfer}{2007}]{levine07}
Levine, D.~K. and Pesendorfer, W. (2007).
\newblock The evolution of cooperation through imitation.
\newblock {\em Games and Economic Behavior}, 58(2):293--315.

\bibitem[\protect\astroncite{Lewis}{1979}]{lewis79}
Lewis, D. (1979).
\newblock Prisoner's dilemma is a {N}ewcomb problem.
\newblock {\em {P}hilosophy and {P}ublic {A}ffairs}, 8(3):235--240.

\bibitem[\protect\astroncite{Moulin and Vial}{1978}]{MouVia}
Moulin, H. and Vial, J. (1978).
\newblock Strategically zero-sum games: the class of games whose completely
  mixed equilibria cannot be improved upon.
\newblock {\em International Journal of Game Theory}, 7(3-4):201--221.

\bibitem[\protect\astroncite{Myerson}{1986}]{myerson86}
Myerson, R.~B. (1986).
\newblock Multistage games with communication.
\newblock {\em Econometrica}, 54(2):323--358.

\bibitem[\protect\astroncite{Nishihara}{1997}]{nishihara97}
Nishihara, K. (1997).
\newblock A resolution of n-person prisoners' dilemma.
\newblock {\em Economic Theory}, 10(3):531--540.

\bibitem[\protect\astroncite{Nishihara}{1999}]{nishihara99}
Nishihara, K. (1999).
\newblock Stability of the cooperative equilibrium in n-person prisoners'
  dilemma with sequential moves.
\newblock {\em Economic Theory}, 13(2):483--494.

\bibitem[\protect\astroncite{Osborne and Rubinstein}{1994}]{OsbRub}
Osborne, M.~J. and Rubinstein, A. (1994).
\newblock {\em A course in game theory}.
\newblock MIT Press.

\bibitem[\protect\astroncite{Pearce}{1984}]{pearce}
Pearce, D. (1984).
\newblock Rationalizable strategic behavior and the problem of perfection.
\newblock {\em Econometrica}, 52(4):1029--1050.

\bibitem[\protect\astroncite{Perea}{2013}]{perea13}
Perea, A. (2013).
\newblock Belief in the opponents' future rationality.
\newblock Maastricht University, mimeo.

\bibitem[\protect\astroncite{Rapoport}{1965}]{RapAl}
Rapoport, A. (1965).
\newblock {\em Prisoner's dilemma}.
\newblock {U}niversity of {M}ichigan {P}ress.

\bibitem[\protect\astroncite{Renou}{2009}]{renou09}
Renou, L. (2009).
\newblock Commitment games.
\newblock {\em Games and Economic Behavior}, 66(1):488--505.

\bibitem[\protect\astroncite{Solan and Yariv}{2004}]{solan04}
Solan, E. and Yariv, L. (2004).
\newblock Games with espionage.
\newblock {\em Games and Economic Behavior}, 47(1):172--199.

\bibitem[\protect\astroncite{Tennenholtz}{2004}]{tennenholtz04}
Tennenholtz, M. (2004).
\newblock Program equilibrium.
\newblock {\em Games and Economic Behavior}, 49(2):363--373.

\bibitem[\protect\astroncite{Van~Damme and Hurkens}{1999}]{vandamme}
Van~Damme, E. and Hurkens, S. (1999).
\newblock Endogenous stackelberg leadership.
\newblock {\em Games and Economic Behavior}, 28(1):105--129.

\end{thebibliography}
